\documentclass[fleqn,usenatbib]{mnras} 

\usepackage{newtxtext,newtxmath} 
\usepackage[T1]{fontenc} 

\usepackage{booktabs}
\usepackage{float}
\usepackage{amsmath}
\usepackage{multirow}
\usepackage{makecell}
\usepackage{rotating}
\usepackage[nolist,nohyperlinks]{acronym}
\usepackage{xcolor}

\newcolumntype{x}[1]{>{\centering\arraybackslash\hspace{0pt}}p{#1}}

\DeclareRobustCommand{\VAN}[3]{#2}
\let\VANthebibliography\thebibliography
\def\thebibliography{\DeclareRobustCommand{\VAN}[3]{##3}\VANthebibliography} 

\usepackage{graphicx}	

\graphicspath{{./}{figures/}}

\newcommand{\multrow}[1]{\begin{tabular}{@{}c@{}} #1 \end{tabular}}

\title[3D code MATINS]{3D code for MAgneto-Thermal evolution in Isolated Neutron Stars, MATINS: thermal evolution and lightcurves} 

\author[S. Ascenzi et al.]{
Stefano Ascenzi,$^{1,2,3}$\thanks{E-mail: stefano.ascenzi@gssi.it}
Daniele Vigano,$^{1,2,4}$
Clara Dehman,$^{1,2,5,6}$
Jos\'e A. Pons,$^{6}$
Nanda Rea,$^{1,2}$
Rosalba Perna$^{7,8}$
\\
$^{1}$Institute of Space Sciences (ICE-CSIC), Campus UAB, Carrer de Can Magrans s/n, Cerdanyola del Vall\`es (Barcelona) 08193, Spain\\
$^{2}$Institut d’Estudis Espacials de Catalunya (IEEC), 08034 Barcelona, Spain\\
$^{3}$Gran Sasso Science Institute (GSSI), Viale F. Crispi 7, L'Aquila, 67100, Italy\\
$^{4}$Institute of Applied Computing \& Community Code (IAC3), University of the Balearic Islands, Palma, 07122, Spain\\
$^{5}$Nordita, KTH Royal Institute of Technology and Stockholm University, Hannes Alfv\'ens v\"ag 12, SE-10691 Stockholm, Sweden \\
$^{6}$Departament de Física Aplicada, Universitat d'Alacant, Ap. Correus 99, E-03080 Alacant, Spain\\
$^{7}$Department of Physics and Astronomy, Stony Brook University, Stony Brook, NY 11794-3800, USA\\
$^{8}$Center for Computational Astrophysics, Flatiron Institute, New York, NY 10010, USA
}

\date{Accepted XXX. Received YYY; in original form ZZZ} 

\pubyear{2023} 

\begin{document}
\label{firstpage}
\pagerange{\pageref{firstpage}--\pageref{lastpage}}
\maketitle 

\begin{abstract}

The thermal evolution of isolated neutron stars is a key element in unraveling their internal structure and composition and establishing evolutionary connections among different observational subclasses. Previous studies have predominantly focused on one-dimensional or axisymmetric two-dimensional models. In this study, we present the thermal evolution component of the novel three-dimensional magnetothermal code \emph{MATINS} (MAgneto-Thermal evolution of Isolated Neutron Star). \emph{MATINS} employs a finite volume scheme and integrates a realistic background structure, along with state-of-the-art microphysical calculations for the conductivities, neutrino emissivities, heat capacity, and superfluid gap models. 
This paper outlines the methodology employed to solve the thermal evolution equations in \emph{MATINS}, along with the microphysical implementation which is essential for the thermal component. We test the accuracy of the code and present simulations with non-evolving magnetic fields of different configurations (all with electrical currents confined to the crust and a magnetic field that does not thread the core), to produce temperature maps of the neutron star surface. Additionally, for a specific magnetic field configuration, we show one fully coupled evolution of magnetic field and temperature. Subsequently, we use a ray-tracing code to link the neutron star surface temperature maps obtained by \emph{MATINS} with the phase-resolved spectra and pulsed profiles that would be detected by distant observers. 
This study, together with our previous article focused on the magnetic formalism, presents in detail the most advanced evolutionary code for isolated neutron stars, with the aim of comparison with their timing properties, thermal luminosities and the associated X-ray light curves.

\end{abstract}



\begin{acronym}
\acrodef{NS}[NS]{neutron star}
\acrodef{TOV}[TOV]{Tolman–Oppenheimer–Volkoff}
\acrodef{EOS}[EOS]{equation of state}
\acrodef{eMHD}[eMHD]{electron magnetohydrodynamics}
\acrodef{PF}[PF]{pulsed-fraction}
\acrodef{LOS}[LOS]{line-of-sight}
\acrodef{CCSN}[CCSN]{Core-Collapse Supernoava}
\acrodefplural{CCSN}[CCSNe]{Core-Collapse Supernovae}
\end{acronym}



\section{Introduction} \label{sec:intro}

\Acp{NS} are born from the gravitational collapse of a stellar core during a supernova explosion. During the first minute of its life as a \emph{proto-\ac{NS}}, the star is hot ($\sim 10^{11}\,\mathrm{K}$), opaque to neutrinos, and undergoes a shrinking of its radius and a deleptonization, while neutrinos diffuse outward \citep{Prakash2001}. The proto-\ac{NS} cools down until, at a temperature of $\sim 10^{10}\,\mathrm{K}$, the matter becomes transparent to neutrinos.
During the subsequent secular evolution, the \ac{NS} temperature drops due to different processes, which are dominated initially by the leakage of neutrinos, during the so-called \emph{neutrino cooling era}, and later by the emission of photons from their hot surface, throughout what is known as \emph{photon cooling era} \citep[e.g.][]{Page2004, Page-etal2006, YakovlevPethick2004}. 

Studying the precise cooling history of \acp{NS} can in principle provide us with a wealth of information. For instance, during the neutrino cooling era, various neutrino emission processes, characterized by a different dependence on the temperature, coexist. The presence of a given mechanism and its relative importance with respect to the others is ultimately determined by the microphysical conditions occurring in the interior of the \ac{NS}, such as the stellar composition, the onset of a superfluid phase transition, or/and the presence of strong magnetic fields. Consequently, the cooling history of \acp{NS} can significantly enhance our understanding of their internal state. 

Moreover, there exists a tension in the fact that the estimated galactic rate of \acp{CCSN} is smaller than the sum of the birth rates of the different \ac{NS} populations \citep{KeaneKramer2008}. This poses a problem for the scenario that attributes the formation of all the \acp{NS}  
to \acp{CCSN}, as they alone could not account for the entire population of \acp{NS} in our Galaxy, if we posit that the presence of phenomenological sub-classes inherently signifies distinct types of \acp{NS}. 
Various solutions have been proposed to this issue, one of which suggests that different \ac{NS} populations are related through evolutionary paths \citep[see e.g.][]{Vigano2013}. If this holds true, we only need to consider a single birthrate for these populations, which helps resolve (or at least alleviate) the tension with \acp{CCSN} rates. Indeed, magnetothermal cooling models are capable of establishing this evolutionary link \citep[see][for a review]{PonsVigano2019}.

However, modeling the thermal evolution of \acp{NS} is not an easy task. The process is determined by several ingredients, which are at present not very well understood, such as the model of superfluidity, the composition of the core, the neutrino processes, and the model and composition of the blanketing envelope, which lies above the outer crust. Moreover, the intense magnetic field in the \acp{NS} interiors influences the transport properties of the star, making the conductivity anisotropic. 
Therefore, an accurate evolution of a strongly magnetized \ac{NS} requires a multi-dimensional approach. 
Furthermore, the dissipation of the electric currents, which leads to a decay in the magnetic field, acts like a heating source in the stellar crust. The magnetic field decay, in turn, is regulated by magnetic diffusivity (in addition to being driven by the Hall effect in the crust and by ambipolar diffusion and other little-understood mechanisms in the core), which is a parameter that depends on the local temperature. It  hence follows that the equations describing the thermal and the magnetic evolution cannot be considered independently, but must be solved together in a coupled magneto-thermal framework (see \citet{PonsVigano2019} for a comprehensive review). 

Extensive investigations spanning several decades have been devoted to unraveling the magnetothermal evolution of \acp{NS}. Initial attempts primarily delved into cooling phenomena within 1D models, with limited consideration for the magnetic field's impact \citep{Tsuruta1965, YakovlevUrpin1981, PageBaron1990, Page2004, Page2009, YakovlevPethick2004, Kaminker2008, PotekhinChabrier2018, BeznogovYakovlev2015a, BeznogovYakovlev2015b}. Advancing beyond this, subsequent studies ventured into axisymmetric, two-dimensional (2D) calculations; however, the first studies did not consider a self-consistent coupling in the magneto-thermal evolution, but rather assumed a prescription for the temperature evolution and solved for the magnetic field one \citep{PonsGeppert2007} or vice versa \citep{Aguilera2008}. 

A step forward occurred with the works by \citet{Pons2009, Vigano2012, Vigano2013}, who presented
a consistent treatment of the coupled magnetothermal evolution in a 2D setting. In particular, the later works successfully accounted for the Hall term in the induction equation, which plays a fundamental role in transferring magnetic energy 
to
smaller spatial scales where the dissipation is more effective \citep{PonsGeppert2007}, but that had posed a numerical challenge for previous codes. 

In parallel, the first simulations in 3D adapted the geo-dynamo code \emph{PARODY} \citep{parody1, parody2} to the \ac{NS} context, albeit with fixed stellar structures and simplified microphysical coefficients \citep{Wood2015, Gourgouliatos2016, Gourgouliatos2018, Gourgouliatos2020}. While these works 
focused on
the evolution of the magnetic field, \citet{DeGrandis2020, DeGrandis2021, Igoshev2021b, DeGrandis2022, Igoshev2023}, applied for the first time \emph{PARODY} to study the coupled magneto-thermal evolution problem.

While all these works have been focused on the magnetic evolution in the crust alone, progress has also been made in the study of the more complex evolution in the core \citep{graber2015,ofengeim2018,gusakov2019,castillo2020,dommes2020,wood2022}. There, a multifluid, multiscale dynamics needs to be studied, which implies the presence of ambipolar diffusion, but with non-trivial complications due to superconductivity and highly uncertain coupling coefficients between the different fluid components. The core evolution is an open problem from a theoretical point of view: 
despite the advances in understanding the relevant timescales and in implementing specific ingredients (such as ambipolar diffusion for non-superfluid/superconducting matter), no fully self-consistent induction equation nor simulations with all the ingredients exist yet.  
Nonetheless,
the general consensus is that the crustal timescales are generally much shorter than the ones for the core. Therefore, while the core evolution will determine the field in Gyr-old neutron stars (like millisecond pulsars and low-mass X-ray binaries), the coupled magneto-thermal evolution in the crust, object of this code, has a much more direct effect on the observables in young neutron stars (magnetars in particular).

The aim of this paper is to introduce the thermal part of \emph{MATINS} (MAgneto-Thermal evolution of Isolated Neutron Star), the magnetic component of which has been already presented in \citet{Dehman2023_MATINS}, with coupled magnetothermal applications in \citet{Dehman2023}. 
\emph{MATINS} differs from \emph{PARODY} in three main aspects: 1) while \emph{PARODY} is a pseudo-spectral code, namely it employs a finite grid in the radial direction and a spherical harmonic expansion in the angular directions, \emph{MATINS} uses a finite volume scheme; 2) while in PARODY the stellar background is accounted for by assuming an analytical radial profile for the electron number density (or the chemical potential, equivalently) and by defining a radius for the star and the crust-core interface, in \emph{MATINS} the stellar background and the composition is obtained by self-consistently solving the \ac{TOV} equation with the option of choosing between different \acp{EOS} among the ones present in the public database CompOSE (CompStar Online Supernovae Equations of State; \citealt{Typel2015, Oartel2017})\footnote{https://compose.obspm.fr/}. Moreover, solving for the structure allows us to compute the spacetime metric within the star, and thus include self-consistently the general relativistic effects in the magnetothermal evolution. These effects in \emph{PARODY} are neglected;   
3) The microphysics in \emph{PARODY} is described by analytic approximations 
which 
describe the relevant quantities (\emph{e.g.} heat capacity, electrical and thermal conductivities, neutrino emissivity) with simplified dependency on the temperature and the magnetic field. In particular, these analytic prescriptions aim to mimic the contribution of the electrons on the conductivity and heat capacity, which however is not always the dominant contribution. In \emph{MATINS} instead we include all the known contributions, through a numerical treatment based on
the public code developed by A. Potekhin\footnote{http://www.ioffe.ru/astro/conduct/}. This code allows us to include a detailed treatment of the relevant quantities including the contribution of different species (\emph{e.g.} electrons, ion lattice, free neutrons) to them. Furthermore, in \emph{MATINS} a proper treatment of the superfluidity is implemented, while in \emph{PARODY} this phenomenon is neglected. 

The paper is organized as follows: in Sec. \ref{sec:thermal_evolution} we introduce the thermal evolution equation in the \emph{MATINS} code, further focusing   
on the grid employed and on the description of the microphysical setup. Sec. \ref{sec:tests} is devoted to the description of the tests we performed to assess the reliability of the code. 
In Sec. \ref{sec:results} we present several runs of \emph{MATINS} with and without the magnetic field evolution. 
In this section, we also introduce a ray-tracing code that takes as input the \ac{NS}'s surface temperature map obtained by \emph{MATINS} to produce phase resolved spectra that a distant observer would detect, and present preliminary applications of this code.
Finally, in Sec. \ref{sec:conclusions} we summarize our work. 

\section{Thermal evolution in magnetized isolated neutron stars} 
\label{sec:thermal_evolution}

In this section, we present the setup of the thermal evolution part of the \emph{MATINS} code. First, we introduce the thermal evolution equation. Then, we briefly describe the cubed sphere grid that we employ in \emph{MATINS}. Finally, we illustrate two 
different microphysical setups that we have included in the code. For specific details on how the thermal evolution equation is discretized and solved in the code, we refer the reader to Appendix \ref{sec:time_discratization} and Appendix \ref{sec:ghost}. The boundary conditions are discussed in Appendix \ref{sec:boundary}.

\subsection{Background Stellar Model}
\label{sec:background}

The stellar model is computed by the equation: 

\begin{equation}
    \frac{dP}{dr} = - \Bigl(\rho + \frac{P}{c^2}\Bigr)\frac{d\nu}{dr},
    \label{eq:TOV}
\end{equation}
which describes hydrostatic equilibrium assuming an internal spherically symmetric and static metric:   
\begin{equation}
    ds^2 = -e^{2\nu(r)}c^2dt^2 + e^{2\lambda(r)}dr^2 +r^2d\Omega~,
    \label{eq:metric}
\end{equation}
where $e^{\nu(r)}$ is the lapse function, describing the gravitational redshift, and $\Omega$ is the solid angle. The functional form of $\lambda$ with radius depends on the function $m(r)$ (which represents the mass distribution inside the star in the Newtonian limit), since 
\begin{equation}
    \lambda(r) = -\frac{1}{2}\ln \Bigl[1 - \frac{2Gm(r)}{c^2r^2}\Bigr]~.
\end{equation}
In the Eq. \ref{eq:TOV}, $P$ and $\rho$ denote the pressure and the density, respectively.
The lapse function, instead, is obtained by solving the $rr$ component of the Einstein field equations, which, assuming the metric in Eq. \ref{eq:metric}, can be written as  
\begin{equation}
    \frac{d\nu}{dr} = \frac{G m(r)}{c^2 r^2}\Bigl(1 + \frac{4\pi r^3 P}{c^2 m(r)}\Bigr)\Bigl(1-\frac{2G}{c^2}\frac{m(r)}{r}\Bigr)^{-1}~,
    \label{eq:lapse_function_equation}
\end{equation}
along with the boundary condition $e^{2\nu(R)} = 1 - 2GM/(Rc^2)$, where $M$ is the star gravitational mass and $R$ is the radius of the star. In all the equations mentioned above, the constants $c$ and $G$ denote the speed of light in vacuum and the gravitational constant, respectively.

Equations \ref{eq:TOV} and \ref{eq:lapse_function_equation} need to be supplemented with the equation for the function $m(r)$:
\begin{equation}
    \frac{d m}{dr} = 4\pi r^2 \rho.
    \label{eq:continuity}
\end{equation}

Equations \ref{eq:TOV}, \ref{eq:lapse_function_equation} and \ref{eq:continuity} are known as \ac{TOV} equations \citep{Tolman1939, Oppenheimer1939}.

Solving the backgroud structure of the star via \ac{TOV} equation requires the use of an \ac{EOS} for cold dense matter, which gives the pressure and composition as a function of the density, along with the central pressure $P_0$. Currently, \emph{MATINS} includes different cold \ac{EOS} from the CompOSE database. The central pressure $P_0$ is an input parameter which regulates the star mass $M$.

Finally, above the crust lies a thin ($\sim 10-100\, \mathrm{m}$) liquid blanketing envelope layer. The steep radial gradients of density, pressure and temperature in this region imply an outward decrease of the local thermal timescales by orders of magnitude, which make it computationally unfeasible to follow the long-term (up to $\textrm{Myr}$) evolution.
For this reason, as in all cooling codes, \emph{MATINS} solves the equations for the core and the crust, down to a threshold pressure given in input and here set (as usual in these studies) to a corresponding density of $\sim 10^{10} \, \mathrm{g/cm^3}$. For the uppermost layers, \emph{MATINS} allows the use of different envelope models (see also Appendix \ref{sec:boundary}), which relate the temperature at the star surface $T_s$ to the temperature at the threshold pressure (the base of the envelope) $T_b$, and that may depend, eventually, also on the local magnetic field $\mathbf{B}$.

The dependence of the results on the envelope models have been presented for instance in \citet{Potekhin2015} and \citet{Dehman2023b}, and we will briefly show some results below.   

\subsection{Heat Diffusion Equation}
\label{sec:heat_diffusion_equation}

The heat diffusion equation determines the evolution of the thermal internal structure and surface luminosity. It can be written as follows:
\begin{equation}
c_\mathrm{v} \frac{\partial (e^\nu T)}{\partial t} + \nabla  \cdot  ( e^{2 \nu} \mathbf{F} ) = e^{2\nu} \dot{\epsilon}~, \label{eq:heat_diffusion}
\end{equation}
where $T$ is the local temperature, $c_\mathrm{v}$ is the heat capacity per unit volume, $\dot{\epsilon}$ represents the sources or losses of energy per unit volume, and $\mathbf{F}$ is the heat flux. 
From the lapse function and the temperature, we can define the redshifted temperature $\tilde{T}$ as $\Tilde{T} \equiv e^\nu T$. 

The source/loss term on the right hand side can be written as the sum of two contributions $\dot{\epsilon} = \dot{\epsilon}_h - \dot{\epsilon}_\nu$. While $\dot{\epsilon}_\nu$ represent the neutrino emissivity per unit volume, the term $\dot{\epsilon}_h$ represent the heating rate per unit volume due to Ohmic dissipation of the electric current (\emph{i.e.} the Joule heating), which writes as: 

\begin{equation}
    \dot{\epsilon}_h =  \frac{\mathbf{||J||^2}}{\sigma},
\end{equation}

where $\sigma$ is the electric conductivity, which has in the crust a  typical value in the range $\sigma \sim 10^{22}-10^{25}\,\mathrm{s}^{-1}$, and  $\mathbf{J}$ is the current density obtained, as standard in magnetohydrodynamics, by the fourth Maxwell's equation neglecting the displacement current:

\begin{equation}
    \mathbf{J} = e^{-\nu} \frac{c}{4\pi}\mathbf{\nabla} \times \bigl(e^\nu \mathbf{B}\bigr).
\end{equation}

It is worth noticing that other heating mechanisms can in principle be at play in the stellar crust, \emph{e.g.} crust-cracking \citep{BaymPines1971, Cheng1992}, non-equilibrium reactions \citep[\emph{e.g.}][]{Haensel1992, Reisenegger1995, RodrigoReisenegger2005, Flores-TulianReisenegger2006, Reisenegger2007, Gonzalez-Jimenez2015}, superfluid vortex creep motion \citep[\emph{e.g.}][]{Alpar1984, Umeda1993, LarsonLink1999, Schaab1999}; however, they are not considered in the present work.

Concerning the loss of energy due to the emission of photons at the surface, we include this contribution by the boundary conditions, as described in Appendix \ref{sec:boundary}.

Since the thermal conductivity is generally dominated by the contribution of electrons, under the presence of strong magnetic fields it becomes anisotropic, with its contribution in the direction orthogonal to the field $k_\perp$ quenched with respect to its contribution along the field lines $k_\parallel$. It is common to treat the problem in the relaxation time approximation, where the ratio between the two quantities is expressed by the following formula \citep{UrpinYakovlev1980}:
\begin{equation}
    \frac{k_\parallel}{k_\perp} \simeq 1 + (\omega_B\tau_0)^2,
    \label{eq:approx}
\end{equation}
where the parameter $\omega_B \tau_0$ is the product between the characteristic electron relaxation time $\tau_0$ and the electron gyrofrequency $\omega_B$, which is written
as: 
\begin{equation}
    \omega_B \equiv \frac{e ||\mathbf{B}||}{m^*_e c}, 
\end{equation}
where $e$ and $m^*_e$ are the electron charge and effective mass, respectively, and $\mathbf{B}$ is the local magnetic field. 
This approximation shows that $\omega_B \tau_0$ is the parameter governing the heat conduction: when $\omega_B \tau_0 \gg 1$, either because the magnetic field is strong or because the electron collision rate is low, the electrons are tightly anchored to the field lines, and slip along them, while the transverse motion is strongly inhibited. This results in a reduction of the thermal conductivity in the direction orthogonal to the field lines. Vice versa, when $\omega_B \tau_0 \ll 1$, the magnetic field can only poorly constrain the electron motion, and the conductivity becomes isotropic ($k_\perp \rightarrow k_\parallel$). 

According to Eq.~\ref{eq:approx} and following \citet{Perez-Azorin2006}, we can write the heat flux as:
\begin{equation}
    e^{\nu(r)}\mathbf{F} = -k_\perp\bigl[\pmb{\nabla}\Tilde{T} + (\omega_B \tau_0)^2 (\mathbf{b}\cdot \pmb{\nabla} \Tilde{T})\mathbf{b}+ (\omega_B\tau_0)(\mathbf{b}\times \pmb{\nabla}\Tilde{T})\bigr],
    \label{eq:flux}
\end{equation}
where $\mathbf{b} \equiv \mathbf{B}/||\mathbf{B}||$ is the unit vector in the direction of the magnetic field. Three terms contribute to the heat flux: a term parallel to the temperature gradient, which is the same appearing in Fourier's law ($\mathbf{F} \propto \pmb{\nabla}T$, valid in the absence of magnetic field); a term parallel to the magnetic field, which is due to the fact that electrons tend to move along the magnetic field lines; and a last term perpendicular to both the previous ones, known as Hall term, which is due to the drifting motion of electrons through field lines due to the presence of the temperature gradient. The magnitude of $\omega_B \tau_0$ regulates which of these terms gives the dominant contribution. In the limit where $\omega_B \tau_0$ is negligible, the first term is the dominant, while when $\omega_B \tau_0 \gg 1$, the second term dominates. 

From Eq.~\ref{eq:flux} we can appreciate that the flux shows a linear dependence with respect to the temperature derivatives, such that it is possible to re-write the previous equation in matrix form: 
\begin{equation}
    e^{\nu(r)}F^i = -K^{ij}\partial_i \tilde{T}, 
    \label{eq:flux_matrix}
\end{equation}
where we assumed the Einstein sum convention of repeating indices. The explicit form of the matrix $K^{ij}$ and its derivation is reported in Appendix \ref{sec:Matrix}.

\subsection{Grid} 
\label{sec:grid} 

\begin{figure}
    \centering
    \includegraphics[width = \columnwidth]{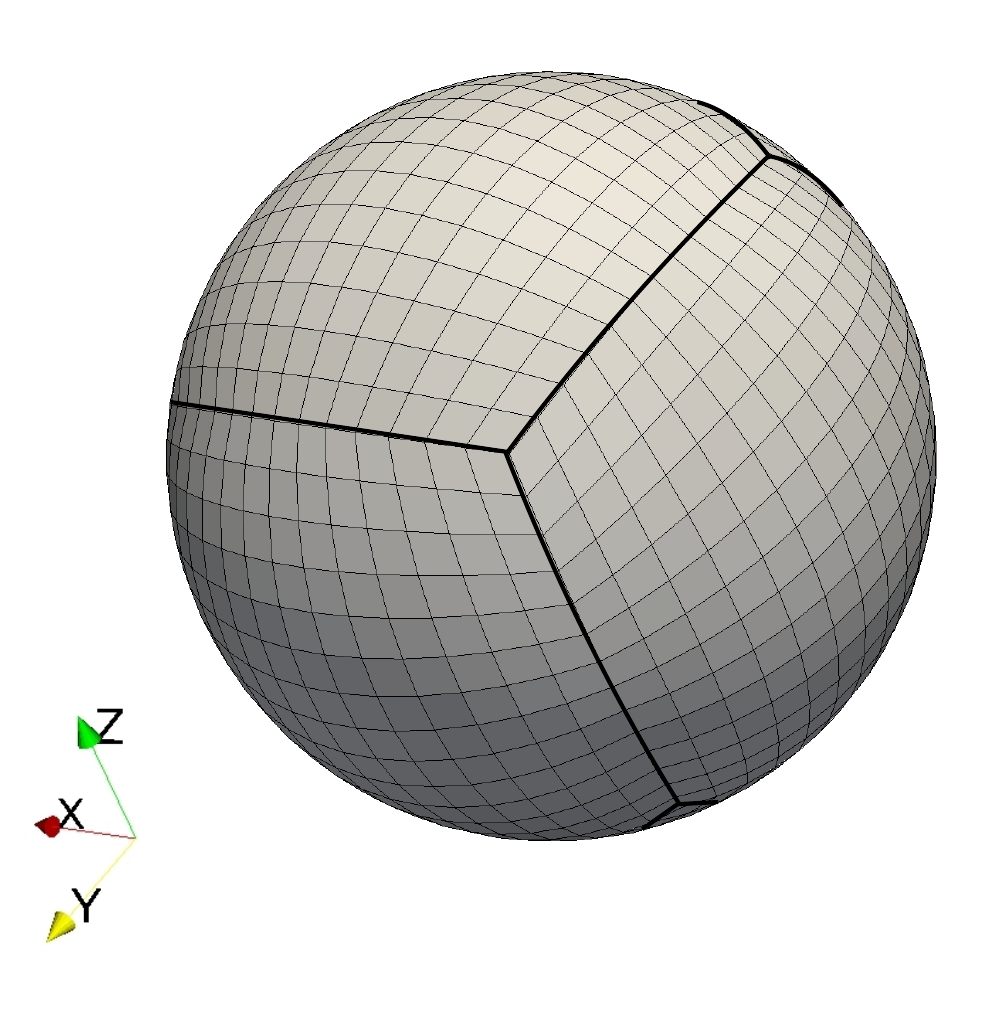}
    \caption{Visualization of the cubed sphere grid at a given radius. The edges between the patches have been highlighted with a bold line. Within each patch, thin lines mark the $\xi$ and $\eta$ directions.}
    \label{fig:cubed_sphere}
\end{figure}

In our code, we use the cubed sphere coordinates as described by \citet{Ronchi1996J}, implemented in \emph{MATINS} by \citet{Dehman2023_MATINS}, and successfully used in several other contexts like geophysics \citep{Breitkreuz2018, DingWordsworth2019, vanDriel2021}, general relativity \citep{Lehner2005, Hebert2018, Carrasco2018, Carrasco2019} and magnetohydrodynamics \citep{Koldoba2002, Fragile2009, HosseinNouri2018, Wang2019,Yin2022}. The spherical volume of the star is characterized using a radial coordinate $r$ and two angular coordinates ($\xi, \eta$). Each spherical surface of constant $r$ is divided into six patches, which can be pictured as the six faces of a cube that have been inflated to adhere to a sphere. The six patches are identical, non-overlapping, and, as in a cube, each of them is bounded by four patches. A representation of the cubed sphere grid is reported in Figure \ref{fig:cubed_sphere}. 

The angular coordinates $\xi$ and $\eta$ are defined differently in each patch, such that they are not singular and assume values in the range [$-\pi/4, \, \pi/4$]. Moreover, $\xi$ and $\eta$ are non-orthogonal everywhere, except at the center of each patch, which implies non-diagonal terms in the metric, \emph{i.e.} additional terms in the mathematical operators used in the discretized equations (gradients, scalar and cross products). This grid presents two main advantages: first, it has a radial coordinate, which is desirable since the background is spherically symmetric (see Sec. \ref{sec:background}). Moreover, the problem we aim to solve is characterized by strong radial gradients, such that a radial coordinate allows us to refine the resolution more in the radial direction than in tangential ones. Second, the coordinates are regular everywhere, in contrast, for example, with the spherical ones, which are singular along the axis. The metric, the line, surface and volume elements and the differential operators for the metric are reported in Appendix \ref{sec:CubedSphere} and in more detail by the accompanying paper \citet{Dehman2023_MATINS}.

In order to consider the derivatives, we use one layer of ghost cells lying behind each edge in the $\xi$ and $\eta$ directions. The values of the quantities at these ghost points are obtained by linear interpolation from the cells of the neighbouring patches. This prescription is explained more extensively in Appendix \ref{sec:ghost}.

Hereafter, the resolution of the grid will be denoted by the number of radial cells $N_r$ and the number of angular cells per-patch $N_a$, where $N_a$ is the same for the $\xi$ and $\eta$ directions. In line with its 2D predecessors \citep{Vigano2012, Vigano2021}, in \emph{MATINS} the thermal grid has half the resolution of the magnetic grid. Specifically, the temperature is evolved only at the center of the thermal grid, while the magnetic field is defined and evolved also at its edges and interface centers. Hence, the magnetic field discretized locations are 8 times more than the temperature ones. This choice is guided by two considerations: firstly, the computational needs, since the cost of the implicit scheme adopted for the thermal evolution (see Appendix \ref{sec:time_discratization}) scales non-linearly with the total number of points, while the magnetic evolution is linear. Secondly, because of the nature of the equations: the non-linear Hall term in the induction equation naturally create small structures and a cascade \citep{Dehman2023_MATINS}, something not present, by definition, in the diffusion of the heat which thus only needs a more moderate resolution.

\subsection{Microphysics}
\label{sec:numerical_microphysics}

The stellar microphysics enters the heat diffusion equation through the heat capacity per unit volume $c_\mathrm{v}$, the source term $\dot{\epsilon}$, the thermal conductivity $k_\perp$ and the $\omega_B \tau_0$ parameter. 
In \emph{MATINS} these quantities are defined at the center of the cells.
To compute the value of $k_\perp$ and $\omega_B \tau_0$ at the cell interfaces, needed to calculate the flux (see Appendix \ref{sec:Matrix}), we interpolate their values at the cell centers. We tested a linear interpolation on the variables
$k_\perp$ and $\omega_B \tau_0$, as well as on their logarithm.  
We chose the latter since we consider it as more suitable to describe quantities that can exhibit orders of magnitude variations among adjacent cells. Moreover, our numerical experiments show slightly higher stability of the code, although, in any case, the differences in the results are negligible and decrease with increasing resolution. 

The microphysics setup employed by the \emph{MATINS} code is the same as the one already used in the previous 2D code \citep{Vigano2013, Vigano2021}, where the microphysical quantities are calculated numerically exploiting the public code released by A.~Potekhin. A detailed review of the microphysics can be found in \citet{Potekhin2015}, while here we provide a summary description of the most important microphysical process contributing to the macroscopic quantities entering in Eq.~(\ref{eq:heat_diffusion}). 

As already described in the previous Sec. \ref{sec:heat_diffusion_equation}, the thermal conductivity under the presence of a magnetic field is anisotropic and it is described by a tensor, whose components are calculated using Potekhin's public code\footnote{http://www.ioffe.ru/astro/conduct/}. In the core, the conductivity, dominated by electrons, is very high. In the crust, the conductivity is lower, and it is mostly influenced by the interaction between electrons, impurities in the lattice, lattice phonons, and phonons of the neutron superfluid and non-superfluid neutrons \citep[see Fig. 4.5 of ][for a comparison between the different contributions]{Vigano2013_thesis}. In the direction of the magnetic field, the electrons are expected to be the main contributor to the heat transfer, while in the orthogonal direction, the phonons can become dominant if the magnetic field is particularly intense.  
Finally, in our analysis, we neglect the quantizing effects due to the gradual fillings of Landau levels, which leads the thermal conductivity to oscillate with varying density around the classical value. Although these effects are considered in the Potekhin code, their inclusion is particularly expensive from a computational point of view, and since their contribution becomes important only at densities lower than that of the crust (such as in the external envelope), we can safely neglect them.  

The heat capacity per unit mass in the outer crust has contributions from the degenerate electron gas and the ion lattice, while in the inner crust also the neutron gas contributes if the temperature is above the neutron superfluidity critical value. For its calculation we employ the public code by Potekhin\footnote{http://www.ioffe.ru/astro/EIP/}, properly designed for a strongly magnetized, fully ionized electron-ion plasma.  

The neutrino emissivity dominates the stellar cooling in the first $10^4-10^5\,\textrm{yr}$ (neutrino cooling era). After that the drop in temperature quenches the neutrino emission and the star cools mainly through the emission of thermal photons from the surface (photon cooling era). In this setup the neutrino emissivity is described by the formulae provided in Table 1 of \citet{Potekhin2015}. 

The microphysical setup of our code implements also superfluidity correction to the previous quantities for neutrons (singlet state) in the crust and for neutrons and protons in the core (triplet and singlet states, respectively). The energy gap and the critical temperature are approximated as a function of Fermi momenta by the parametrization provided in \citet{Kaminker2001}, whose fit parameters depend on the chosen superfluid gap model. In \emph{MATINS}, we provide the possibility to choose between the same gap models present in the 2D code of \citet{Vigano2021}. The models and the relative parameter values are listed in Table II of \citet{ho2015}. Throughout this work, we employ the same gap model choice, corresponding to the model indicated by \citet{ho2015} as SFB \citep{Schwenk2003} for the neutron in the crust, TToa \citep{Takatsuka2004} for the neutrons in the core, and CCDK \citep{Chen1993, Elgaroy1996} for the protons in the core.  

\subsection{Timestep and workflow}
 
 Since the cooling timescale is highly dependent on the temperature itself (mostly due to the neutrino processes), it increases by orders of magnitude in the long term. Therefore, in \emph{MATINS} the timestep is set proportional to the time itself. Users can customize this timestep by specifying the constant of proportionality, with default value set at $0.1$.
Additionally, the timestep has a minimum and a maximum value, that typically we set as $\Delta t_{\mathrm{min}} = 10^{-2}$ yr and $\Delta t_{\mathrm{max}} = 10^4$ yr, respectively. With this choice of parameters, \emph{MATINS} completes a simulation with $1\,\mathrm{Myr}$ as total physical time in $\sim 250$ thermal evolution timesteps. 

The workflow of \emph{MATINS} closely resembles the one presented by \citet{Vigano2021} in the predecessor 2D magnetothermal code. For a detailed visual representation, we refer the reader to Fig. 2 in their work. At the beginning of the simulation, we import the input parameters and utilize them to construct the stellar background. Additionally, we initialize the temperature and the magnetic field. Subsequently, we initiate the thermal evolution loop, wherein, at each timestep, the microphysics of the star (thermal and electric conductivities, heat capacity, neutrino emissivity) is updated using the new values of temperature and magnetic field.

The magnetic evolution loop is nested within the thermal evolution loop. During each thermal timestep, the magnetic field undergoes evolution, involving the calculation of currents and the Joule heating term, which enters the source term on the right-hand side of Eq.~\ref{eq:heat_diffusion}. If the magnetic field evolution is disabled, the Joule heating is switched off. Once the nested magnetic loop is complete (typically, a magnetic timestep of fraction of year, meaning hundreds or thousands per each thermal timestep), the temperature is updated, and a new cycle begins. 

\section{Tests} \label{sec:tests}

In this section, we describe two problems employed to test \emph{MATINS}: a test with a simplified microphysics and a known analytical solution (hereafter \emph{benchmark test}), and a test with realistic microphysics.

\subsection{A benchmark test}
In this section, we describe a benchmark test to \emph{MATINS}: a problem with a known analytical solution presented in \citet{Perez-Azorin2006}. It consists of a non-stratified spherical shell with inner radius $r_{in}$ and outer radius $r_{out}$, with a uniform, non-evolving magnetic field oriented along a given axis. The model assumes a uniform conductivity and heat capacity. Additionally, the general relativistic corrections are neglected. No source term is included, thus the problem tests the anisotropy of the heat diffusion. This problem admits the following analytical solution:

\begin{equation}
    T(r, \theta, t) = T_0\Bigl(\frac{t_0}{t}\Bigr)^{3/2}\exp{\Bigl[-\frac{c_\textrm{v}\, r^2}{4 k_\perp t}\Bigl(\sin^2{\theta}+ \frac{\cos^2{\theta}}{1+(\omega_B \tau_0)^2}\Bigr)\Bigr]},
    \label{eq:PA_analytical}
\end{equation}
 where $T_0$ and $t_0$ are the initial values of temperature and time, $\theta$ is the polar angle calculated from the axis centered on the star and parallel to the magnetic field direction (see \citet{Ronchi1996J, Dehman2023_MATINS} for details about transformation between spherical and cubed-sphere coordinates). The problem is expressed in dimensionless physical units, where $c_\mathrm{v} = k_\perp = t_0 = 1$, $r_{in} = 5$, $r_{out} = 10$, $T_0 = 100$, $\omega_B \tau_0 = 10$. 
With this choice of units the diffusion timescales, defined as $\tau_{\rm diff} \equiv c_\mathrm{v} (r_{out} - r_{in})^2/k_\parallel$, is $\tau_{\rm diff} = 0.25$. We choose the simulation timestep $\Delta t = 0.01 \ll \tau_{\rm diff}$. Given the unconditional stability of the implicit method we use to solve the heat diffusion equation, the timestep is not subjected to the Courant condition or any similar limitation. As such, the timestep is the same for all the resolutions employed here.  
For this test, instead of the usual boundary conditions (described in Appendix \ref{sec:boundary}), we impose, at each time $t$, the analytical value of $T$ given by Eq. (\ref{eq:PA_analytical}) at the innermost and outermost radii.

\begin{figure}
    \centering
    \includegraphics[width=\columnwidth]{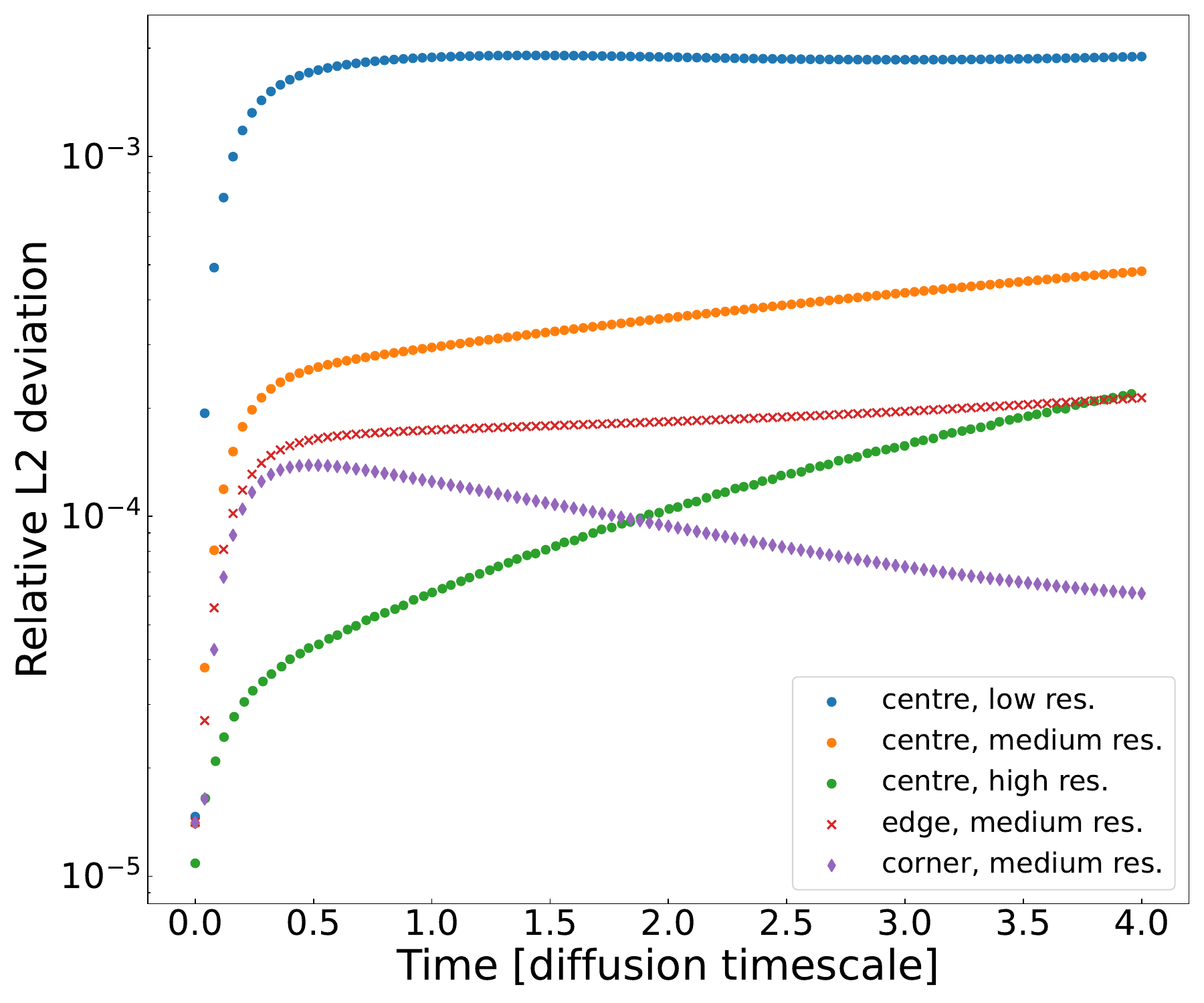}
    \caption{Relative L2 deviation between the analytic and numerical solution as a function of time (measured in diffusion timescale units). The circles represent simulations with the magnetic field oriented along the z-axis with low ($N_a = 7, N_r = 20$; blue), medium ($N_a = 14, N_r = 20$; orange), and high ($N_a = 28, N_r = 20$; green) resolution. Red crosses and purple diamonds are simulations at medium resolution with different magnetic field orientations.}
    \label{fig:PA_3res}
\end{figure}

First, we tested our code on this problem by choosing the orientation on the uniform magnetic field along the $z$-axis, such that the magnetic field is perfectly radial at the centre of the north and south patches of our cubed-sphere grid. 
We performed three simulations at three different angular resolutions: a low ($N_a = 7$), medium ($N_a = 14$), and high resolution ($N_a = 28$). In all these cases the radial resolution is held fixed, at a value of $N_r = 20$.  
In Figure~\ref{fig:PA_3res} we show at each time step the average L2 relative error, quantifying the volume-averaged deviations from the analytical solution, defined as 
\begin{equation}
    \textrm{Avg L2}(t) \equiv   \frac{\sum^{\rm cells}_i[T^{\rm num}_i - T(r_i,\theta_i, t)]^2}{\sum^{\rm cells}_iT^2(r_i,\theta_i, t)},
    \label{eq:L2}
\end{equation}
where the sum is performed all over the computational domain. Here $T^{\rm num}$ denotes the numerical solution, while $T(r_i,\theta_i, t)$ denotes the analytical solution calculated at the center of the i-th cell. 

The results are reported with circle symbols in Figure \ref{fig:PA_3res}. The label "centre" indicates that in these simulations the magnetic field is radial at the center of a patch. As 
one
can see, improving the angular resolution 
by a 
factor of 2, improves the average L2 by almost an order of magnitude. With respect to the three cases with a radial field at the patch's centre (blue, orange and green curves), we note that, although L2 decreases with increasing angular resolution, the increase of L2 in time is faster at higher resolution. This behavior can be ascribed to the fact that at higher resolution the size of the cell in the transverse direction becomes comparable to its radial size (which is fixed in the three cases), causing the radial resolution to constrain the numerical performance. Additionally, the error accumulates non-linearly with increasing resolution, so that we do not expect the three curves to exhibit the same slope. Nevertheless, despite their faster growth, the L2 at medium and high resolution also exhibit a negative second derivative, indicating a deceleration in their growth rates over time, which is more pronounced in the high resolution case. 

\begin{figure*}
    \centering
    \includegraphics[width = \textwidth]{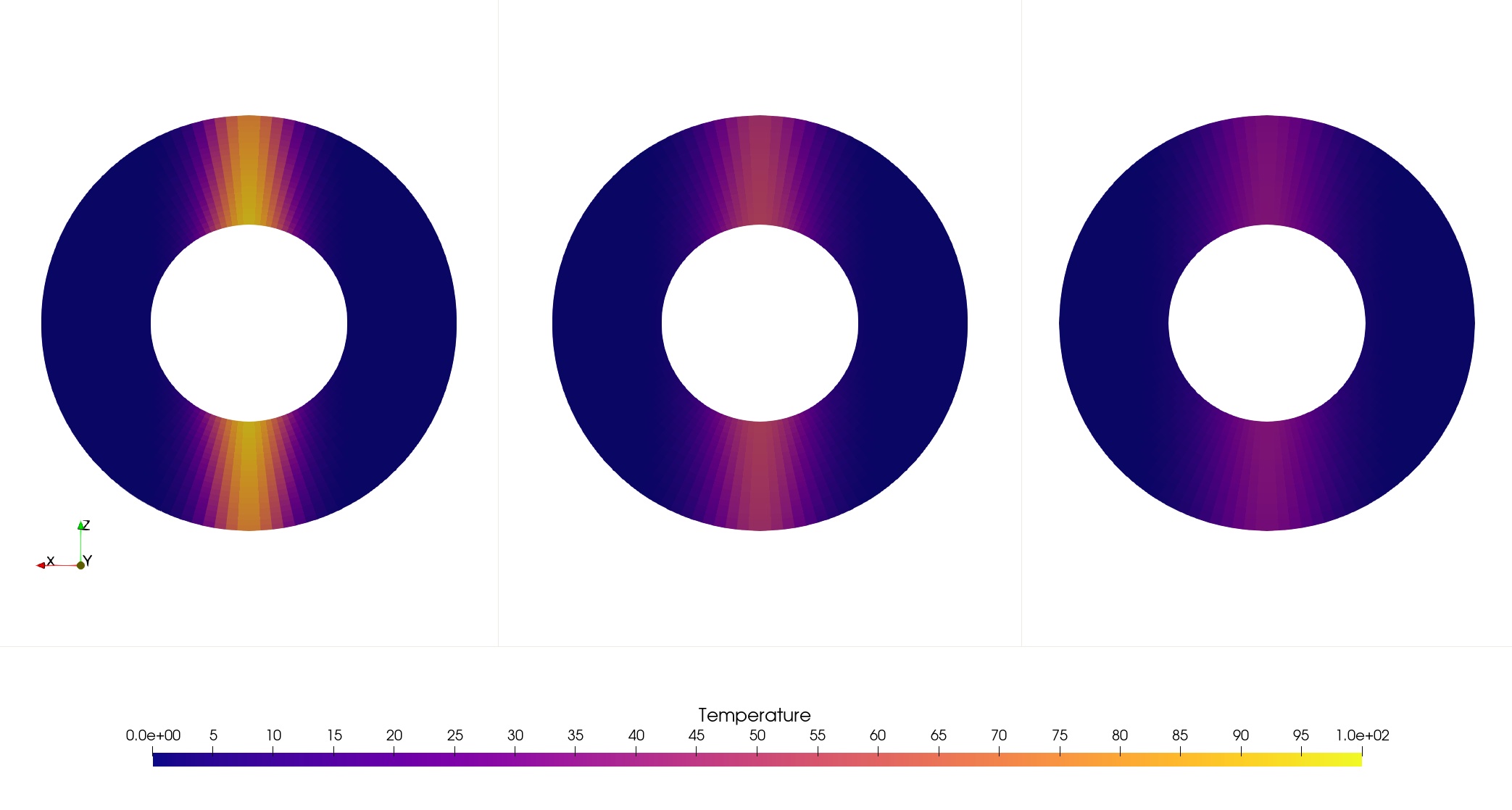}
    \caption{Meridional cut of the internal temperature profile (in dimensionless units) of the benchmark test. In this test, the microphysics is uniform, with dimensionless value of $c_\mathrm{v} = k_\perp =1$, the magnetic field is uniform and oriented along the $z$-axis, and with $\omega_B \tau_0 = 10$. The initial temperature profile has a cylindrical symmetry around the magnetic field axis and evolves in time according to Eq. \ref{eq:PA_analytical}. The figure shows the run with high resolution ($N_r = 20$, $N_a = 28$) at three different times ($t = 0\tau_{\rm diff}$, $t = 2 \tau_{\rm diff}$ and $t = 4\tau_{\rm diff}$).}
    \label{fig:PA_3D_solution}
\end{figure*}

\begin{figure}
    \centering
    \includegraphics[width= \columnwidth]{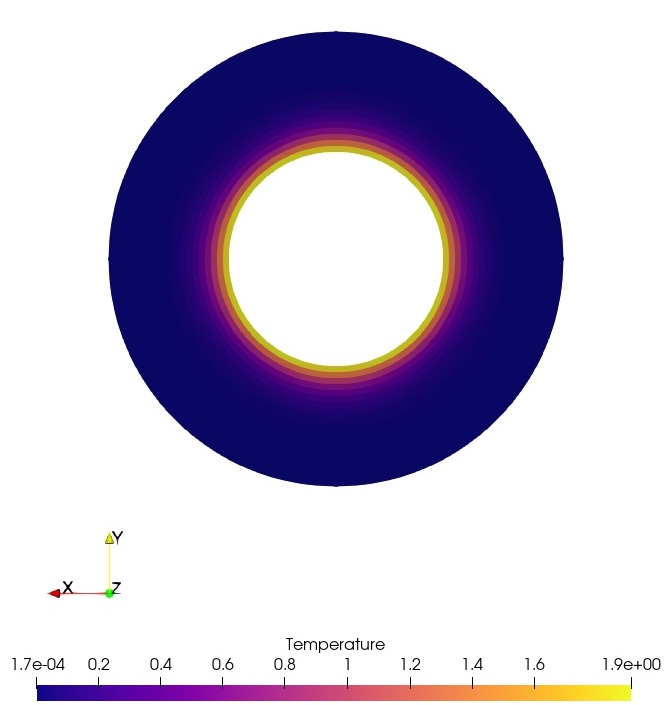}
    \caption{Equatorial cut of the internal temperature profile (in dimensionless units) of the benchmark test with the field oriented along the z-axis and high resolution ($N_r =20$, $N_a = 28$) at time $t = 4\tau_{\rm diff}$. }
    \label{fig:PA_3D_equatorial}
\end{figure}

The result of this simulation for the high-resolution case is reported in Fig. \ref{fig:PA_3D_solution} for three different times. As expected, we 
notice that the temperatures diffuse mainly along the direction of the magnetic field, while the diffusion in the orthogonal direction is suppressed. In Figure \ref{fig:PA_3D_equatorial} we report also the equatorial cut at the final timestep of the simulation. From this figure, we can appreciate how the axial symmetry is maintained during the simulation. 

\begin{figure}
    \centering
    \includegraphics[width= \columnwidth]{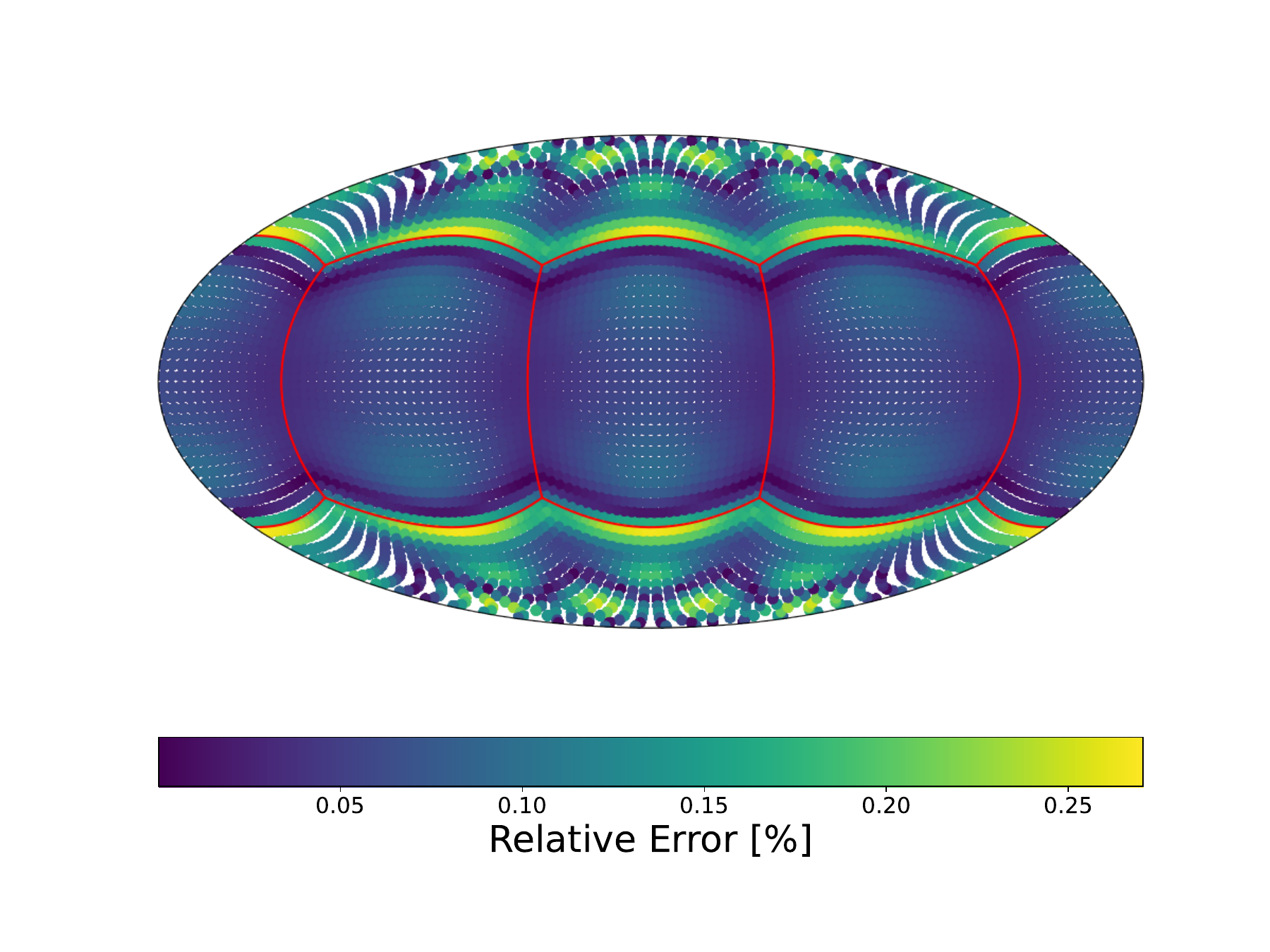}
    \caption{Map of the relative error in the benchmark test on a spherical shell at $r= 0.25(r_\mathrm{out}+r_\mathrm{in})$. The patches' edges are underlined with a red curve.}
    \label{fig:relative_error_PA}
\end{figure}

Finally, we repeat the same test but orienting the symmetry axis of the problem in different directions, in order to check if we are able to obtain the same evolution of the system. We perform four additional simulations at medium resolution, in which the field is oriented along the $x$-axis, the $y$-axis, the center of the edge between an equatorial patch and the north pole patch, and the corner between two equatorial patches and the north pole patch. The result in terms of average L2 is shown in Figure \ref{fig:PA_3res}, along with the result for the cases with the field along the $z$-axis discussed before. Here, the errors for the field orientated along the patch centres (namely along the $x$, $y$, or $z$-axis) perfectly overlap and correspond to the orange curve. In the other two cases where the field is radial at the centre of an edge between two patches (red crosses) or at the corner between three patches (purple diamond), the error is, at the end of the simulation, a factor $\sim\,2\,$ and a factor $\sim\,8\,$ smaller than in the previous cases with the same resolution, respectively. Additionally, also the curve's slopes appear different, even if in all cases we note a flattening in the final steps. 
The reason for these differences can be, at least in part, ascribed to the differences in the cells, both in terms of size and shape, of the cubed sphere grid within the patch. The higher error in the "centre" case, for example,  can be motivated by the fact that the cells at the centre of the patch are the largest ones. Therefore, the hot-column (see Figure \ref{fig:PA_3D_solution}), in the "centre" case (orange curve in Figure \ref{fig:PA_3res}) becomes less resolved than the other two cases (red and purple curves), leading to a larger L2 error\footnote{The L2 error, as defined in Eq. \ref{eq:L2} weights more the cells with higher temperature}. On the other hand, the "corner" case (purple curve) is associated with a lower error than the "edge" case (red curve), even though the corner cells are slightly larger than the cells at the edge's centre ($\sim 77\%$ and $\sim 70\%$ of the size of the cell on the patch centre, respectively). In this case, we ascribe the lower error to the difference in the shape of the cells at the corner compared to those at the edge's center. Indeed, the former are more suitable to approximate the circular section of the hot column, than the latter. We thus conclude that the difference in the error among the three cases is non-trivial, and is determined by a combination of different factors.


Note that, regardless of the (low) resolutions employed here and the magnetic orientation, the relative uncertainty, estimated by the square root of the L2 error, is in the range $1-4 \%$ and decreases with resolution, therefore perfectly acceptable. 

Finally, with the aim of assessing how the accuracy of our numerical solution varies with respect to the angular position on the cubed sphere grid, we report in Figure \ref{fig:relative_error_PA} a map of the relative uncertainty, calculated at the final timestep, on a spherical surface cut at the (arbitrarily chosen) radius $r = 0.25(r_\mathrm{out}+r_\mathrm{in})$. We define the relative error as $[T^\mathrm{num}_i - T(r_i, \theta_i, t)]/\mathrm{max}[T(r_i, \theta_i, t)]$, where the denominator is the maximum  temperature on the chosen spherical cut. We use the maximum of the temperature in the definition of the relative error, instead of the value point-by-point, in order to avoid large error values where the temperature is low. We observe that the error is largest at the interfaces between the equatorial and the polar patches, but not at the edges between one equatorial patch and the next one. 

\subsection{Tilting the field in a realistic model}

\begin{figure}
    \centering
\includegraphics[width=\columnwidth]{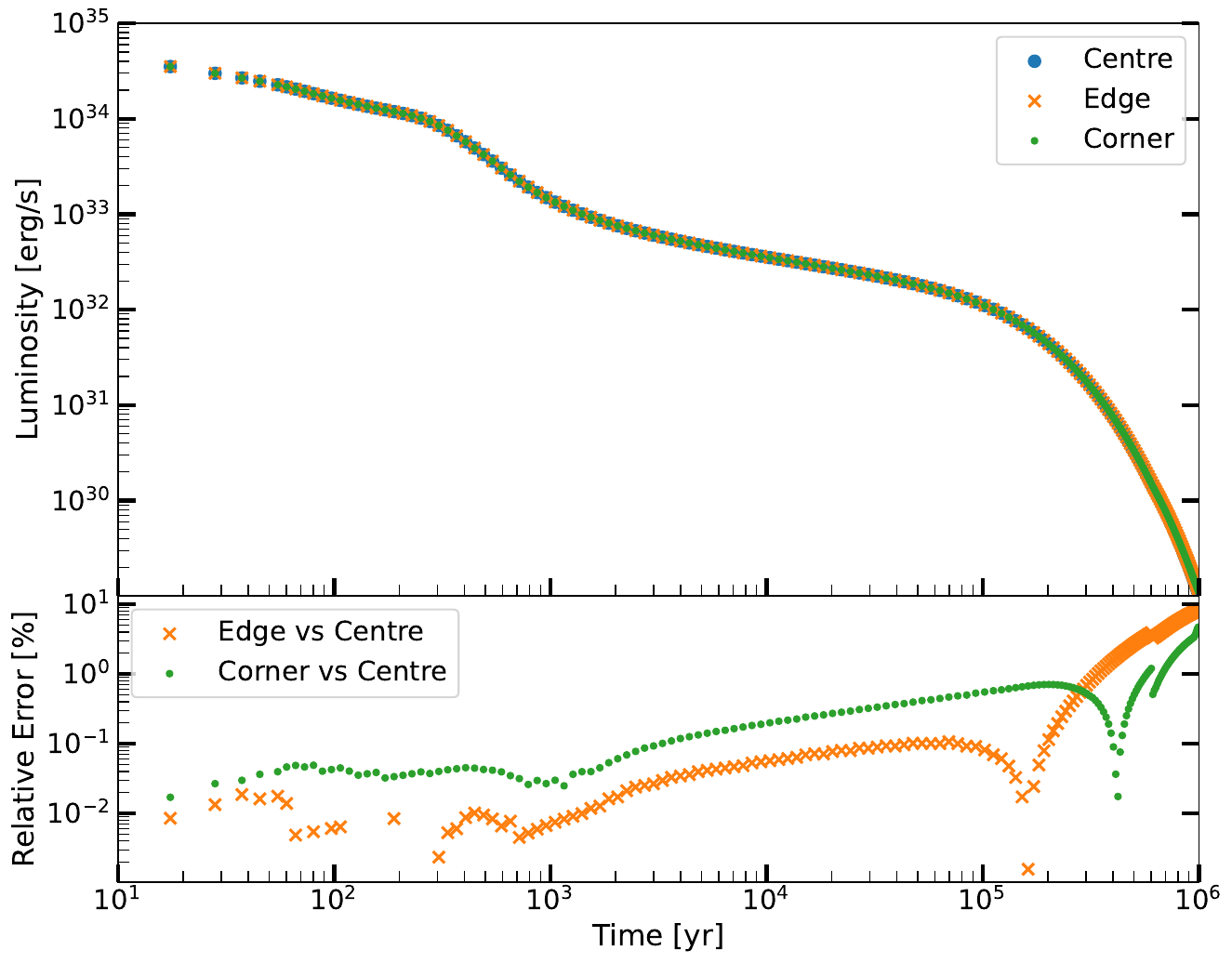}
    \caption{Top: Cooling curve for three different configurations of the \emph{Sly4-M1.4-B14-L1} model, where the magnetic axis falls on the center of the patch (blue), on the edge between two patches (orange) and on the corner between three patches (green). The resolution of all the simulations is $N_a = 25$ and $N_r = 30$. Bottom: relative error of the "edge configuration" and the "corner configuration" with respect to the "centre configuration". }
    \label{fig:tilt_realistic}
\end{figure}

In the previous section we studied the effect of tilting the orientation of the magnetic field in the benchmark test. In this section, we repeat the exercise with a realistic model, which, unlike the benchmark test, includes a physical \ac{NS} background, neutrino emission processes, a non-uniform magnetic field, general relativistic corrections, and the boundary conditions described in Appendix \ref{sec:boundary}. The model employed for this study is  \emph{Sly4-M1.4-B14-L1} reported in Table \ref{tab:Bfield_models}, characterized by a crust-confined dipolar poloidal magnetic field configuration. This configuration is characterized by the presence of two antipodal magnetic poles and it is symmetric for rotation around the magnetic axis, defined as the axis passing through the centers of the poles and the center of the star. The temperature field is initially uniform, with a value of $T=10^{10}\,\textrm{K}$. Further details on realistic configurations are provided in the next section. Here we limit ourselves to analyzing the impact of the different orientations of the magnetic axis on the results. This is shown in Figure \ref{fig:tilt_realistic}, where, in the top panel, the bolometric thermal luminosity of the star is shown for three different configurations: one configuration in which the magnetic axis passes through the center of the patches (blue points), one in which it passes through the center of the edge between two patches (orange points), and one in which it passes through the corner between three patches (green points). From this figure, we can appreciate how the three curves appear indistinguishable. In the lower panel, we plot the relative error of the second and third configurations, taking the first one as a reference. We can see that before $\sim 3\times 10^5\, \textrm{yr}$ for the edge configuration and $\sim 6\times 10^5\,\textrm{yr}$ for the corner configuration, the error remains below $1\,\%$, for a total simulation time of $1\,\times 10^6\,\textrm{yr}$. Beyond this time, the increase of the relative error, which almost reaches $10\,\%$ at the end of the simulation, is likely caused by the rapid decrease in luminosity. Even including the increased error at  very late times, the simulations are in good agreement with each other well within the observational uncertainties. This test shows that the results keep their consistency under rotation of the magnetic field not only in the simplified benchmark test, but also for a realistic model.

\section{Results}
\label{sec:results}

Let's now turn to the application of \emph{MATINS} to a realistic \ac{NS} structure, under different magnetic field configurations. All the initial magnetic field configurations here considered are confined to the crust as discussed more in detail in \citet{Dehman2023_MATINS}.
Unless stated otherwise, we run our simulations for a total time of $1\, \mathrm{Myr}$. 
As a benchmark for our study, we consider a \ac{NS} characterized by the unified (\emph{i.e.} crust + liquid core) \ac{EOS} \emph{SLy4} \citep{Dauchin2001}, which assumes a minimal composition of neutrons, protons, electrons and muons in the core and is obtained in a tabulated form from the CompOSE database. 
Finally, in most of our simulations, we use the simple, classical envelope model by  \citet{Gudmundsson1983} (see Eq. \ref{eq:Gudm} in Appendix \ref{sec:boundary}). This decision is based on its simplicity and its adequacy for the objectives of the current study. We also discuss one case with a magnetized iron envelope model from \citet{Potekhin2015}. 

Since here we mostly focus on the thermal evolution, in most simulations we keep the magnetic field fixed. However, we will also present one simulation (up to $10^5\, \mathrm{yr}$ due to the substantially higher computational cost) with a fully coupled magnetothermal evolution, where also the Joule heating is accounted for. We refer to \citet{Dehman2023_MATINS} for the evolution of the magnetic field with simplified treatment of temperature adopted from \citet{Yakovlev2011} and to \citet{Dehman2023} for the first fully coupled magnetothermal study with \emph{MATINS}. We show below the results in terms of thermal evolution (\emph{i.e.} cooling curve) and the corresponding pulsed profile expected from a given temperature map on the stellar surface. Notice that the total (photon) luminosity of the source at a given time that we report in the cooling curve has been calculated by integrating Eq. \ref{eq:StefanBoltzmann} over the entire stellar surface. For \ac{NS} that are not too old, this luminosity is emitted mainly in the X-ray band, and thus this quantity can be compared with observations, if the distance and the interstellar absorption below $\sim 1 \, \mathrm{keV}$ can be well estimated. Moreover, ideally, one can infer the temperature map on the surface of the star and the emission model (here assumed as a blackbody for simplicity) by a simultaneous fit of light curve in different energy bands.

\subsection{Thermal evolution}

\begin{table*}
    \centering
    \begin{tabular}{|c|c|c|c|c|c|c|c|}
    \hline
    Name & $M\, [M_\odot]$ & EOS & $B_{avg}$ [G] & $l_p$ & $m_p$ &  $l_t$ & $m_t$  \\
    \hline
    \hline
        BSk24-M1.4-B0 & 1.47 & BSk24  &0&/&/&/ & /\\
    \hline
        BSk24-M1.8-B0  & 1.87 & BSk24 & 0 &/ & / & / & / \\
    \hline
        Sly4-M1.4-B14-L1 & 1.40 & SLy4 & $6\times10^{14}$ &1  & 0 & /& /\\
    \hline
     Sly4-M1.6-B14-L1 & 1.60 & SLy4 & $6\times10^{14}$ &1  & 0 & /& /\\
    \hline
        SLy4-M1.4-B14-L2 & 1.40 & SLy4 & $2\times 10^{14}$   & \multrow{1 \\ 2} & \multrow{-1,0,1  \\ -2,-1,0,1,2 }  & \multrow{1 \\ 2} & \multrow{-1, 0, 1 \\ -2,-1,0,1,2}\\
    \hline
        SLy4-M1.4-B14-L5& 1.40 & SLy4 & $2\times 10^{14}$  & \multrow{1 \\ 2 \\ 3 \\ 5}
        & \multrow{-1,0,1 \\-1,0,1,2\\-1,0,1,2,3\\ -1,0,1,2,3} & \multrow{1 \\ 2 \\ 3 \\ 5} & \multrow{0, 1 \\ 0,1,2 \\ 0,1,2,3 \\ 0,1,2,3}\\
    \hline
        SLy4-M1.4-B14-L10& 1.40 & SLy4 & $2\times 10^{14}$  & \multrow{ 1 \\ 2 \\ 9 \\ 10}
        & \multrow{-1,0,1 \\ -2,-1,0,1,2 \\ -7,-2,0,2,5 \\ -6,-5,0,3,10} & \multrow{1 \\ 2 \\ 10} & \multrow{ -1, 0, 1 \\ -2, -1, 0,1,2 \\ -1,0,1}\\
         \hline
        SLy4-M1.4-B14-L2-alt & 1.40 & SLy4 & $3\times10^{14}$ & \multrow{1 \\ 2} & \multrow{-1,0,1 \\ -2,-1,0,1,2} &  \multrow{1 \\ 2} & \multrow{-1, 0, 1 \\ -2,-1,0,1,2}\\
    \hline
    \end{tabular}
    \caption{Simulations setup. The first four columns report the name of the simulation, the mass of the star, the \ac{EOS}, and the magnetic field intensity averaged over the crustal volume. The last four columns describe the initial magnetic field configuration. This is constructed by decomposing the magnetic field into poloidal and  toroidal components, which are obtained from a poloidal and a toroidal scalar functions as described in Eq. \ref{eq:poloidal_toroidal_decomposition} of the Appendix \ref{sec:initial_Bfield_configuration}. The poloidal and toroidal scalar functions are defined in turn by a spherical harmonic expansion, with the expansion coefficients as input parameters. The columns report the degree $l$ and the order $m$ of the non-vanishing spherical harmonic weights of the initial poloidal and toroidal field expansion. For more details on the initial magnetic field configuration we refer the reader to Appendix \ref{sec:initial_Bfield_configuration} or to \citet{Dehman2023_MATINS}. The input parameters that define the configurations reported here are summarized in the supplementary Table \ref{tab:Bfield_config_input}.}
    \label{tab:Bfield_models}
\end{table*}

We run different simulations, with and without magnetic field, with different \acp{EOS} and different field configurations. Our runs are summarized in Table \ref{tab:Bfield_models}. The first column denotes the name of the simulation, the second the gravitational mass of the star, and the third reports the \ac{EOS} that we used. The remaining columns specify the configuration of the magnetic field. 
In \emph{MATINS} the initial magnetic field configuration is set up by defining the multipolar weights of the expansion of the poloidal and toroidal scalar functions $\Phi$ and $\Psi$, respectively, as described in detail in Appendix B of \citet{Dehman2023_MATINS} (see their Eq. B3).  
The fourth column reports the average value of the magnetic field in the crust, the fifth and sixth (seventh and eight) columns report the degree $l$ and the order $m$ of the non-vanishing multipoles in the expansion of the poloidal function $\Phi$  (toroidal function $\Psi$), respectively. As in virtually any existing magneto-thermal simulation in literature, in all these simulations we impose potential boundary condition at the surface. The radial component of the magnetic field at the crust-core interface is kept at zero. We refer to  \cite{Dehman2023_MATINS} for more details about boundary conditions, and to the Appendix \ref{sec:initial_Bfield_configuration} and  \cite{Dehman2023_MATINS} for more details on the initial magnetic field configuration.

\begin{figure*}
    \centering
    \includegraphics[width = \textwidth]{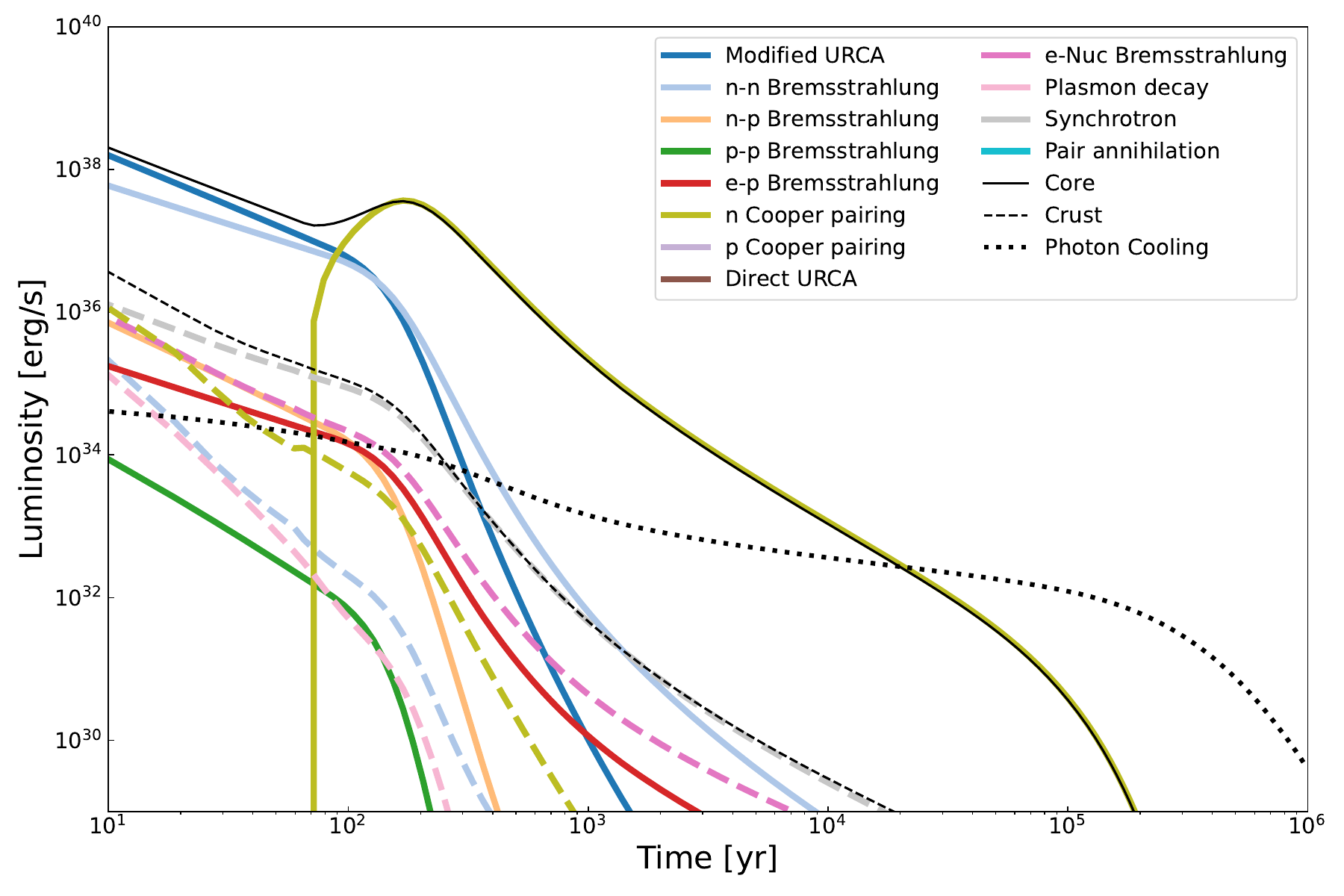}
    \caption{Evolution of neutrino and photon luminosities of all the different emission processes over the \ac{NS} history for the simulation \emph{Sly4-M1.6-B14-L1}. This run, which uses the SLy4 \ac{EOS}, is characterized by a mass of $M = 1.6 \, M_\odot$ and a dipolar poloidal magnetic field with a crust-averaged value of $B_{\rm avg} = 6\times 10^{14}$ G. The chosen resolution is $N_a=25$, $N_r = 30$. Solid-colored lines represent different neutrino emission processes occurring in the core. Dashed lines represent different neutrino emission processes occurring in the crust. The same process (marked with a given color) may involve both the core and the crust.  Black lines represent the total neutrino luminosity for processes involving the core (continuous line) and the crust (dashed line), respectively. The black-dots report the surface photon luminosity. It is worth noticing that while the legend includes all the possible neutrino processes included in the simulation, some of them are not effectively active in this particular simulation, as such, they do not appear in the figure. Moreover, in case  magnetic fields are present in the core, additional processes can become relevant \citep{Kantor2021}.}
    \label{fig:cooling_curve_benchmark}
\end{figure*}

The first simulation that we show, the \emph{Sly4-M1.6-B14-L1}, involves a magnetized star. 
The result of this run is presented in Figure \ref{fig:cooling_curve_benchmark}. Here the luminosity of the different neutrino emission processes is represented with colored lines. Continuous lines denote processes occurring in the core of the star and the dashed line denotes processes in the crust (the same process may involve both the crust and the core and in those cases, the same color appears in the plot with both a continuous and a dashed line). The continuous and the dashed black light denote the total neutrino luminosity for processes in the core and in the crust, respectively. Finally, the dotted black line represents the luminosity in photons. From Fig. \ref{fig:cooling_curve_benchmark}, it is evident that the cooling process is primarily governed by neutrino emission until $\sim 25 \mathrm{kyr}$, beyond which photon emission from the surface becomes the dominant cooling mechanism.

With respect to the neutrino emission, we can appreciate how the core provides the main contribution to the neutrino energy loss being 2-3 orders of magnitude larger than the crustal one, at all times. This reflects that most of the mass (typically $\sim 99\%$) is in the core, and that it is denser than the crust, although we need to acknowledge that the processes in the two regions are rather different, and as such they cannot be directly compared. In the core, the dominant mechanisms are represented by the Modified URCA processes and the neutron-neutron 
Bremsstrahlung until $\sim 100 \, \mathrm{yr}$, when the core becomes superfluid. After that, neutron Cooper pair formation is the main neutrino cooling mechanism. The onset of the superfluidity in the core is also followed by an increase in the slope in the photon luminosity. The crust, instead, is characterized by a much lower neutrino luminosity with respect to the core. Here the dominant mechanism is the neutrino-synchrotron emission, by virtue of the strong fields present in the crust with this setup. We stress that in the legend we indicate all the neutrino processes included in the simulation, even if some of them are not effectively active in this particular simulation and as such, they do not appear in the figure. For example, the neutrino pair-production is a process that becomes active at low densities and high temperature (it dominates at $\rho \lesssim 10^{10}\, \mathrm{g/cm^3}$ and $T>10^9 \, \mathrm{K}$, see Fig. 3 of \citet{Potekhin2015}), but it becomes strongly suppressed, due to the reduced number of positrons, when the temperature drops below the Fermi value. The absence of direct URCA in the core can be instead ascribed to the low abundance of protons in the core that characterizes this setup. In Appendix \ref{sec:different_EOS} we show the activation of this important cooling process for a simulation employing a different \ac{EOS}, which allows direct URCA above a given threshold mass. 

These results do not depend on the 3D nature of the code and are in agreement with our current knowledge about the cooling of \acp{NS}. Nevertheless, they show how the interplay among different neutrino emission processes plays an important role in the modeling of the cooling curve of a \ac{NS}. In this sense, to our knowledge, \emph{MATINS} is the most advanced 3D code in terms of including all of these essential contributions. 

\subsection{Simulations with different magnetic configurations}

\begin{figure}
    \centering
    \includegraphics[width = \columnwidth]{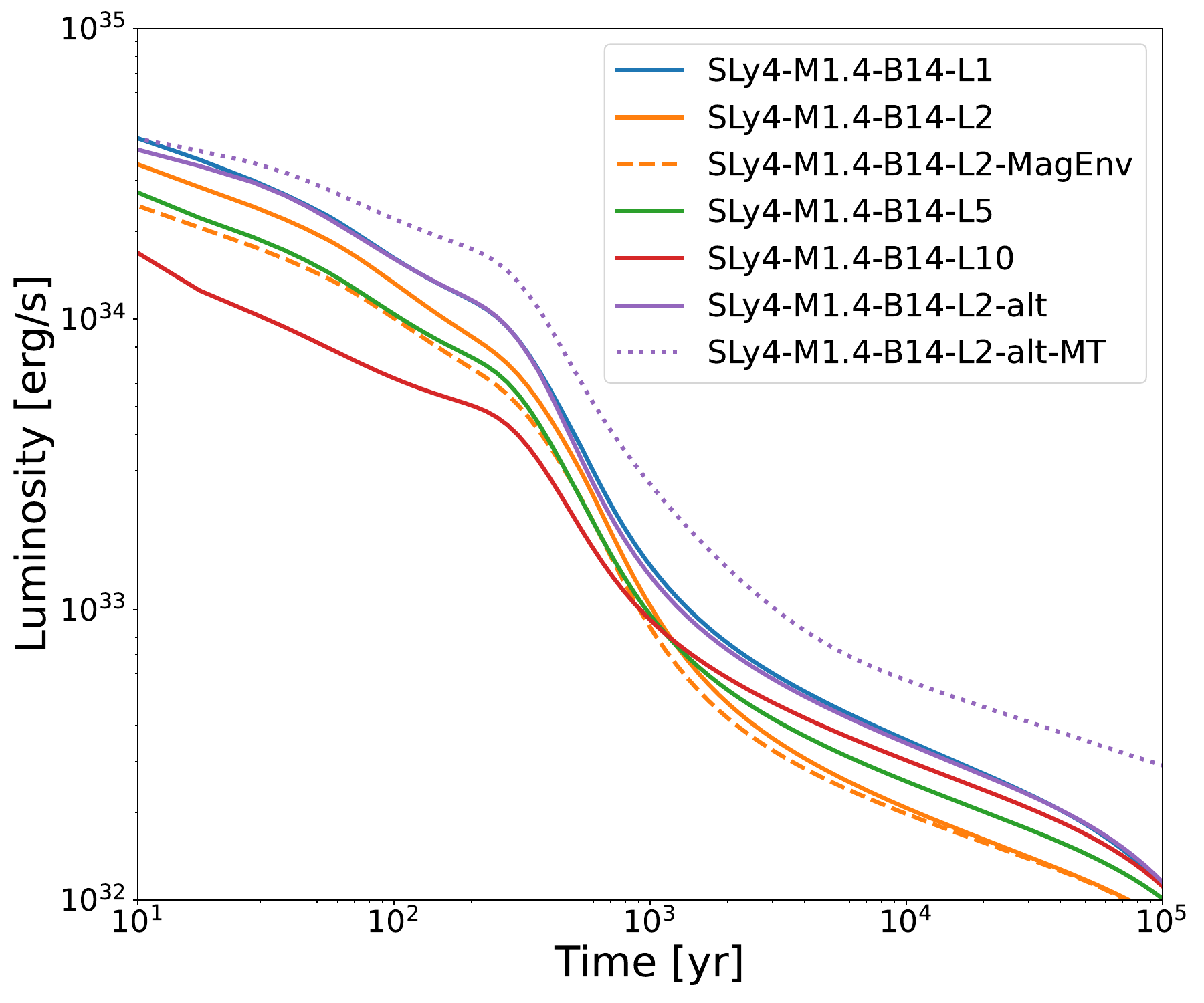}
    \caption{Cooling curves of different models characterized by different magnetic field configurations. The case \emph{Sly4-M1.4-B14-L1 } is characterized by an axisymmetric dipolar poloidal field. The cases \emph{Sly4-M1.4-B14-L2} (quadrupole), \emph{Sly4-M1.4-B14-L5} and \emph{Sly4-M1.4-B14-L10} have a non-axysymmetric field configuration. All the simulations are characterized by the same mass of the \ac{NS} ($M=1.4\,M_\odot$), the same \ac{EOS} (SLy4) and the same resolution of $N_a = 31$ and $N_r = 30$. The magnetic field is not evolved. }
    \label{fig:comparison_LC}
\end{figure}

Starting with the aforementioned setup for the \ac{EOS}, the superfluidity, and the envelope model, we study the thermal evolution of a \ac{NS} characterized by a mass of $M = 1.40\, M_\odot$, varying the configuration of its magnetic field as described in Table \ref{tab:Bfield_models}. 

Except for the dipolar cases \emph{Sly4-M1.6-B14-L1} and \emph{Sly4-M1.4-B14-L1}, where we aimed at an axisymmetric poloidal configuration, all other non-axisymmetric configurations have been chosen arbitrarily. For a realistic initial magnetic field configuration inspired by dynamo simulations representing the \emph{proto-\ac{NS}} stage \citep{reboul2021}, we direct the reader to \citet{Dehman2023}. 

\begin{figure*}
    \centering
    \includegraphics[width = \textwidth]{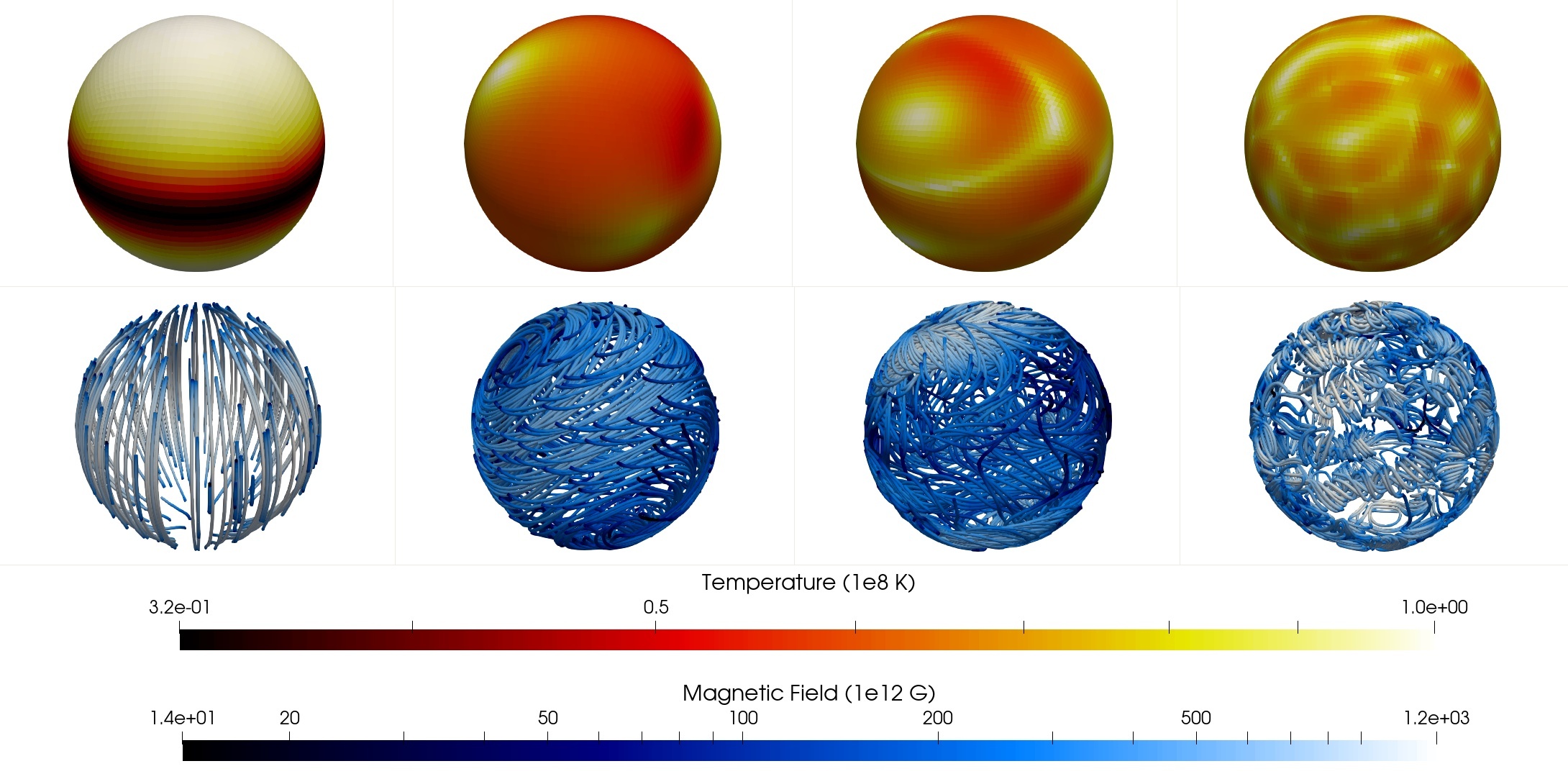}
    \caption{3D rendering of the models \emph{SLy4-M1.4-B14-L1}, \emph{SLy4-M1.4-B14-L2}, \emph{SLy4-M1.4-B14-L5}, and \emph{SLy4-M1.4-B14-L10} (from left to right) at time $t = 2726\,\mathrm{yr}$. 
    Top: temperature map at the base of the envelope $T_b$. Bottom: Magnetic field lines. The color denotes the intensity of the field. All the simulations use the same mass of $M = 1.4\,M_\odot$ and the \ac{EOS} SLy4.  The chosen resolution is $N_a = 31$, $N_r = 30$. In all these simulations the magnetic field is not evolved. 
    }
    \label{fig:3D_rendering}
\end{figure*}

In Figure \ref{fig:comparison_LC} we report the cooling curves of the case \emph{Sly4-M1.4-B14-L1 } (dipole), characterized by an axisymmetric poloidal field configuration, and the non-axisymmetric cases \emph{Sly4-M1.4-B14-L2} (quadrupole), \emph{Sly4-M1.4-B14-L5} and \emph{Sly4-M1.4-B14-L10}. 
As one can see, all the configurations have a similar cooling curve. This is due to the fact that the cooling processes are the same for all the configurations, while the Joule heating, which is determined by the (time-dependent) magnetic field intensity and geometry, is naturally absent since the field is constant in time. 

A 3D rendering of the magnetic field lines and the temperature at the crust-envelope interface $T_b(\theta, \phi)$ evaluated after $2726\,\mathrm{yr}$ of evolution is reported in Figure \ref{fig:3D_rendering} for the four different configurations just described\footnote{For the cases \emph{Sly4-M1.4-B14-L2} (central left), \emph{Sly4-M1.4-B14-L5} (central right) and \emph{Sly4-M1.4-B14-L10} (right) we report here a case with enhanced angular resolution $N_a = 41$.}. From the figure, it is possible to appreciate the impact of the configuration of the magnetic field on the temperature distribution. The cold areas on the map correspond to regions with a stronger magnetic field oriented perpendicular to the radial directions. In these areas, the magnetic field acts as an insulating barrier between the outer crust's surface and the hotter interior. As a result, while these regions lose thermal energy through neutrino and photon emission, there is no replenishment by heat diffusion, leading to the formation of cold spots. Vice versa, when the magnetic field is weaker, or it has a significant radial component, the outer regions of the crust are coupled and the heat lost can be replenished by diffusion, thus generating a hot area. Clearly, the complexity of the field reflects the complexity of the temperature map: a magnetic field characterized by small-scale structures leads to smaller hot-spots at the crust-envelope boundary.

\begin{figure*}
    \includegraphics[width = \textwidth]{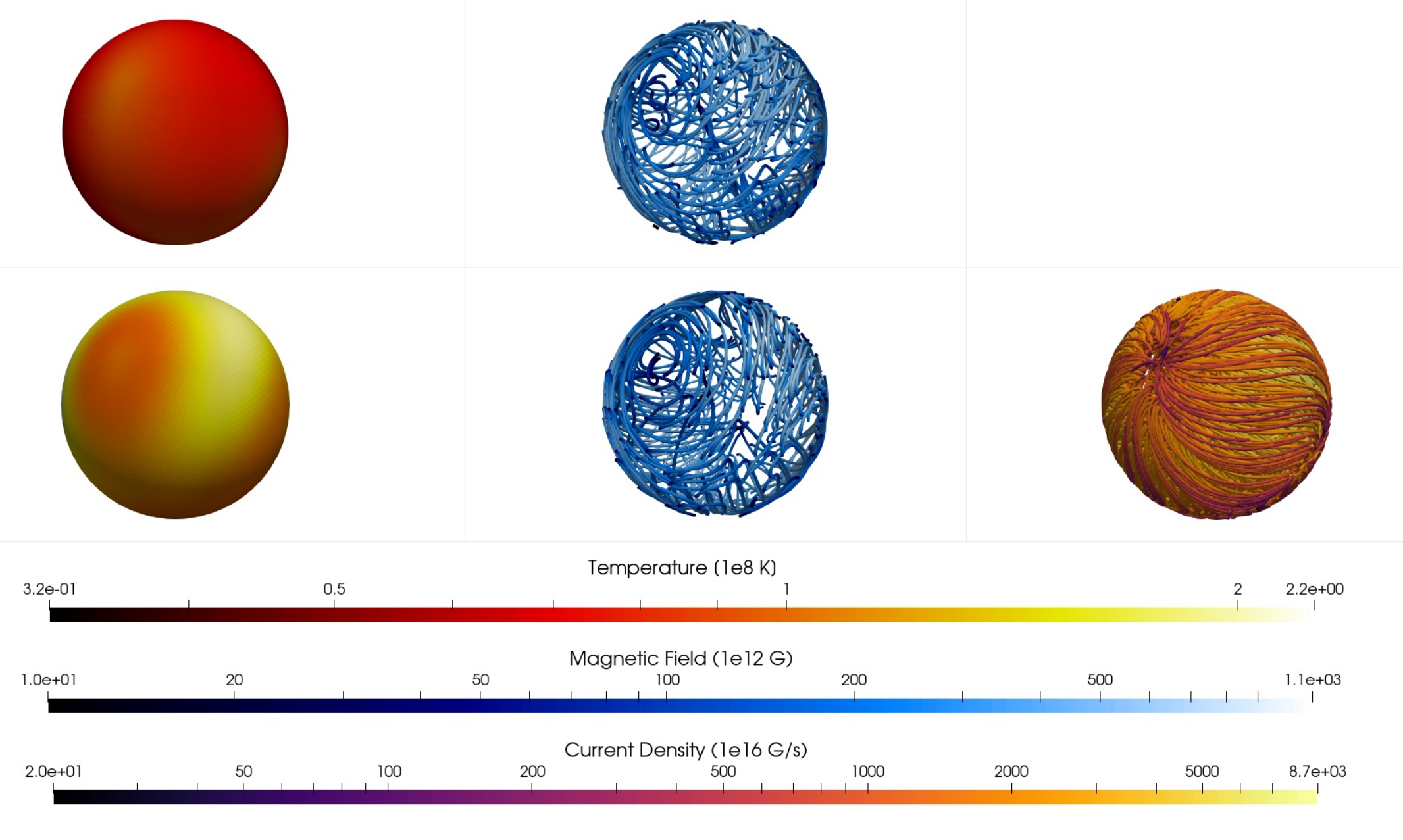}
    \caption{3D rendering at 2726 yr of the model \emph{SLy4-M.1.4-B14-L2-alt}.  Top: Simulation with non-evolving magnetic field. The two panels show, from left to right, the temperature and the magnetic field lines. Bottom: full magnetothermal simulation, which includes also the evolution of the magnetic field and Joule heating. The three panels show, from left to right, the temperature, the magnetic field lines, and the electric currents lines (related to the Ohmic heating). All the simulations adopt a mass of $M = 1.4\,M_\odot$, the \ac{EOS} SLy4, and a resolution of $N_r = 30$ and $N_a = 31$.  }
    \label{fig:3D_render_MT} 
\end{figure*}

In Fig. \ref{fig:comparison_LC} we have also included (orange dashed line) a variation of the run \emph{SLy4-M.1.4-B14-L2}, where the envelope described in Appendix B of \citet{Potekhin2015} is used instead of Gudmundsson's model. Similarly to Gudmundsson's model, the envelope in \citet{Potekhin2015} comprises heavy elements such as iron; however, unlike the former, the latter also incorporates the effect of the magnetic field in the modeling. As depicted in the figure, the cooling curves for the two models are comparable, with the magnetized envelope resulting in a slightly lower luminosity. This outcome will be discussed in more detail in Sec. \ref{sec:lightcurve}. 

Finally, we present the run \emph{SLy4-M.1.4-B14-L2-alt}, which has the same multipole components of the run \emph{SLy4-M.1.4-B14-L2}, but with a different field configuration, in which the contribution of the toroidal multipoles with $l=2$ has been enhanced. This run is presented in two versions: a purely thermal one, where the evolution of the magnetic field is frozen like in the cases previously studied, and a magnetothermal one \emph{SLy4-M.1.4-B14-L2-alt-MT}, where the magnetic field is evolved and Joule heating is correspondingly included. Since a full magneto-thermal simulation is considerably more expensive from a computational point of view than a simulation with a non evolving magnetic field, we restricted this run to a maximum time of $10^5\, \textrm{yr}$ only. 

In the magneto-thermal case, the effect of Joule heating can be appreciated in Figure \ref{fig:comparison_LC}. Here, the brown curves represent the magnetothermal run \emph{SLy4-M.1.4-B14-L2-alt-MT}, which can be compared with the purple curve representing the same configuration with a non-evolving magnetic field. In this case, the heating produced by the dissipation of the currents has the effect of increasing the luminosity of the star by a factor $\sim 2-3$ after the first $\sim 100 \, \textrm{yr}$ of evolution.  

A 3D rendering of this run (with an enhanced angular resolution of $N_a = 41$) at time $2726 \, \mathrm{yr}$ is reported in the bottom row of Fig. \ref{fig:3D_render_MT}.  
The figure represents, from left to right, the temperature $T_b$, the magnetic field, and the electric currents. The upper row represents the same run with the non-evolving magnetic field (\emph{i.e.} \emph{SLy4-M.1.4-B14-L2-alt}) at the same evolutionary time, with the same spacial orientation and with the same color range for the represented quantities, so that the images in the upper and bottom rows are directly comparable. 
As we can see, the temperature at the base of the envelope, in this case, is higher than in all the previous cases, due to the presence of the heating source. The other point to note is that, apart from an overall scaling, the angular distribution of $T_b$ is radically different with respect to the equivalent case with the non-evolving magnetic field. It appears in fact that the two distributions are almost inverted, with the hottest regions of one roughly corresponding to the coldest regions of the other. The reason may rely on the fact that the coldest regions in the case with non-evolving magnetic field, which are those with an intense non-radial field that provides an insulating effect, are also the ones that in the magnetothermal case develop stronger currents and consequently more Joule heating. This effect is well known from previous 2D studies like \citet{Pons2009} and \citet{Vigano2013}. This deep change in the $T_b$ angular distribution is particularly important because it is expected to have a major impact on the pulsed profile observed from the source, as we will see in the next section.

\subsection{{Light curves}}
\label{sec:lightcurve}

As we saw from the previous results, the magnetic-induced anisotropy in the thermal conductivity generates an inhomogeneous temperature distribution on the surface of the \ac{NS}, presenting hot spots and colder regions. While the star undergoes rotation, the observer's view of the hot spots on the stellar surface may become periodically obstructed. This results in a periodic modulation of the stellar X-ray flux, which is eventually observable and is indeed observed in several sources \citep[see \emph{e.g.}][]{KaspiBeloborodov2017, Esposito2021}.

Starting from our simulations, we compute the energy dependent flux (or, equivalently the phase-resolved spectrum) as detected by a distant observer. This is an important step because it allows us to link the results of our simulations with quantities that can be directly observed.

To this aim, we employ the ray-tracing code presented in \citet{PernaGotthelf2008} and \citet{Vigano2014} and generalized to work with a non-axisymmetric temperature map. 
The ray-tracing code takes as input a temperature map from \emph{MATINS}, and maps it via interpolation onto an internal, equally-spaced spherical coordinate grid.
The temperature map is used to calculate the local emission spectrum on the surface of the \ac{NS}. The phase dependent spectrum is computed (neglecting absorption from the interstellar medium) by integrating the specific intensity allover the stellar surface \citep{Page1995}:
\begin{align}
    F(E_\infty, \gamma) = \frac{2\pi}{ch^3}\frac{R^2_\infty}{D^2}E^{2}
     \int^1_0 2 x \, dx \int^{2\pi}_0 \frac{d\phi}{2\pi}B_\nu[T_s(\theta(x),\phi), E].
    \label{eq:phase_resolved_spectrum}
\end{align}
Here $E$ and $E_\infty$ 
denote, respectively, the energy of photons at the star surface and the redshifted energy detected by an observer at infinity. $R_\infty$ is the radius of the star at infinity, $D$ is the distance from the source, while the parameter $\gamma$ represents its rotational phase. We assume the local spectrum to be a blackbody $B_{\nu}(T)$, and also that the emission at the surface is isotropic. Both assumptions neglect all the possible spectral and anisotropic effects introduced by the passage of radiation trough the atmosphere of the \ac{NS}, which further complicates the modelling (see \emph{e.g.} Sec 2.3 of \citealt{Ozel2013}). While we make this assumption for the sake of simplicity in this first work and in order to highlight the dependence of the lightcurve on the temperature map, we note that the spectral and anisotropic distortions introduced by the atmosphere are second order effects. The angles $\theta$ and $\phi$ in Eq. (\ref{eq:phase_resolved_spectrum}) are the latitude and longitude angles on the star surface measured relatively to the axis aligned with the \ac{LOS}, while
$x \equiv \sin \delta$, where $\delta$ is the emission angle of the photon with respect to the normal to the stellar surface. At a given latitude $\theta$ only the photons emitted with a given angle $\delta$ will reach the observer. In our ray-tracing code, the relation between $\theta$ and $\delta$ is provided by the formula derived by \citet{Beloborodov2002}, which is valid for a Schwarzschild spacetime,
\begin{equation}
    1-\cos \theta = (1-\cos\delta)\Bigl(1 - \frac{R_s}{R}\Bigr)^{-1},
\end{equation}
where $R_s$ and $R$ are the Schwarzschild and the \ac{NS} radius, respectively. It is important to acknowledge that the equation presented above serves as an approximation of the geodesic equation, and its accuracy may diminish for significant values of $\delta$ (refer to the detailed discussion in \citealt{LaPlaca2019}). Nevertheless, within the context of our specific application, where the range of $\delta$ is confined to $\delta \le \pi/2$, this equation maintains an accuracy level 
better
than $1\%,$ which proves to be satisfactory for our intended purposes, while significantly reducing the computational cost of the code.

The flux in Eq. (\ref{eq:phase_resolved_spectrum}) depends on the orientation of the temperature map with respect to the \ac{LOS}, which also depends on the phase $\gamma$. This orientation (hereafter \emph{geometry}) is defined by two angles: $\psi$, that is the angle between the \ac{LOS} and the rotation axis, and $\chi$, that is the angle between the z-axis in \emph{MATINS} and the rotation axis. The phase-resolved spectrum can be calculated once $\psi$ and $\chi$ are defined. 

Here we present the phase-resolved spectrum for three selected cases: \emph{Sly4-M1.6-B14-L1}, \emph{Sly4-M1.4-B14-L2} and \emph{Sly4-M1.4-B14-L10}. For all the cases we chose the temperature map at $t = 2726\, \mathrm{yr}$, a geometry defined by $\psi = \chi = \pi/2$, and a reference distance of $D = 4\, \mathrm{kpc}$. In all the ray-tracing simulations, the resolution of the grid is set to 150 points both in $\theta$ and $\phi$ directions. The temperature map obtained by \emph{MATINS} is interpolated on this grid via a barycentric coordinate interpolation.

\begin{figure}
    \centering
    \includegraphics[width = 0.45\textwidth]{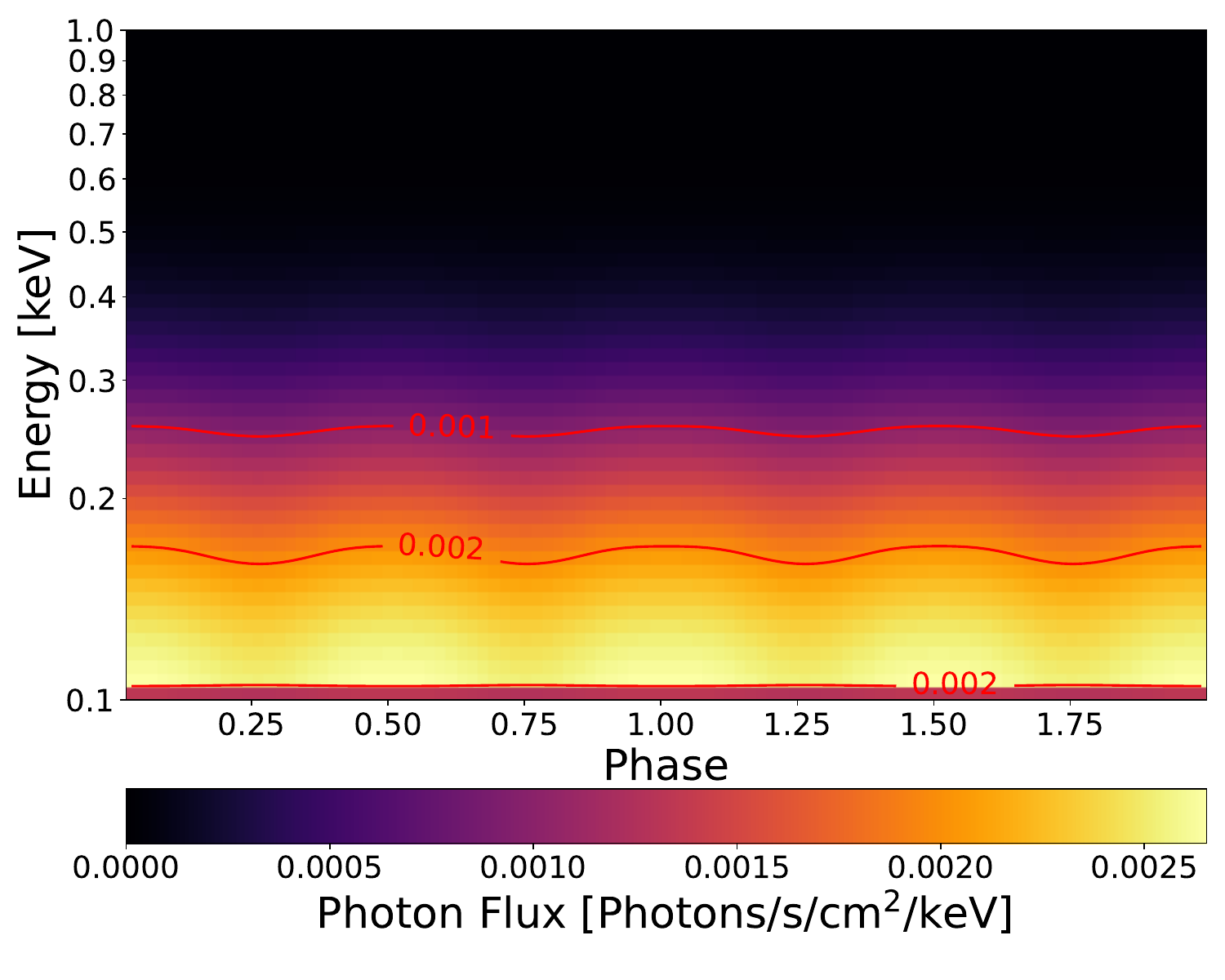}
    \includegraphics[width = 0.45\textwidth]{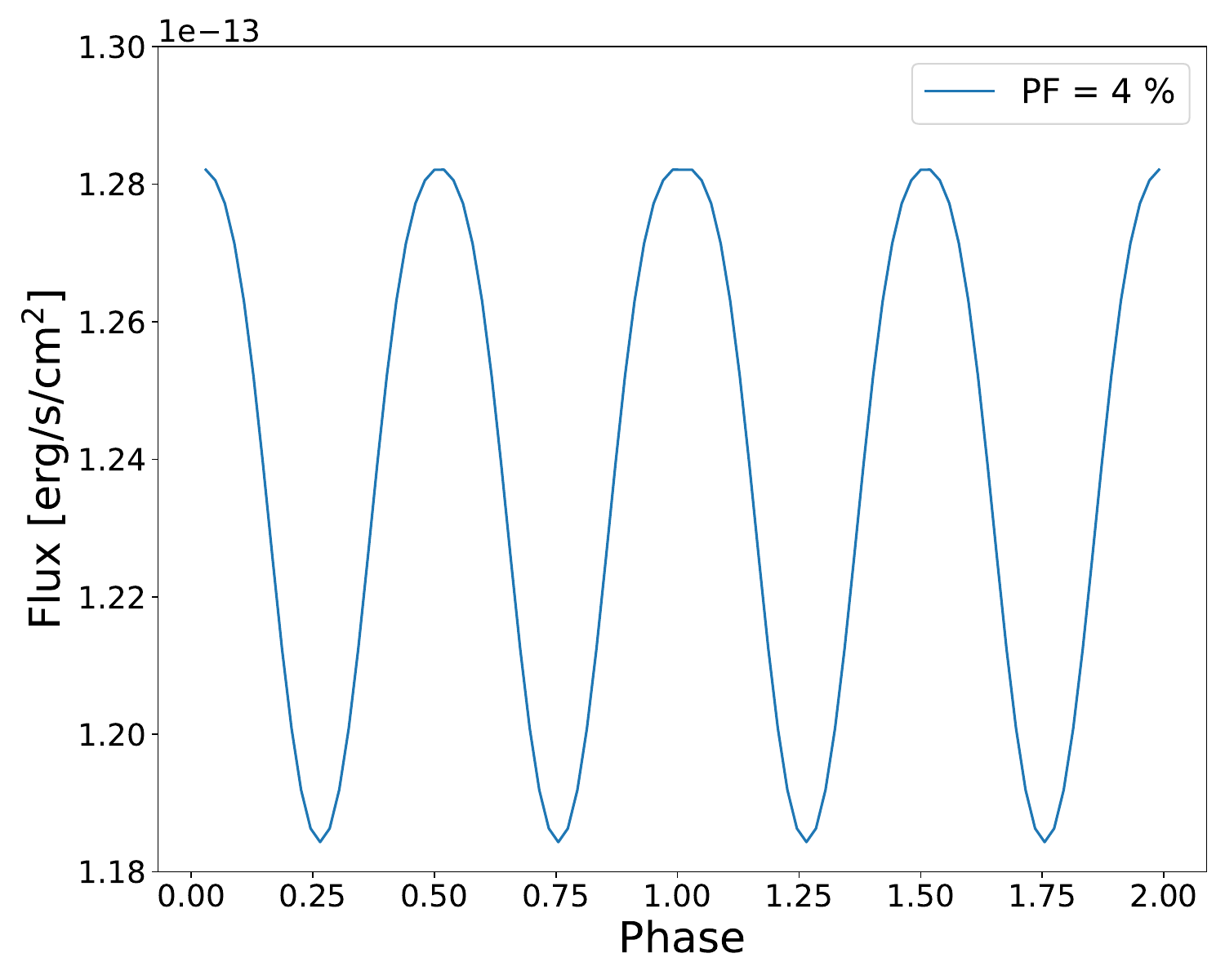}
    \caption{Simulated observed emission from the \emph{MATINS} run \emph{Sly4-M1.4-B14-L1} evaluated at $2726 \, \mathrm{yr}$, for a source located at $4 \, \mathrm{kpc}$, negligible $N_H$, and a geometry defined by the angles $\chi=\psi=90^\circ$. \emph{Top}: phase-energy diagram. The color code denotes the photon flux. Red lines mark isocontour lines. \emph{Bottom}: bolometric flux as a function of the phase measured in erg/(s $\mathrm{cm}^2$). The label marks the \ac{PF}.}
    \label{fig:dipole_fiammelle}
\end{figure}

 The first case that we present in Fig.~\ref{fig:dipole_fiammelle}
 is the one referring to the simulation \emph{Sly4-M1.4-B14-L1}. Here the 
 top
 panel displays the phase-resolved spectrum, in an energy-phase plane. The color code represents the specific flux measured in $\mathrm{photons/(s\, cm^2\, keV)}$, and the red continuous lines mark iso-flux contours. The bottom panel shows the flux integrated over the energy range $0.1-10 \, \mathrm{keV}$, measured in $\mathrm{erg/(s \,cm^2)}$. Since most of the emission occurs in this energy window, this flux coincides approximately with the bolometric thermal flux. The label here denotes the pulsed fraction (PF), that is the magnitude of the pulsed emission, computed as the difference between the maximum and the minimum of the flux divided by the sum of 
 these.

 This run is characterized by a dipolar configuration of a purely poloidal magnetic field. Here the magnetic dipolar moment is coincident with the $z$-axis in \emph{MATINS}. As such, the angle $\chi$, in this case, is the inclination angle \emph{i.e.} the angle between the rotation and the magnetic moment. The temperature map is characterized by a cold equatorial belt and hotter poles, which acts like two broad antipodal hot spots, centered on the magnetic axis. These features, along with the chosen geometry, are reflected on the pulsed profile, which shows two symmetric peaks within a single rotational phase.  With the chosen geometry, which is the one that maximizes the \ac{PF}, the \ac{LOS} passes through the poles at $\gamma = 0, 0.5$, where we have the maximum of the emission, while the minima at $\gamma = 0.25, 0.75$ coincide with the \ac{LOS} pointing at the equatorial cold belt. Note that in this specific case, due to the peak symmetry given by the magnetic configuration and inclination, the spin period inferred by the X-ray light curves would be half the real value.

\begin{figure}
    \centering
    \includegraphics[width = 0.45\textwidth]{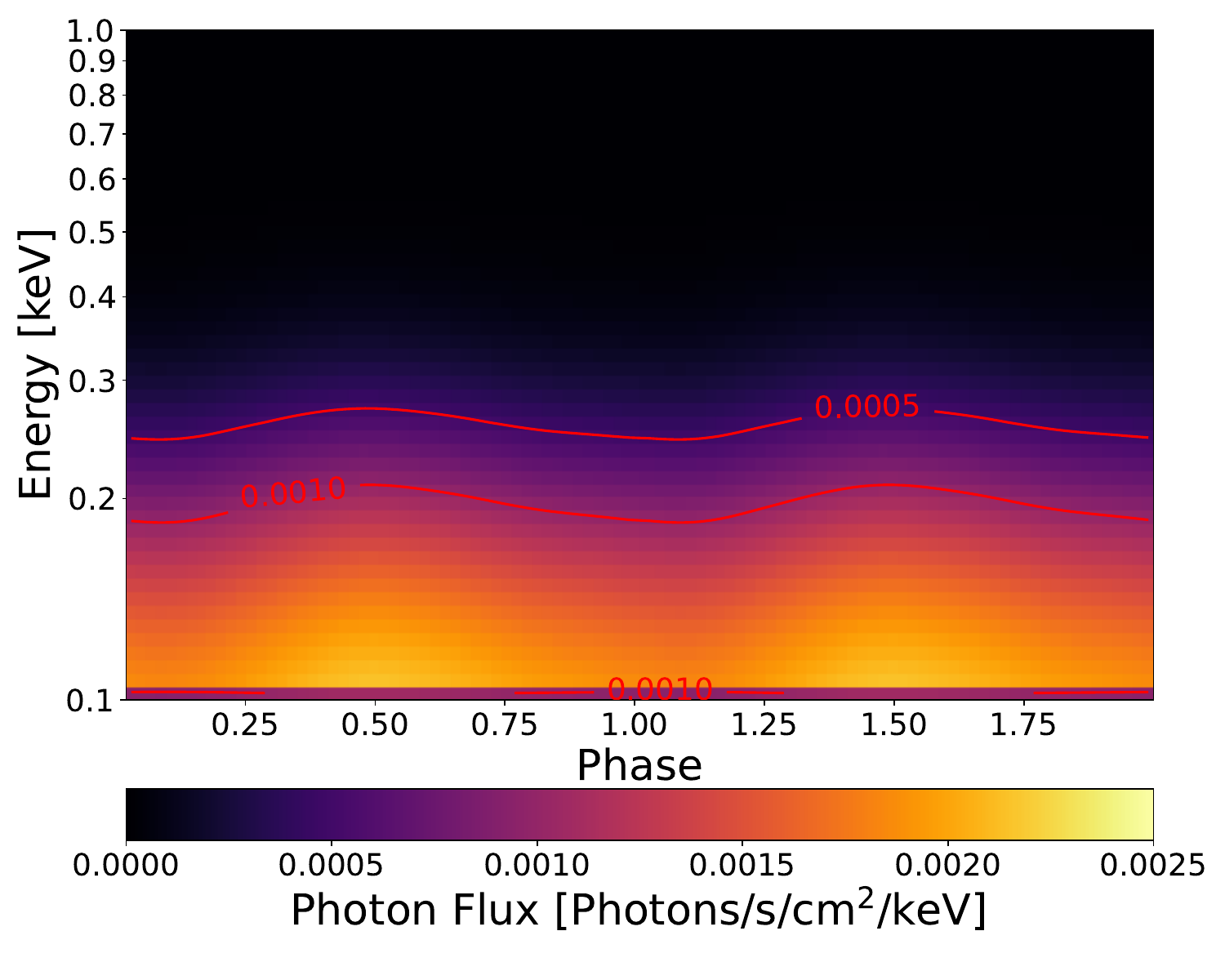}
    \includegraphics[width = 0.45\textwidth]{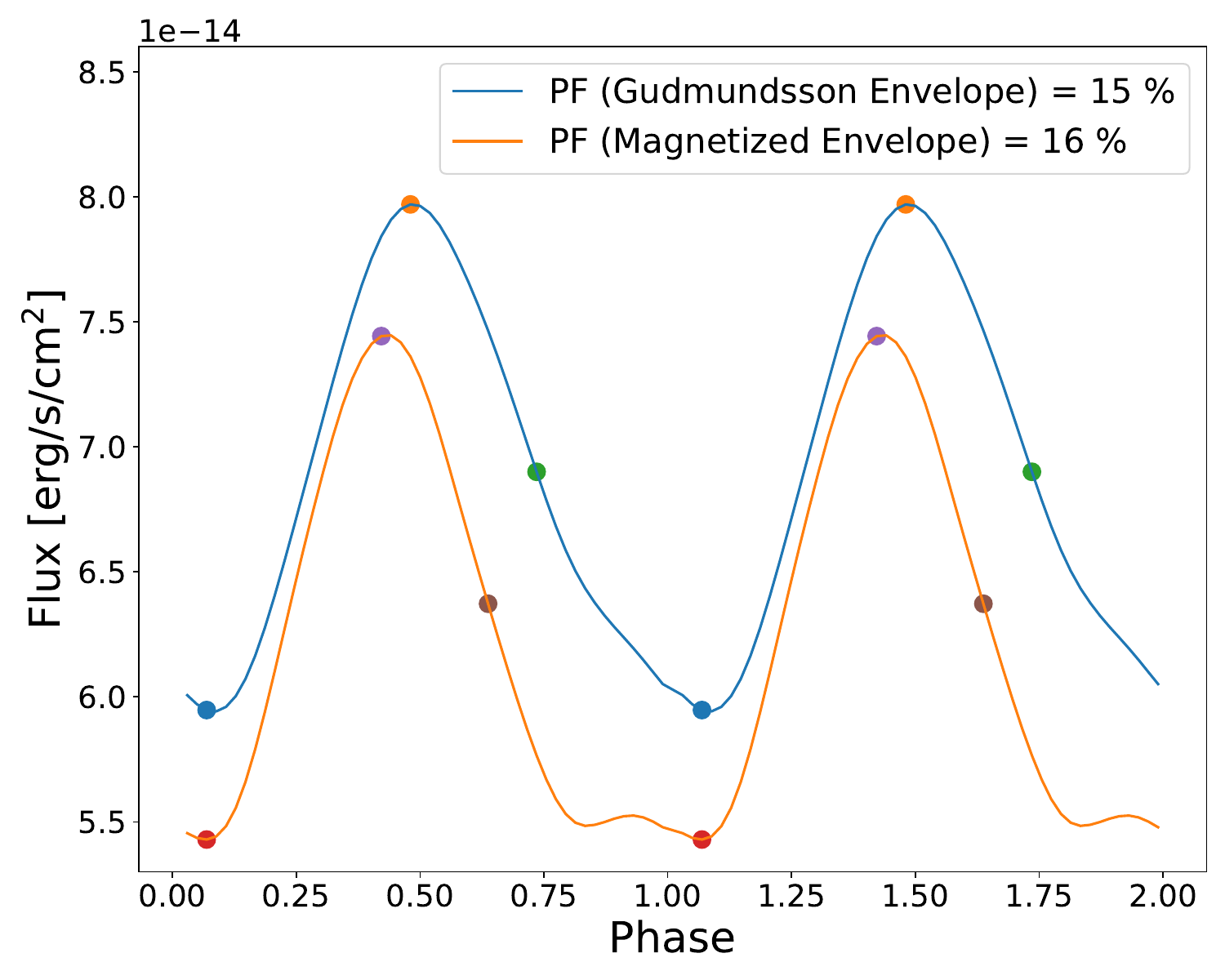}
    \caption{Same as Figure \ref{fig:dipole_fiammelle} but for the simulation \emph{Sly4-M1.4-B14-L2}. In the bottom panel colored dots mark three different selected phases, coincident with the minimum (blue), the maximum (orange), and an intermediate (green) value of the flux. The orange curve represents the same run using an iron envelope model instead of the Gudmundsson envelope model. The colored dots mark the minimum (red), maximum (purple) and an intermediate (brown) value of the flux. }
    \label{fig:quadrupole_fiammelle}
\end{figure}

\begin{figure*}
    \centering
    \includegraphics[width = \textwidth]{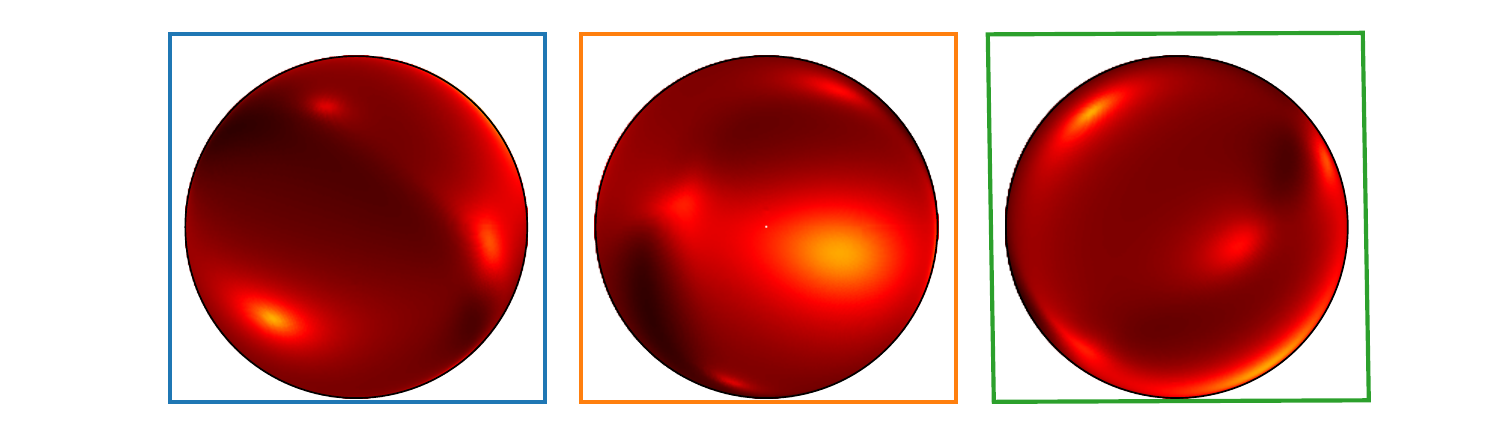}
    \includegraphics[width = \textwidth]{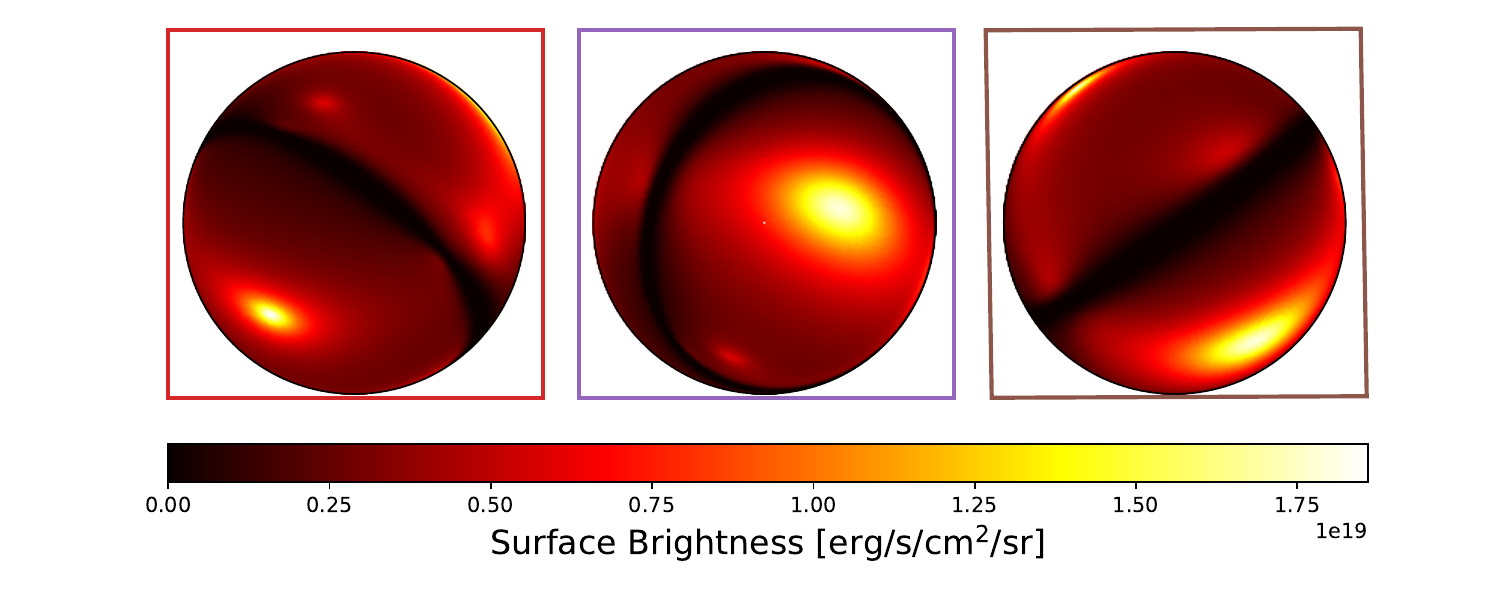}
    \caption{Surface brightness of the star projected on the observer plane at three different rotational phases. Top: simulation with Gudmundsson envelope. Bottom: simulation with the iron magnetized envelope. The rotational axis is horizontal in each image. The images refer to the three (per simulation) rotational phases highlighted in the right panel of Fig. \ref{fig:quadrupole_fiammelle}. The color of the frame around each image matches the color of the dot in the previous figure to which the image refers. Note that the phases in the top row and in the bottom row are close but not exactly equal (refer to Fig. \ref{fig:quadrupole_fiammelle}).}
    \label{fig:surface_brightness_quadrupole}
\end{figure*}

In Figure \ref{fig:quadrupole_fiammelle}, we display the same quantities as in Figure \ref{fig:dipole_fiammelle}, but for the \emph{MATINS} simulation \emph{Sly4-M1.4-B14-L2}. 
This case is characterized by a more complex magnetic field, which results in a temperature map that is no longer axisymmetric like the previous case. As we can see from the figure, the phase-resolved spectrum and the pulsed profile show only a single slightly-asymmetric peak during the rotational phase. The \ac{PF} is higher than the previous case, reaching a value of $\sim 15 \%$. In the bottom panel of the figure we also show (orange curve) the pulsed profile for the variant of the same run, computed with the magnetized envelope of \citet{Potekhin2015}. For this case we note that, although the overall flux is lower, the \ac{PF} is slightly higher, and a second small peak appears in proximity of the flux minimum. 

In order to illustrate the origin of the pulsed profiles, which are less straightforward to interpret than the previous one, we report in Fig. \ref{fig:surface_brightness_quadrupole} the surface brightness of the star at three selected phases corresponding to the minimum (left), the maximum (center), and an intermediate value of the flux (right). The surface brightness has been computed by integrating the black-body specific intensity into the energy interval $0.1-10 \, \mathrm{keV}$ at each point of the visible surface of the star and then projecting each point into the observer taking into account the propagation of photons in a Schwarzschild spacetime. In the top row we report the result for the standard run, while in the bottom row we report the result for the variant with the magnetized envelope. From the figure, we can see that the temperature map is characterized by several small spots of semi-aperture angle $\sim 10-15^\circ$ and a bigger, brighter spot of semi-aperture angle $\sim 35-45^\circ$. The latter, due to its enhanced size and brightness, dominates the emission producing the peak at $\gamma \sim 0.5$, when it is oriented face-on with respect to the observer. The minimum of the emission in turn occurs at $\gamma \sim 0$, when the big spot is completely out of sight. Comparing the two runs characterized by different envelopes we can appreciate how, in the case of the magnetized envelope, the gradients in surface brightness (\emph{i.e.} in the surface temperature) are larger with respect to the Gudmundsson envelope case; this is in agreement with the slightly higher \ac{PF} observed in this case. Where the field is tangential to the surface, the magnetized envelope model provides an enhanced screening; viceversa when the field is radial, the surface temperature is higher with respect to the Gudmundsson case, as also reported in \citet{Dehman2023b}. This is the case of the two quasi-antipodal hot spots visible in the bottom row of Fig. \ref{fig:surface_brightness_quadrupole}, which are hotter with respect to the equivalent regions in the top row. It is worth noticing that the magnetized envelope enhances the temperature only for these two spots and not for the other existing hot regions. The reason is that these two spots are characterized by a field with a substantial radial component, and as such they are poorly screened by the magnetic envelope. On the other hand, the other hot areas have a high temperature not because of the field orientation but due to its weak local intensity; the field however is tangential to the surface in those locations, such that they are subjected to a higher screening when a magnetized envelope is used.

Note that, unlike the dipolar case, in this case the $z$-axis does not pass through the hottest region. As such it is not guaranteed that this geometry ($\psi = \chi = 90^\circ$) is the one that maximizes the \ac{PF}. Nevertheless, we can see from Fig. \ref{fig:surface_brightness_quadrupole} that since the widest and hottest region passes close to the center of the figure at the maximum of the pulse profile, while it completely disappears at the its minimum, it is reasonable to expect that this configuration may not be too far from the one with highest \ac{PF}.

\begin{figure}
    \centering
    \includegraphics[width = 0.45\textwidth]{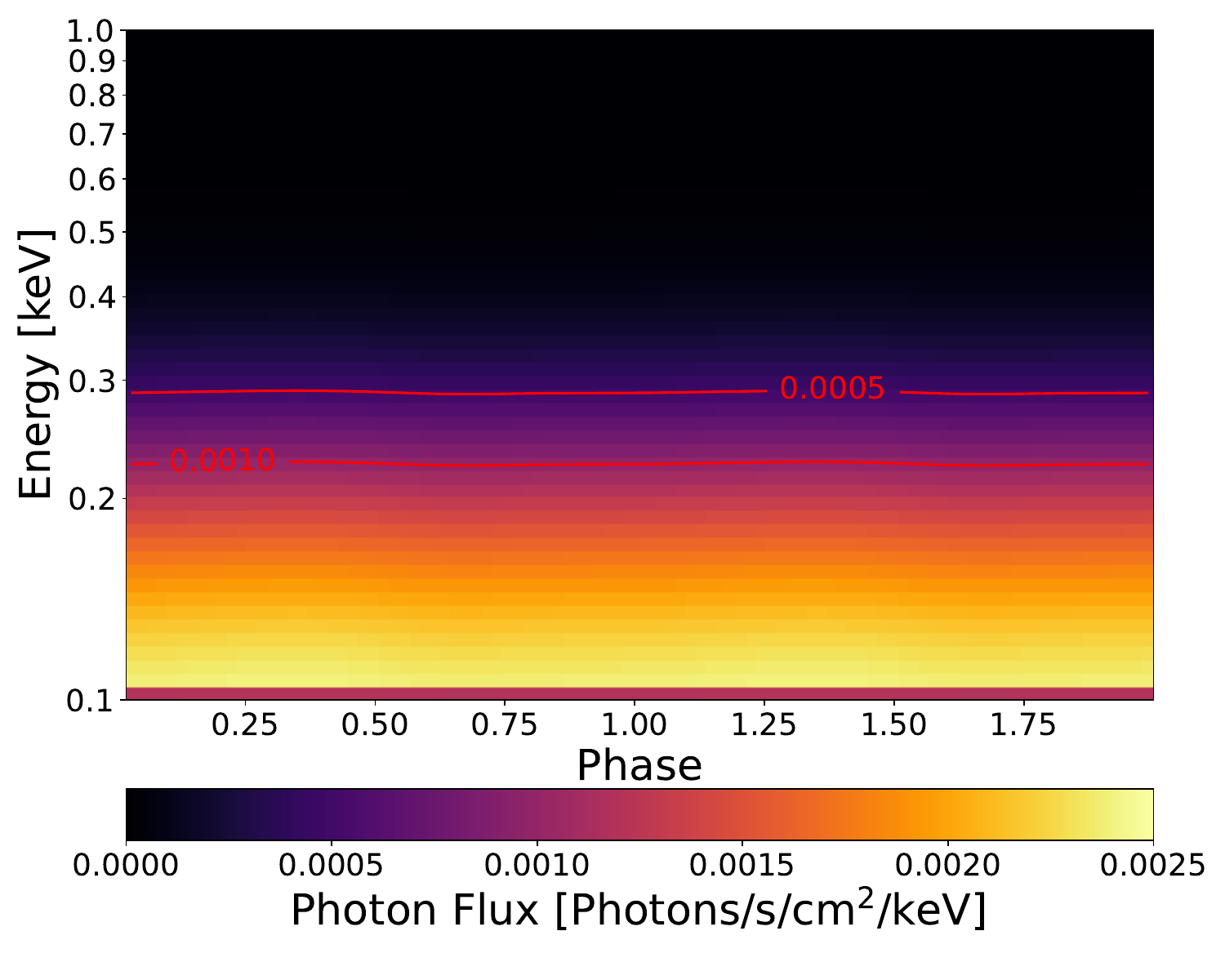}
    \includegraphics[width = 0.45\textwidth]{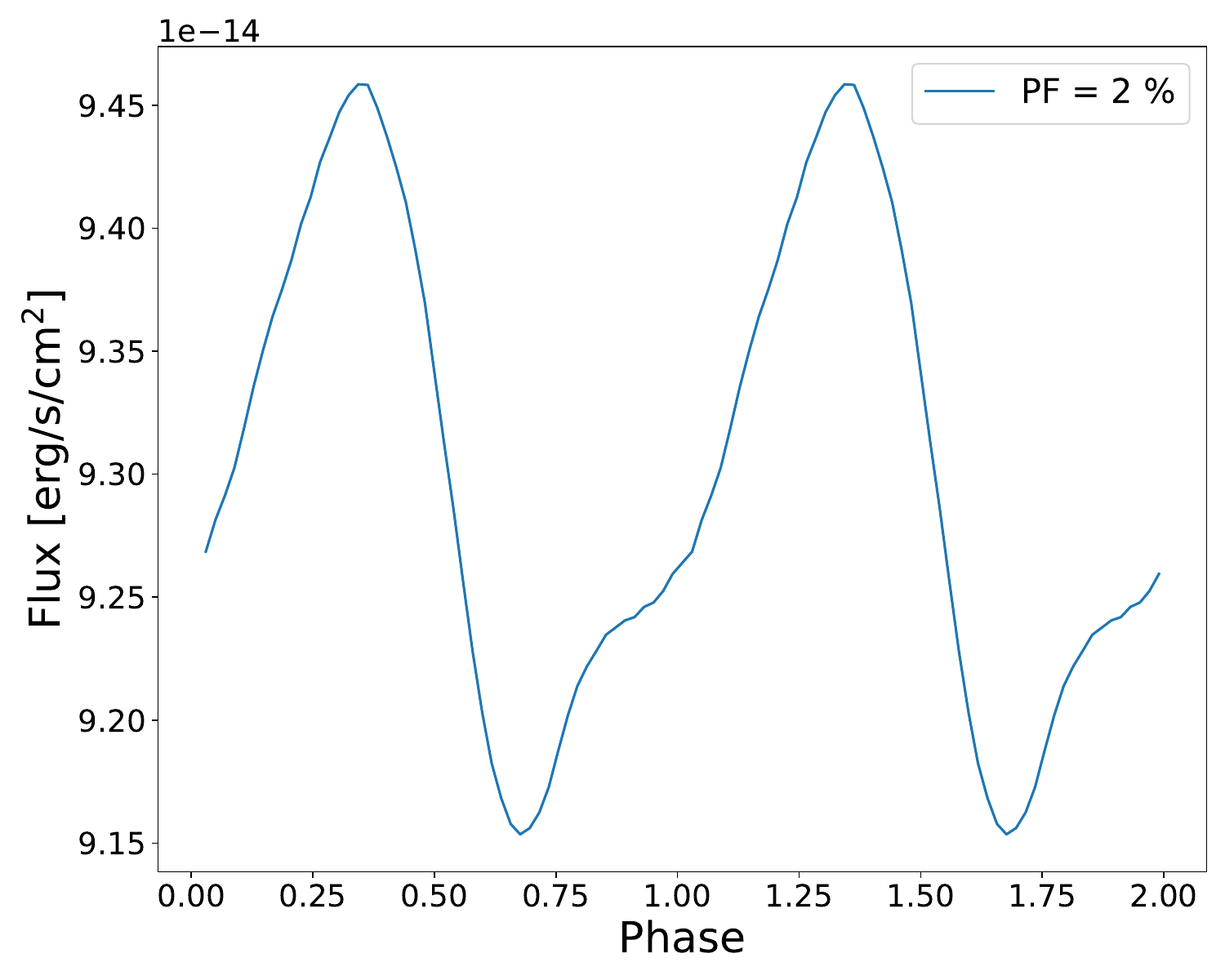}
    \caption{Same as Figure \ref{fig:dipole_fiammelle} and Figure \ref{fig:quadrupole_fiammelle} for the run \emph{Sly4-M1.4-B14-L10}.}
    \label{fig:L10_fiammelle}
\end{figure}

The final case, depicted in Fig. \ref{fig:L10_fiammelle}, corresponds to the model \emph{Sly4-M1.4-B14-L10}. Similarly to the quadrupolar case, this particular scenario exhibits one peak that appears within a single rotational phase. However, in this case the shape of the profile is more complex, reflecting the more complex shape of the hot region, which can also be appreciated in the top right panel of Fig. \ref{fig:3D_rendering}. Around $\gamma \sim 0.8$ we notice a step-like shape which has a shift in phase of $\Delta \gamma \sim 0.5$, with respect to the peak. This particular profile suggests the presence of two hot regions positioned at almost antipodal positions and characterized by different brightness levels. 

Comparing the \acp{PF} of the three cases presented here, it is worth mentioning that the \emph{Sly4-M1.4-B14-L10} configuration has the lowest \ac{PF}. This is due to the fact that this temperature map is characterized by a high number of hot regions, so that while some of them are occulted by the star rotation some other are revealed, thus reducing the \ac{PF}. For this reason, even if also in this case, like for the quadrupole one, the choice of $\psi=\chi=90^\circ$ may not necessarily be the one maximizing the \ac{PF}, we argue that the maximum  \ac{PF} is probably not too far from the value reported here.

The second case with the lowest \ac{PF}, comparable to \emph{Sly4-M1.4-B14-L10}, is the dipole. In this case the low value of the \ac{PF} can be ascribed to the wideness of the polar hot regions which can never be completely occulted, unlike for the quadrupolar case.

\begin{figure}
    \centering
    \includegraphics[width = \columnwidth]{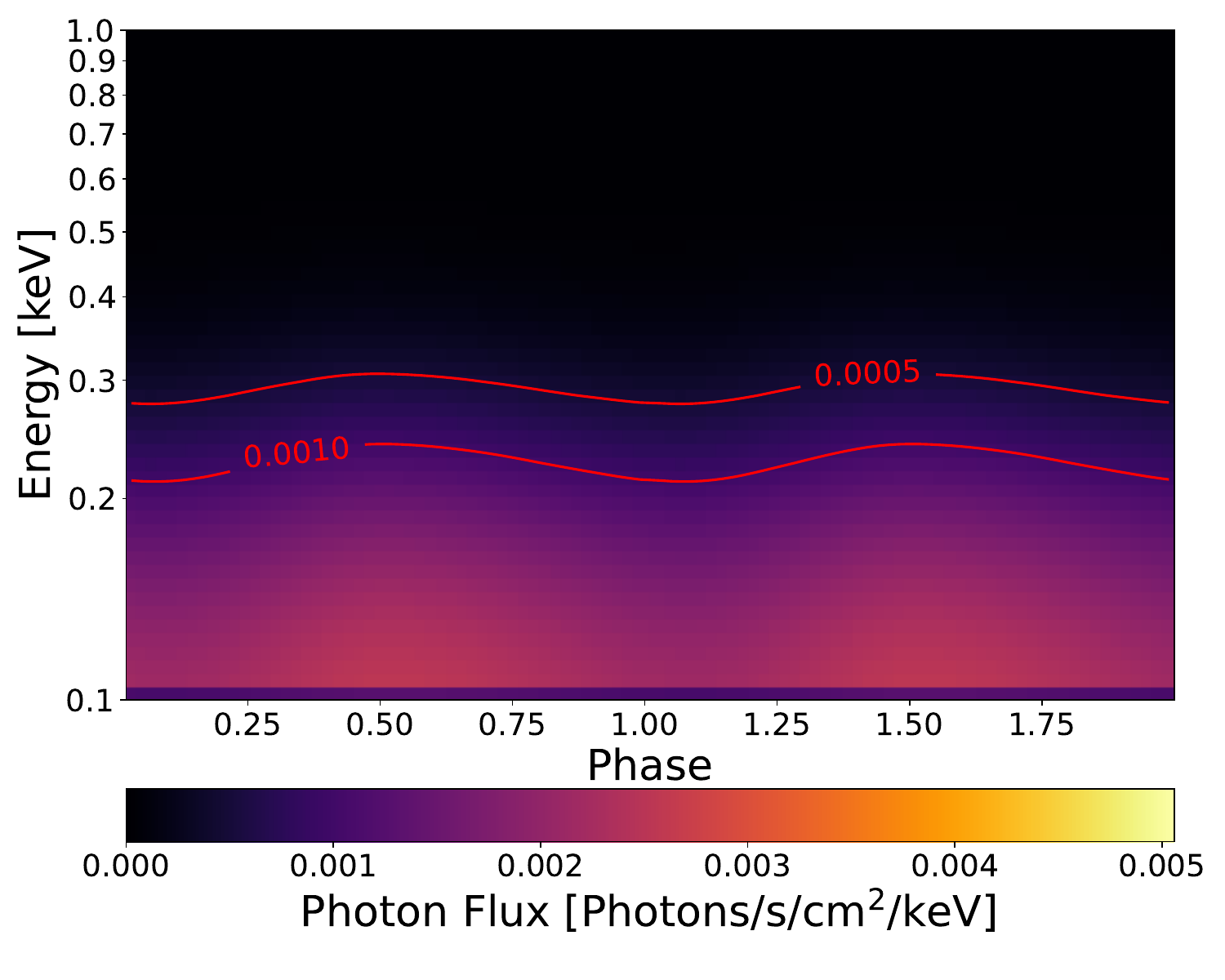}
    \includegraphics[width = \columnwidth]{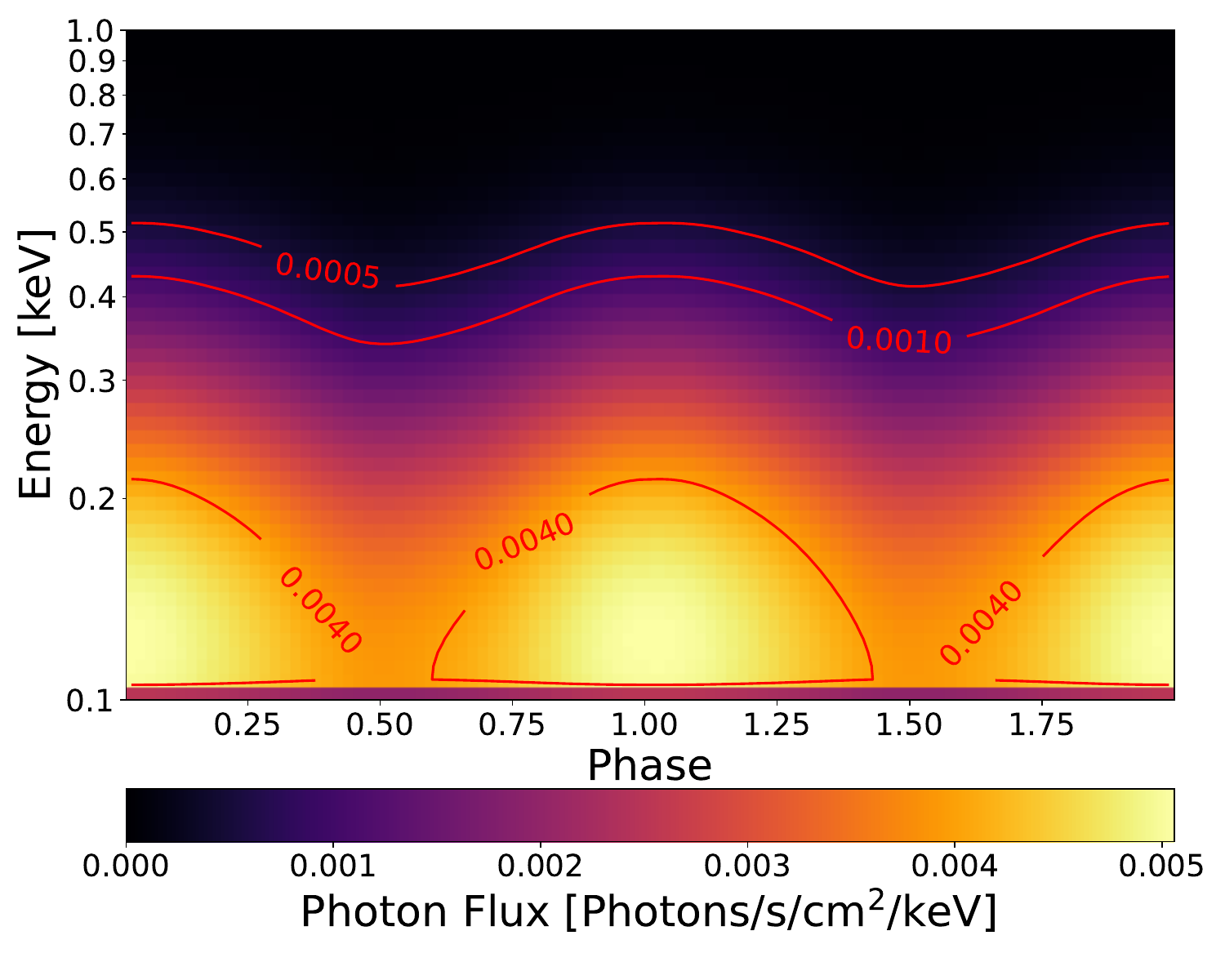}
    \includegraphics[width = \columnwidth]{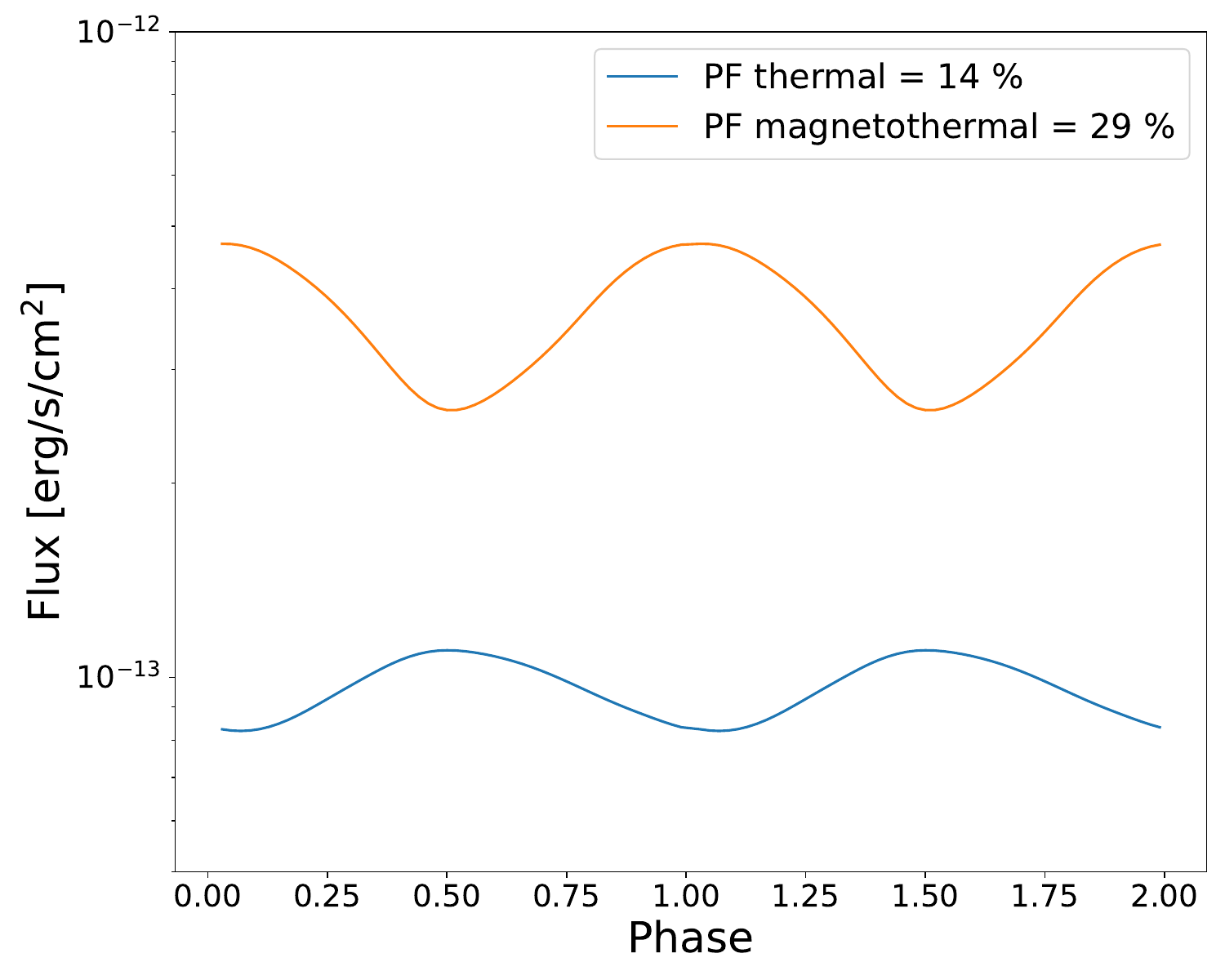}
    \caption{Phase-resolved spectrum and pulsed profile at $t=2726\,\mathrm{yr}$ of the simulation \emph{Sly4-M1.4-B14-L10} with the thermal evolution only and the coupled magnetothermal problem. {\em Top}: phase-resolved spectrum with the non-evolving magnetic field. {\em Middle}: phase-resolved spectrum with evolving magnetic field. {\em Bottom}: pulsed profiles with non-evolving (blue) and evolving (orange) field.}
    \label{fig:T_MT_L2_fiammelle}
\end{figure}

To conclude, we consider the run \emph{Sly4-M1.4-B14-L10}, with the inclusion of the full magnetothermal evolution. Also for this run, we consider a temperature map at \emph{i.e.} $t = 2726\,\mathrm{yr}$, which gives enough time for the magnetic field to experience some Hall and Ohmic evolution. 
The result is presented in Fig. \ref{fig:T_MT_L2_fiammelle}. The top and the middle panel represent the phase-resolved spectrum expressed in photon counts for the case with non-evolving (top) or evolving (middle) magnetic field. The bottom panel represents the pulsed profile of the two cases. Hereafter we will refer to the first case as \emph{thermal} case, and to the second case as \emph{magnetothermal} case. In this plot, we can notice two interesting features: first of all, as expected, the Ohmic dissipation contributes to increasing the surface temperature, and consequently the flux, which increases by almost a factor 5, while the peak of the spectrum moves to higher energies. Secondly, there is a phase-shift of $\Delta \gamma \sim 0.5$ of the profile maximum between the two cases, such that the peak of the pulsed profile in the case with the non-evolving field coincides with the minimum of the  magnetothermal profile. This shift denotes a switch in the hot regions with and without the Ohmic dissipation, as already noted in the previous section. Finally, we note that the magnetothermal case is characterized by a \ac{PF} of $\sim 30\,\%$, which is a value higher than the \acp{PF} of all the other cases studied here, where the field is not evolving.  

\section{Summary}
\label{sec:conclusions}

The present work, together with \citet{Dehman2023_MATINS} introduces MATINS (MAgneto-Thermal evolution of Isolated Neutron Star), a three-dimensional code designed to investigate the secular magnetothermal behavior of a \ac{NS}, featuring a self-consistent coupling between thermal and magnetic processes, a realistic star structure based on different possible cold dense matter EOS, and detailed local microphysics. While the previous study by \citet{Dehman2023_MATINS} described how the code solves the magnetic evolution equation with a particular emphasis on the cubed-sphere grid, our focus here is to illustrate the formalism, testing and some applications of the thermal evolution part. The results from the fully coupled induction and heat diffusion equations have already been presented in \citet{Dehman2023}, which considered a realistic magnetic field configuration derived from a \emph{proto-\ac{NS}} dynamo simulation \citep{reboul2021}, along with the most updated envelope model \citep{Potekhin2015}.

The MATINS code utilizes a finite volume scheme to solve the thermal evolution (energy conservation) equation in the cubed sphere grid coordinate system \citep{Ronchi1996J}. 
The equation is solved via an implicit scheme, which is optimal to treat the stiffness of the neutrino emissivity, and allows us to use larger time steps, due to the unconditional stability of the method.   

We included in our code a numerical implementation of the microphysics that exploits the public code by A. Potekhin\footnote{http://www.ioffe.ru/astro/conduct/}, which provides a detailed description of microphysical parameters in the outer and inner crust, and in the core of the \ac{NS}, along with the inclusion of all the relevant neutrino emission processes in the crust and in the core. The structure of the \ac{NS} is obtained by solving the \ac{TOV} equation at zero temperature. This requires the choice of an \ac{EOS} that can be selected from the public online database CompOSE\footnote{https://compose.obspm.fr/}. 



We applied MATINS to various scenarios, including different \acp{EOS} and masses, and also to a case with a fully coupled magnetothermal evolution. 
We observed the impact of different magnetic field configurations on the distribution of temperature inside the star and in particular on the interface between the envelope and the crust, which is directly related to the temperature map at the surface of the star. We appreciated how the increasing complexity of the field translates into the complexity of the temperature map, where field configurations with small-scale structures result in the formation of smaller hot spots. 

Moreover, we coupled the output of MATINS to a general-relativistic ray-tracing code to calculate the phase-dependent spectrum observed by a distant observer from a thermally emitting \ac{NS}. This code is essential to link the results provided by MATINS in term of surface temperature maps with the observations. In particular, we showed how the simulations discussed in this paper produce different phase-resolved spectra, pulsed profiles, and pulsed fractions for a given orientation of the \ac{LOS} with respect to the temperature map and star's rotation axis (geometry). The few cases presented in the paper represent just a proof of concept of the synergy between our ray-tracing code and MATINS, and a more systematic study considering a wider sample of simulations and different geometries will be left for future work.

In conclusion, by considering the interplay between thermal and magnetic processes, MATINS provides a more comprehensive and realistic understanding of the complex thermal histories of \ac{NS}. The versatility of the code in handling various scenarios, together with its successful application in testing and simulations, offers promising potential for advancing our knowledge of \ac{NS} characteristics and their evolutionary history. 
In particular, it is an essential tool to simulate and fit simultaneously the timing properties (controlled by the dipolar surface field, \citealt{Vigano2013}), the thermal luminosity and the X-ray light curves.

\section*{acknowledgments}
SA, NR, CD are supported by the European Research Council (ERC) under the European Union’s Horizon 2020 research and innovation programme (ERC Consolidator Grant "MAGNESIA" No. 817661, PI: NR), and acknowledge partial support from grant SGR2021-01269 (PI: Graber/Rea). CD acknowledges the Nordita Visiting PhD Fellow program.
DV is supported by the European Research Council (ERC) under the European Union’s Horizon 2020 research and innovation programme (ERC Starting Grant "IMAGINE" No. 948582, PI: DV). JAP acknowledges support from the Generalitat Valenciana grants ASFAE/2022/026 (with funding from NextGenerationEU PRTR-C17.I1) and the AEI grant PID2021-127495NB-I00. SA, DV, CD and NR acknowledge support from ``Mar\'ia de Maeztu'' award to the Institut de Ciències de l'Espai (CEX2020-001058-M). 
RP acknowledges support by NSF award AST-2006839.
CD's work has been carried out within the framework of the doctoral program in Physics of the Universitat Autònoma de Barcelona. We thank Carlos Palenzuela and Borja Miñano (UIB-IAC3) for the computational support. We thank Davide de Grandis for the useful discussions. The authors acknowledges the anonymous referee for the suggestions that
help us to improve the presentation of our manuscript.

\section*{Data Availability}
Data available on request.

\vspace{5mm}

\bibliographystyle{mnras}
\bibliography{sample631} 

\appendix

\section{Cubed sphere grid: metric and operators}
\label{sec:CubedSphere}

In this section, we report some of the geometric elements and operators for the cubed sphere grid that have been used in the calculations throughout the present work. All these results were originally derived by \citet{Ronchi1996J} and further complemented by \cite{Dehman2023_MATINS}, where we also adapted them to incorporate relativistic corrections owing to spacetime curvature, manifesting in the form of the function $e^{\lambda(r)}$. This function is the usual $\sqrt{g_{rr}}$ of a static spherically symmetric spacetime, whose metric writes as: 

\begin{equation}
    ds^2 = -e^{2\nu(r)}c^2dt^2 + e^{2\lambda(r)}dr^2 +r^2d\Omega,
\end{equation}
where $e^{\nu(r)}$ is the lapse function and $\Omega$ is the solid angle. The dependence of $\lambda$ with the radius depends on the mass distribution inside the star $m(r)$, since: 

\begin{equation}
    \lambda(r) = -\frac{1}{2}\ln \Bigl[1 - \frac{2Gm(r)}{c^2r^2}\Bigr].
\end{equation}

The lapse function instead is determined by solving the \ac{TOV} equation \citep{Tolman1939, Oppenheimer1939,PonsVigano2019}.

As in \citet{Ronchi1996J}, we considered a unitary vector basis $\mathbf{\hat{e}_r}$, $\mathbf{\hat{e}_\xi}$ and $\mathbf{\hat{e}_\eta}$.
The metric of the cubed sphere writes as: 

\begin{equation} 
g_{ij}=
\begin{pmatrix}
1 &0  & 0\\
0 & 1 &  - \frac{XY}{CD} \\
0 &  - \frac{XY}{CD} & 1
\end{pmatrix},
\label{eq:metric_matrix}
\end{equation}

while the inverse metric as:
\begin{equation} g_{ij}^{-1}=
\begin{pmatrix}
1 &0  & 0\\
0 & \frac{C^2 D^2}{\delta} &   \frac{CDXY}{\delta} \\
0 &  \frac{CDXY}{\delta} & \frac{C^2 D^2}{\delta}
\end{pmatrix}
\label{eq:inverse_metric}
\end{equation}
where X, Y, C, D and $\delta$ are functions of the angular coordinates $\xi, \eta$ defined as:

\begin{align}
    X(\xi) = \tan(\xi)\, ,\notag\\
    Y(\eta) = \tan(\eta)\, ,\notag\\
    C(\xi) = \sqrt{1 + X^2(\xi)} = 1/\cos(\xi)\, ,\\
    D(\eta) = \sqrt{1 + Y^2(\eta)} = 1/\cos(\eta)\, ,\notag\\
    \delta(\xi, \eta) = 1 + X^2(\xi) + Y^2(\eta)\notag\,.
\end{align}

The (contravariant) components of the length infinitesimal element writes as: 

\begin{align}
 &dl^r = e^{\lambda(r)}dr\,, \notag\\
 &dl^\xi = \frac{2rD}{\delta} \frac{d\xi}{\cos^2\xi}\,,\\
 &dl^\eta = \frac{2rC}{\delta} \frac{d\eta}{\cos^2\eta}\,. \notag
\end{align}
The (covariant) components of the area elements are:

\begin{align}
    &dA_r = \frac{4r^2}{\delta^{3/2}} C^2D^2 d\eta d\xi\,, \notag \\
    &dA_\xi = \frac{2re^\lambda D }{\delta^{1/2}}dr d\eta\,, \\
    &dA_\eta = \frac{2re^{\lambda}C }{\delta^{1/2}} dr d\xi\,, \notag
\end{align}
while the volume is:
\begin{equation}    
     dV = e^{\lambda} \frac{4r^2 C^2D^2}{\delta^{3/2}} dr d\eta d\xi. 
\end{equation}

The scalar product writes as: 

\begin{equation}
    \mathbf{a} \cdot \mathbf{b} = a^{r} b^{r}  + a^{\xi} b^{\xi}+ a^{\eta} b^{\eta} - \frac{XY}{CD} \big(a^{\xi} b^{\eta}+ a^{\eta} b^{\xi} \big),
\end{equation}
where the last term reflects the non-orthogonality of the grid. Finally, the vector product between two contravariant vector is another contravariant vector, that reads:
\begin{align}
    (\mathbf{a} \times \mathbf{b})^{l} =& a^{i} b^{j} g^{kl} e_{ijk} = \frac{\delta^{1/2}}{ CD} \big( a^{\xi} b^{\eta} - a^{\eta} b^{\xi} \big) \mathbf{\hat{e}_r}
              \notag\\
      & +
      \frac{1}{\delta^{1/2}} \bigg(CD \big( a^{\eta} b^{r} - a^{r} b^{\eta} \big)+ XY  \big( a^{r} b^{\xi} - a^{\xi} b^{r} \big) \bigg) \mathbf{\hat{e}_\xi}
      \notag\\
      & + \frac{1}{\delta^{1/2}} \bigg( XY \big( a^{\eta} b^{r} - a^{r} b^{\eta} \big)+ CD  \big( a^{r} b^{\xi} - a^{\xi} b^{r} \big)  \bigg) \mathbf{\hat{e}_\eta}\,,
          \label{eq:cross_product}
    \end{align}
where $e_{ijk} = \sqrt{g}\varepsilon_{ijk}$ is the the covariant Levi-Civita tensor, $\varepsilon_{ijk}$ is the Levi-Civita symbol and $g = det(g_{ij})$.

The only spacial differential operator needed in the present work is the gradient, which writes in his contravariant form as:

\begin{align}
    \pmb{\nabla}f = & e^{-\lambda(r)}\partial_r f\mathbf{\hat{e}_r} + \frac{1}{r}\Bigl(D\partial_\xi f + \frac{XY}{D}\partial_\eta f\Bigr)\mathbf{\hat{e}_\xi} +\notag \\
    & + \frac{1}{r}\Bigl(\frac{XY}{C}\partial_\xi f + C\partial_\eta f\Bigr)\mathbf{\hat{e}_\eta} \,.
    \label{eq:gradient}
\end{align}


\section{Time discretization: implicit method}
\label{sec:time_discratization}

In this appendix we discuss the discretization of the heat diffusion equation and the technique employed to solve it in \emph{MATINS}. Let us consider Eq. \ref{eq:heat_diffusion} for each of our $(i,p,j,k)$ cells, where hereafter the four indexes denote the radial distance, the patch and the $\xi$ and $\eta$ directions, respectively. Applying Gauss' theorem in the cell volume $V^{i,j,k}$ (note that the latter does not depend on the patch $p$), we can write the equation in its discretized form, approximating the volume-integrals with the central values times the volume:
\begin{align}
&V^{i,j,k}\frac{c_{v;i,p,j,k}}{\Delta t}  (\tilde{T}_{i,p,j,k}^{n+1}-\tilde{T}_{i,p,j,k}^{n}) + \notag\\& + \Phi_{i,p,j,k}(e^{2\nu_i}\mathbf{F}_{i,p,j,k}) =  V^{i,j,k} e^{2\nu_i}\dot{\epsilon}_{i,p,j,k}
    \label{eq:heat_diffusion_disc}
\end{align}
where $\Delta t = t^{n+1} - t^n$ is the discretized timestep leading from step $n$ to $n+1$, and the specific heat and neutrino emissivities are evaluated locally. Note that the redshift factors only depend on the radial direction. 
Here $\Phi_{i,p,j,k}(e^{2\nu_i}\mathbf{F}_{i,p,j,k})$ represents the net flux across the volume and it is evaluated along the cell surface $A_{i,p,j,k}$, as follows:

\begin{align}
\Phi_{i,p,j,k} = &\int_{A_{i,p,j,k}} e^{2\nu} \mathbf{F}\cdot \mathbf{dA} \simeq \notag\\ &\simeq  
 e^{2\nu_{i+1/2}}F^r_{i+1/2, p,j,k} A_{r; i+1/2, j, k} + \notag\\&- e^{2\nu_{i-1/2}}F^r_{i-1/2, p,j,k} A_{r; i-1/2, j, k}+\notag\\
    &+e^{2\nu_i}\bigl(F^\xi_{i,p, j+1/2,k} A_{\xi; i, j+1/2, k}+\notag\\&-F^\xi_{i,p,j-1/2,k} A_{\xi; i, j-1/2, k}\bigr)+\notag\\
    &+e^{2\nu_i}\bigl(F^\eta_{i,p, j,k+1/2} A_{\eta; i, j, k+1/2}+\notag\\
    &-F^\eta_{i,p,j,k-1/2} A_{\eta; i, j, k-1/2}\bigr),
\label{eq:net_flux}
\end{align}
where the half-integer values of the indices denotes quantities evaluated at the cell interfaces.
Once discretized, it is possible to notice that the net flux can be expressed as a linear combination of the temperature within the cell and the temperature in the first neighbouring cells, such that we can express it in matrix form: 

\begin{equation}
    \Phi_{i,p,j,k} = C^{\alpha,p,\beta,\gamma}_{i,p,j,k}\tilde T_{\alpha,p,\beta,\gamma},
    \label{eq:net_flux_discrete}
\end{equation}
where the indexes $\alpha, \beta, \gamma$ can assume the values $\alpha \in [i-1, i, i+1]$, $\beta \in [j-1, j, j+1]$ and $\gamma \in [k-1, k, k+1]$. 

Eq. (\ref{eq:heat_diffusion_disc}) is solved using an implicit backward Euler method, which is particularly suitable to handle the stiff terms appearing in the neutrino cooling and to increase the timestep preserving the stability of the solution. This implies that the flux is written in terms of the updated temperatures, namely those calculated at the timestep $n+1$:
\begin{align}
&\tilde{T}^{n+1}_{i,p,j,k} + \frac{\Delta t}{c_{v;i,p,j,k}}\frac{\Phi_{i,p,j,k}(\tilde{T}_{\alpha, p, \beta,\gamma}^{n+1})}{V^{i,j,k}} = \notag\\ &= \tilde{T}^n_{i,p,j,k} + \frac{\Delta t}{c_{v;i,p,j,k}} e^{2\nu_i}\dot{\epsilon}_{i,p,j,k}.
\label{eq:discretized_implicit_general}
\end{align} 
Rearranging the cells' temperature in a $6N_r(N_a+2)^2$-dimensional\footnote{$N_a +2$ instead of $N_a$ because the temperature of the ghost cells (2 layers of them in each angular direction) must be included.} vector $\tilde{T}^{n+1}_l$, we can re-write Eq. \ref{eq:discretized_implicit_general} in a matrix form as:
\begin{equation}
{\hat m}^\alpha_l~\tilde{T}_\alpha^{n+1} = v_l(\tilde{T}^n)\,,
\label{eq:implicit_scheme}
\end{equation}
where the source $v_l$ contains terms of the discretized equation which depend only on the local old temperatures, and whose explicit form, along with that of the matrix ${\hat m}^\alpha_l$, is reported in Appendix \ref{sec:Matrix}. This corresponds to expressing our differential equation as a system of $6N_r(N_a+2)^2$ algebraic equations, whose unknown values are the temperatures in each cell at the timestep $n+1$. In our code we solve the system with the aid of the public library LAPACK \citep{lapack}.

The right hand side of Eq. (\ref{eq:implicit_scheme}) corresponds to the old temperatures plus the source term (as in the right-hand-side of Eq.~\ref{eq:discretized_implicit_general}), which includes the Joule heating and the neutrino emissivity.
The latter contribution has a steep dependence with temperature, making the source term a stiff term in our equation. In order to handle it, it is convenient to perform a linearization, writing:

\begin{equation}
    \dot{\epsilon}_{i, p, j, k}(T^{n+1}_l) \simeq \dot{\epsilon}_{i, p, j, k}(T^{n}_l) + \Bigl(\frac{d\dot{\epsilon}}{dT}\Bigr)_{i, p, j, k}e^{-{\nu_i}}\bigl(\tilde{T}_l^{n+1} - \tilde{T}_l^n\bigr)\,
\end{equation}
where the derivatives are calculated at the old temperatures. The terms multiplying $T^{n+1}$ in the previous equation are added to the matrix diagonal:

\begin{equation}
    \hat{m}^\alpha_l \rightarrow \hat{m}^\alpha_l - \delta^\alpha_l\frac{\Delta t}{c_{v; i, p, j, k}}e^{\nu_i}\Bigl(\frac{d\dot{\epsilon}}{dT}\Bigr)_{i, p, j, k}~,
\end{equation}
where $\delta^\alpha_l$ is a Kronecker delta, while all the other terms are included in the vector $v_l(\tilde{T}^n)$, which is written as 
\begin{align}
    &v(\tilde{T}_l^n) = \Bigl[1 - \frac{\Delta t}{c_{v; i, p, j, k}}e^{\nu_i}\Bigl(\frac{d\dot{\epsilon}}{dT}\Bigr)_{i, p, j, k}\Bigr]\tilde{T}_l^n + \notag\\ &+\frac{\Delta t}{c_{v; i, p, j, k}}e^{2\nu_i}\dot{\epsilon}_{i, p, j, k}(T_l^n).
\end{align}

\section{Ghost Cells and Patch Edges}
\label{sec:ghost}

As we mentioned in Sec. \ref{sec:grid} we introduced a layer of ghost cells at each patch interface. 
The temperature of these extra cells is used to calculate the heat flux through the cells next to the patch edges; consequently they must enter in Eq. \ref{eq:implicit_scheme}. While the other temperatures are determined by a physical equation (Eq. \ref{eq:heat_diffusion}), the temperatures in the ghost cells are determined by a linear interpolation with the temperature of two cells in the neighbouring patch. This results in an equation that is actually a constrain and writes as:

\begin{equation}
    T^{n+1}_{\text{ghost}, j} - \Bigl(1 - W_j\Bigr)T^{n+1}_j - W_j T^{n+1}_{j'} = 0\,.
    \label{eq:edge_ghost_constraint}
\end{equation}

Here $T^{n+1}_{\rm ghost, j}$ is the temperature of a ghost cell a given patch, $T^{n+1}_j$ and $T^{n+1}_{j'}$ are the temperatures of two non-ghost cell in the neighbouring patch. The index $j$ is the index of the coordinate parallel to the edge, while $j' = j+1$ for $j \le N_a/2$ and $j' = j-1$ for $j > N_a/2$. $W_j$ are the weights defined as:

\begin{equation}
    W_j = 
    \begin{cases}
    \frac{1}{2}\Bigl(1-\frac{p^m_{j-1} - p^o_{j+1}}{p^o_{j+1} - p^o_{j-1}}\Bigr) \qquad &j<(N_a +1)/2\\
    0 \qquad &j = (N_a +1)/2\\
    \frac{1}{2}\Bigl(1-\frac{p^m_{j+1} - p^o_{j-1}}{p^o_{j+1} - p^o_{j-1}}\Bigr)&j>(N_a +1)/2,
    \end{cases}
\end{equation}
where $p^m_j$ and $p^o_j$ are the ghost cell coordinates, defined in Sec. 2.6 of \citet{Dehman2023_MATINS}.
 Eq. \ref{eq:edge_ghost_constraint} basically states that the temperature of the ghost cell of a given patch is the linear combination of the temperature of the cell at the edge of the neighbouring patch with the same parallel (to the edge) coordinate and the temperature of the cell next to it towards the patch center (to have an insight the reader can refer to Fig. 2 of \citet{Dehman2023_MATINS}). Finally, the term $W_j$ represents the weights of the linear interpolation and it belong to the range $[0,1]$, where $W_j = 0$ at the center of the edge ($j = N_a/2, N_a/2 +1$) and $W_j = 1$ at the edge extremes ($j=1,N_a$). In this way, at the center of the edge $T^{n+1}_{ghost, j}$ takes the contribution only from a single cell, while at the extremes it correspond to the numerical average of the two cells.

A separate treatment is necessary for the cells with $j, k = 0, N_a+1$, which corresponds to the ghost corner of the patch. These cells are special because they are ghost cells with respect to two different edges, namely, in our approach, their temperature must be coupled to temperature of two different patches. Similarly to Eq. \ref{eq:edge_ghost_constraint} the temperature of the corner ghost cell is written as:

\begin{equation}
    T^{n+1}_{\text{ghost, corner}} - \frac{1}{2}T^{n+1}_1 - \frac{1}{2}T^{n+1}_{2} = 0,
    \label{eq:corner_ghost_constraint}
\end{equation}

where $T^{n+1}_1$ and $T^{n+1}_{2}$ are temperatures of the non-ghost cells at the corner of the two patches adjacent to the original patch at the given corner. For example, if we are considering the ghost corner $j, k = N_a + 1$ of patch $I$, $T^{n+1}_1$ and $T^{n+1}_{2}$ are the temperatures in the cells $j = N_a, k = 1$ of patch $V$ and $j =1, k = N_a$ of patch $II$, respectively. 

\section{Boundary Conditions}
\label{sec:boundary}

Our setup requires the definition of boundary conditions on the innermost and the outermost radial layers, represented by the interfaces with the core and the envelope, respectively. 
The core is described by a single radial layer because, due to its high conductivity, it becomes perfectly isothermal (precisely, $Te^{\nu(r)}$ is constant) after at most only a couple of centuries, a timescale that is irrelevant for our purposes and without observational constraints (see \emph{e.g.} \citealt{Vigano2021}). For this reason, in our code, we consider a single redshifted temperature for the core, which evolves according to the ratio between the volume-integrated sources and heat capacity, and accounting for the core-crust flux:
\begin{align}
    \frac{d }{dt} (e^\nu T_{\textrm{core}}) =& \frac{\int^{R_\textrm{core}}_0\dot{\epsilon}_{\nu}(T_{\textrm{core}})e^{2\nu}r^2 dr}{\int^{R_\textrm{core}}_0 c_\mathrm{v}(T_{\textrm{core}})r^2 dr} + \notag\\& - \frac{\Phi_{CC}}{4\pi\int^{R_\textrm{core}}_0 c_\mathrm{v}(T_{\textrm{core}})r^2 dr},
    \label{eq:core_diffusion}
\end{align}
where the neutrino emissivity and the specific heat are integrated over the entire core volume, exploiting the fact that they depend only on the radial coordinate. The flux term $\Phi_{CC}$ is the net flux at the core-crust interface, 
written as: 
\begin{equation}
    \Phi^{CC}(\Tilde{T}^{n}_\textrm{core}) = \sum_{jkp}F^r_{3/2,p,j,k}A_{r;3/2,p,j,k}\,,
\end{equation}
which is Eq. \ref{eq:net_flux} integrated along the whole crust-core interface, where we consider only the radial contribution of the flux at this boundary. 

Eq.~\ref{eq:core_diffusion} is solved by an implicit backward Euler method. However, it is worth noticing that the flux $\Phi^{CC}$ is treated explicitly. The reason is that in this term the flux does not depend only on the temperature of the core, but also on the temperature of the cells in the innermost crustal layer ($i=2$). 
We observed empirically by refining the temporal resolution that this approach, alas not fully self-consistent, has a negligible effect on the solution.   

From an operational point of view we treat the cells in the innermost layer as ghost cells, in the sense that like the ghost cells at the border of the patches (see Appendix \ref{sec:ghost}), their temperature is determined by a constrain, which is written as: 
 \begin{equation}
    T^{n+1}_{1, p, j, k} - T^{n+1}_{\textrm{core}} = 0,
\end{equation}
whre $T^{n+1}_{\textrm{core}}$ is determined by the solution of Eq. \ref{eq:core_diffusion}.

The external boundary condition consists in the addition of a source term, describing the emission from the stellar surface. Such emission is assumed for simplicity to be a blackbody with surface temperature $T_s$ and is given by the Stefan-Boltzmann law (written in a discretized form)
\begin{equation}
    S(T^{n+1}_{b}) = \sigma_{sb} A_{r; N_r+1/2, p, j, k} T_s^4(T^{n+1}_{b})\,,
    \label{eq:StefanBoltzmann}
\end{equation}
where $S$ is the luminosity emitted by the radiating area $A$ at temperature $T_s$, and $\sigma_{sb}$ is the Stefan-Boltzmann constant.

 It is worth noticing that here the surface temperature $T_s$ is not the temperature in the outermost layer of the computational domain, corresponding to the crust-envelope interface and denoted by $T_b$. $T_s$ is instead the temperature at the top of the envelope. The envelope is not part of our computational domain, since in this region the gradients of temperature and density are so steep and the timescales so short that including this zone makes the thermal evolution computationally prohibitive to tackle. Instead, in \emph{MATINS} we utilize envelope models, which are analytic effective $T_s = T_s(T_b, \mathbf{B})$ relations, obtained by fitting stationary solutions of heat diffusion in the envelope models computed with different $T_b$ and $\mathbf{B}$. Most of the simulations presented in this paper utilize the envelope model provided by \citet{Gudmundsson1983}: 
 \begin{equation}
    T_s = 10^6 \mathrm{K} \times \Bigl[ \frac{g^{0.455}_{14} (T_{b}/10^8\,\mathrm{K})}{1.288}\Bigr]^{1/1.82},
    \label{eq:Gudm}
 \end{equation}
 where $g_{14}$ is the surface gravity in units of $10^{14} \, \mathrm {g\,cm/s^2}$. We chose this model because it is particularly simple, not inclusive of the dependence on the magnetic field, which makes it straightforward to interpret the cooling curve in terms of the $T_b$ profile. Nevertheless, \emph{MATINS} accounts also for several other models, including magnetized envelopes with heavy element or light element composition. In addition to Gudmundsson model, we provide also a run characterized by the magnetized iron envelope presented in \citet{Potekhin2015}. To see the impact of the envelope model on the cooling curve we refer to \citet{Dehman2023b}.

The term in Eq. \ref{eq:StefanBoltzmann} is treated implicitly similarly to the neutrino cooling: the source term, evaluated at $T^{n+1}$, is linearized and the terms dependent on $T^{n}$ are included in the source vector $v_l$, while those dependent on $T^{n+1}$ are included in the diagonal elements of the matrix $\hat{m}^\alpha_l$: 

\begin{align}
    v(\tilde{T}_l^n) \rightarrow & v(\tilde{T}_l^n) - \frac{\Delta t}{c_{v; i, p, j, k}V_{i, j, k}}e^{\nu_i}\Bigl(\frac{dS}{dT}\Bigr)_{i, p, j, k} \tilde{T}_l^n+ \notag\\&+ \frac{\Delta t}{c_{v; i, p, j, k}V_{i, j, k}}e^{2\nu_i}S(T_l^n)
\end{align}
and:
\begin{equation}
    \hat{m}^\alpha_l \rightarrow \hat{m}^\alpha_l - \delta^\alpha_l\frac{\Delta t}{c_{v; i, p, j, k}V_{i, j, k}}e^{\nu_i}\Bigl(\frac{dS}{dT}\Bigr)_{i, p, j, k}.
\end{equation}
It is worth noticing that while such implicit treatment is essential for the neutrino emissivity, due to the strong stiffness of this term, for the photon luminosity it just represent a second order correction that gives little difference compared to an explicit treatment.


\section{Matrix elements}
\label{sec:Matrix}
In this Appendix we write extensively the matrix elements $K^{ij}$ and $m^\alpha_l$ and we show how to derive them. 

First of all we start from Eq. \ref{eq:flux} and we explicit all the three term on the right hand side of the equation. We have first the temperature gradient, which according to Eq. (\ref{eq:gradient}) writes in cubed sphere grid as:
\begin{align}
    \pmb{\nabla}\Tilde{T} = & e^{-\lambda(r)}\partial_r \Tilde{T}\mathbf{\hat{e}_r} + \frac{1}{r}\Bigl(D\partial_\xi\Tilde{T} + \frac{XY}{D}\partial_\eta\Tilde{T}\Bigr)\mathbf{\hat{e}_\xi} +\notag \\
    & + \frac{1}{r}\Bigl(\frac{XY}{C}\partial_\xi\Tilde{T} + C\partial_\eta \Tilde{T}\Bigr)\mathbf{\hat{e}_\eta}.
\end{align}

The second term of interest is the scalar product $\mathbf{b}\cdot \pmb{\nabla}\Tilde{T}$, which writes as:
\begin{align}
    (\mathbf{b}\cdot \pmb{\nabla} \Tilde{T}) = & b^r e^{-\lambda(r)}\partial_r \Tilde{T}+ \frac{1}{r}\Bigl[D - \Bigl(\frac{XY}{C}\Bigr)^2\frac{1}{D}\Bigr]b^\xi\partial_\xi\Tilde{T} + \\ &+ \frac{1}{r}\Bigl[C - \Bigl(\frac{XY}{D}\Bigr)^2\frac{1}{C}\Bigr]b^\eta\partial_\eta\Tilde{T}.
\end{align}

Finally, we report the three component of the vector product $\mathbf{b}\times \pmb{\nabla}\Tilde{T}$ in the last term:
\begin{align}
    (\mathbf{b} \times \pmb{\nabla}\Tilde{T})^r = & \frac{\sqrt{\delta}}{CDr}\Bigl[\Bigl( b^\xi\frac{XY}{C} - b^\eta D\Bigr)\partial_\xi \Tilde{T} + \notag\\
    &+ \Bigl(b^\xi C - b^\eta\frac{XY}{D} \Bigr)\partial_\eta \Tilde{T}\Bigr]\\
    (\mathbf{b} \times \pmb{\nabla}\Tilde{T})^\xi = &\frac{1}{\sqrt{\delta}}\Bigl[\Bigl(CDb^\eta - XYb^\xi\Bigr)e^{-\lambda(r)}\partial_r \Tilde{T} -\frac{\delta}{rD}b^r\partial_\eta\Tilde{T}\Bigr]\\
    (\mathbf{b} \times \pmb{\nabla}\Tilde{T})^\eta = &\frac{1}{\sqrt{\delta}}\Bigl[\Bigl(XYb^\eta -CDb^\xi\Bigr)e^{-\lambda(r)}\partial_r \Tilde{T} + \frac{\delta}{rC}b^r\partial_\xi\Tilde{T}\Bigr]
\end{align}

It is useful to write the three terms in eq. \ref{eq:flux} as follows:
\begin{equation}
    k_\perp \pmb{\nabla}\Tilde{T} = A\partial_r\Tilde{T}\mathbf{e_r} + (B\partial_\xi\Tilde{T} + E\partial_\eta \Tilde{T})\mathbf{\hat{e}_\xi} + (F\partial_\xi \Tilde{T} + G\partial_\eta \Tilde{T} )\mathbf{\hat{e}_\eta}
     \label{eq:flux_term1}
\end{equation}
\begin{align}
    k_\perp(\omega_B\tau_0)^2(\mathbf{b}\cdot \pmb{\nabla}\Tilde{T})\mathbf{b} = &(Hb^r\partial_r\Tilde{T} + Ib^r\partial_\xi \Tilde{T} + Jb^r\partial_\eta \Tilde{T})\mathbf{e_r} + \notag \\
    & + (Hb^\xi\partial_r\Tilde{T} + Ib^\xi\partial_\xi \Tilde{T} + Jb^\xi\partial_\eta \Tilde{T})\mathbf{\hat{e}_\xi} +\notag \\
    & + (Hb^\eta\partial_r\Tilde{T} + Ib^\eta\partial_\xi \Tilde{T} + Jb^\eta\partial_\eta \Tilde{T})\mathbf{\hat{e}_\eta}
     \label{eq:flux_term2}
\end{align}
\begin{align}
    k_\perp(\omega_B\tau_0)(\mathbf{b}\times\pmb{\nabla}\Tilde{T}) = & (K\partial_\xi \Tilde{T} + L\partial_\eta \Tilde{T})\mathbf{e_r} + \notag\\
    &+(M\partial_r \Tilde{T} - N\partial_\eta \Tilde{T})\mathbf{\hat{e}_\xi} + \notag \\
    & + (O\partial_r\Tilde{T} + P\partial_\xi \Tilde{T})\mathbf{\hat{e}_\eta}
    \label{eq:flux_term3}
\end{align}

where the terms $A, B, E, F,G, H, I, J, K, L, M, N, O, P$ are the following functions:

\begin{align}
    &A = k_\perp e^{-\lambda(r)}
    &B = \frac{k_\perp D}{r}\notag\\
    &E = \frac{k_\perp XY}{rD}
    &F = \frac{k_\perp XY}{rC} = \frac{ED}{C}\notag\\
    &G = \frac{k_\perp C}{r}
    &H = k_\perp (\omega_B\tau_0)^2e^{-\lambda(r)}b^r\notag\\
    &I = k_\perp(\omega_B\tau_0)^2\frac{b^\xi}{r}\Bigl[D- \Bigr(\frac{XY}{C}\Bigl)^2\frac{1}{D}\Bigr]\notag\\
    &J = k_\perp(\omega_B\tau_0)^2\frac{b^\eta}{r}\Bigl[C - \Bigl(\frac{XY}{D}\Bigr)^2\frac{1}{C} \Bigr]\notag\\
    &K = \frac{\sqrt{\delta}k_\perp(\omega_B\tau_0)}{CDr}\Bigl( b^\xi\frac{XY}{C} - b^\eta D\Bigr)\notag\\
    &L = \frac{\sqrt{\delta}k_\perp(\omega_B\tau_0)}{CDr}\Bigl(b^\xi C - b^\eta\frac{XY}{D}\Bigr) \notag\\
    &M = \frac{k_\perp(\omega_B\tau_0)}{\sqrt{\delta}e^{\lambda(r)}}(CDb^\eta - XYb^\xi)
    &N =  k_\perp (\omega_B\tau_0)\frac{\sqrt{\delta}}{rD}b^r \notag \\
    &O = \frac{k_\perp(\omega_B\tau_0)}{\sqrt{\delta}e^{\lambda(r)}}(XYb^\eta - CDb^\xi)
    &P = k_\perp(\omega_B\tau_0)\frac{\sqrt{\delta}}{rC}b^r
\end{align}

It is straightforward to show that the matrix elements $K^{ij}$ in Eq. \ref{eq:flux_matrix} are: 

\begin{equation}
    K^{ij} = 
    \begin{pmatrix}
    A + Hb^r & Ib^r + K & Jb^r + L\\
    Hb^\xi + M & B + Ib^\xi & E - N + Jb^\xi\\
    Hb^\eta + O & F + P + Ib^\eta & G + Jb^\eta
    \end{pmatrix}.
    \label{eq:matrixK}
\end{equation}
Concerning the matrix elements $m^\alpha_l$ in Eq. (\ref{eq:implicit_scheme}), the first step is to consider the net flux in Eq. (\ref{eq:net_flux}) and make explicit its dependence on the temperature. In its discretized form this equation writes as:

\begin{align}
    \Phi_{i,p,j,k} = \, & 
 e^{2\nu_{i+1/2}}F^r_{i+1/2, p,j,k} A_{r; i+1/2, j, k} + \notag\\&- e^{2\nu_{i-1/2}}F^r_{i-1/2, p,j,k} A_{r; i-1/2, j, k}+\notag\\
    &+e^{2\nu_i}\bigl(F^\xi_{i,p, j+1/2,k} A_{\xi; i, j+1/2, k}+\notag\\&-F^\xi_{i,p,j-1/2,k} A_{\xi; i, j-1/2, k}\bigr)+\notag\\
    &+e^{2\nu_i}\bigl(F^\eta_{i,p, j,k+1/2} A_{\eta; i, j, k+1/2}+\notag\\
    &-F^\eta_{i,p,j,k-1/2} A_{\eta; i, j, k-1/2}\bigr)\,,
    \label{eq:netflux}
\end{align}
where the six terms on the right hand side are the contributions from the six faces of the cell. We can now write explicitly the discretized fluxes in the previous equation exploiting Eq. (\ref{eq:flux_matrix}). We evaluate the temperature derivatives as cell centered difference between the values at the first neighbours. They are located either at the centers (for the derivatives associated to the direction normal to the surface, namely the diagonal terms of the conductivity) or at the middle of the edges of the 3D cells (associated to the transverse conductivity, namely the off-diagonal terms). The latter temperatures are obtained as the average between the values at the centers of the four closest cells. The expressions that we obtain are the following:

\begin{align}
    &e^{\nu_{i+1/2}}F^r_{i+1/2, j,k} =\notag\\& - K^{rr}_{i+1/2,j,k}\frac{\tilde T_{i+1, j,k} - \tilde T_{i,j,k}}{r_{i+1}-r_{i}} + \notag \\
    &- K^{r\xi}_{i+1/2, j,k}\frac{\tilde T_{i+1,j+1, k} + \tilde T_{i, j+1, k} - \tilde T_{i+1, j-1, k} -\tilde T_{i, j-1, k}}{4(\xi_{j+1/2} - \xi_{j-1/2})} +\notag \\
    &-K^{r\eta}_{i+1/2, j, k} \frac{\tilde T_{i+1,j, k+1} + \tilde T_{i, j, k+1} - \tilde T_{i+1, j, k-1} -\tilde T_{i, j, k-1}}{4(\eta_{k+1/2} - \eta_{k-1/2})}\label{eq:f1}
\end{align}

\begin{align}
    &e^{\nu_{i-1/2}}F^r_{i-1/2, j,k} =\notag\\ & - K^{rr}_{i-1/2,j,k}\frac{\tilde T_{i,j,k} - \tilde T_{i-1,j,k}}{r_{i} - r_{i-1}} + \notag \\
    &- K^{r\xi}_{i-1/2, j,k}\frac{\tilde T_{i,j+1, k} + \tilde T_{i-1, j+1, k} - \tilde T_{i, j-1, k} -\tilde T_{i-1, j-1, k}}{4(\xi_{j+1/2} - \xi_{j-1/2})} +\notag \\
    &-K^{r\eta}_{i-1/2, j, k} \frac{\tilde T_{i,j, k+1} + \tilde T_{i-1, j, k+1} - \tilde T_{i, j, k-1} -\tilde T_{i-1, j, k-1}}{4(\eta_{k+1/2} - \eta_{k-1/2})}
\end{align}

\begin{align}
    &e^{\nu_i}F^\xi_{i, j+1/2,k} =\notag \\ & - K^{\xi r}_{i,j+1/2,k}\frac{\tilde T_{i+1,j+1, k} + \tilde T_{i+1, j, k} - \tilde T_{i-1, j+1, k} -\tilde T_{i-1, j, k}}{4(r_{i+1/2} - r_{i-1/2})} + \notag \\
    &- K^{\xi \xi}_{i, j+1/2, k}\frac{\tilde T_{i,j+1, k} - \tilde T_{i, j, k}}{\xi_{j+1} - \xi_{j}} +\notag \\
    &-K^{\xi \eta}_{i, j+1/2, k} \frac{\tilde T_{i,j+1, k+1} + \tilde T_{i, j, k+1} - \tilde T_{i, j+1, k-1} -\tilde T_{i, j, k-1}}{4(\eta_{k+1/2} - \eta_{k-1/2})}
\end{align}

\begin{align}
    &e^{\nu_i}F^\xi_{i, j-1/2,k} = \notag \\ & - K^{\xi r}_{i,j-1/2,k}\frac{\tilde T_{i+1,j, k} + \tilde T_{i+1, j-1, k} - \tilde T_{i-1, j, k} -\tilde T_{i-1, j-1, k}}{4(r_{i+1/2} - r_{i-1/2})} + \notag \\
    &- K^{\xi \xi}_{i, j-1/2, k}\frac{\tilde T_{i,j, k} - \tilde T_{i, j-1, k}}{\xi_{j} - \xi_{j-1}} +\notag \\
    &-K^{\xi \eta}_{i, j-1/2, k} \frac{\tilde T_{i,j, k+1} + \tilde T_{i, j-1, k+1} - \tilde T_{i, j, k-1} -\tilde T_{i, j-1, k-1}}{4(\eta_{k+1/2} - \eta_{k-1/2})}{}
\end{align}

\begin{align}
    &e^{\nu_i}F^\eta_{i, j,k+1/2} = \notag \\ & - K^{\eta r}_{i,j,k+1/2}\frac{\tilde T_{i+1,j, k+1} + \tilde T_{i+1, j, k} - \tilde T_{i-1, j, k+1} -\tilde T_{i-1, j, k}}{4(r_{i+1/2} - r_{i-1/2})} + \notag \\
    &- K^{\eta \xi}_{i, j, k+1/2}\frac{\tilde T_{i,j+1, k+1} + \tilde T_{i, j+1, k} - \tilde T_{i, j-1, k+1} -\tilde T_{i, j-1, k}}{4(\xi_{j+1/2} - \xi_{j-1/2})} +\notag \\
    &-K^{\eta \eta}_{i, j, k+1/2} \frac{\tilde T_{i,j, k+1} - \tilde T_{i, j, k}}{\eta_{k+1} - \eta_k}
\end{align}

\begin{align}
    &e^{\nu_i}F^\eta_{i, j,k-1/2} = \notag\\ & - K^{\eta r}_{i,j,k-1/2}\frac{\tilde T_{i+1,j, k} + \tilde T_{i+1, j, k-1} - \tilde T_{i-1, j, k} -\tilde T_{i-1, j, k-1}}{4(r_{i+1/2} - r_{i-1/2})} + \notag \\
    &- K^{\eta \xi}_{i, j, k-1/2}\frac{\tilde T_{i,j+1, k} + \tilde T_{i, j+1, k-1} - \tilde T_{i, j-1, k} -\tilde T_{i, j-1, k-1}}{4(\xi_{j+1/2} - \xi_{j-1/2})} +\notag \\
    &-K^{\eta \eta}_{i, j, k-1/2} \frac{\tilde T_{i,j, k} - \tilde T_{i, j, k-1}}{\eta_{k} - \eta_{k-1}}\label{eq:f6}
\end{align}

From these expressions, we can see that the net flux in a given cell $\Phi_{i,j,k}$ depends on the temperatures of the cell itself (with indices $(i,j,k)$), at the 6 first neighbors (\emph{e.g.} $(i+1, j, k)$, $(i, j-1, k)$ etc.), at the 12 second neighbors (\emph{e.g.} $(i-1, j, k+1)$, $(i, j-1, k-1)$), but not on the third neighbors (\emph{e.g.} $(i-1, j+1, k+1)$, $(i-1, j-1, k-1)$ etc.) or even further cells (\emph{e.g} $(i+2, j,k)$ etc.); such that in total each equation to advance $T_{i,j,k}$ couples 19 values of temperatures. 

We now introduce the following quantities for the seek of a compact notation: 

\begin{eqnarray}
  && Z^{rr}_{i,j,k} = \frac{e^{\nu_{i+1/2}}K^{rr}_{i+1/2, j,k}A_{r; i+1/2,j,k}}{r_{i+1} - r_{i}} \\
  && Z^{\xi\xi}_{i,j,k} = \frac{e^{\nu_{i}}K^{\xi\xi}_{i,j+1/2,k}A_{\xi; i,j+1/2,k}}{\xi_{j+1} - \xi_{j}} \\
  && Z^{\eta\eta}_{i,j,k} = \frac{e^{\nu_{i}}K^{\eta\eta}_{i, j,k+1/2}A_{\eta; i,j,k+1/2}}{\eta_{k+1} - \eta_{k}} \\
  && Z^{r\xi}_{i,j,k} =  \frac{1}{4}\frac{e^{\nu_{i+1/2}}K^{r\xi}_{i+1/2, j,k}A_{r; i+1/2,j,k}}{\xi_{j+1/2} -\xi_{j-1/2}} \\
  && Z^{r\eta}_{i,j,k} =  \frac{1}{4}\frac{e^{\nu_{i+1/2}}K^{r\eta}_{i+1/2, j,k}A_{r; i+1/2,j,k}}{\eta_{k+1/2} -\eta_{k-1/2}} \\
  && Z^{\xi r}_{i,j,k} =  \frac{1}{4}\frac{e^{\nu_{i}}K^{\xi r}_{i,j+1/2,k}A_{\xi; i,j+1/2,k}}{r_{i+1/2} -r_{i-1/2}} \\
  && Z^{\xi\eta}_{i,j,k} =  \frac{1}{4}\frac{e^{\nu_{i}}K^{\xi\eta}_{i, j+1/2,k}A_{\xi; i,j+1/2,k}}{\eta_{k+1/2} -\eta_{k-1/2}} \\
  && Z^{\eta r}_{i,j,k} =  \frac{1}{4}\frac{e^{\nu_{i}}K^{\eta r}_{i,j,k+1/2}A_{\eta; i,j,k+1/2}}{r_{i+1/2} -r_{i-1/2}} \\    
  && Z^{\eta\xi}_{i,j,k} =  \frac{1}{4}\frac{e^{\nu_{i}}K^{\eta\xi}_{i, j,k+1/2}A_{\eta; i,j,k+1/2}}{\xi_{j+1/2} -\xi_{j-1/2}}
\end{eqnarray}
and the common prefactor appearing in Eq. (\ref{eq:discretized_implicit_general}):
\begin{equation}
  H_{i,j,k} = \frac{\Delta t}{c_{v;i,j,k} V^{i,j,k}}.
\end{equation}
In this way, the matrix elements $m^{\alpha}_l$ in Eq. (\ref{eq:implicit_scheme}) are written as: 
\begin{eqnarray}
  && m_{i,p,j,k} = 1 + H_{i,j,k}(Z_{i,j,k}^{rr} + Z_{i-1,j,k}^{rr} + Z_{i,j,k}^{\xi\xi} + \nonumber \\&& +Z_{i,j-1,k}^{\xi\xi} + Z_{i,j,k}^{\eta\eta} + Z_{i,j,k-1}^{\eta\eta}) \nonumber\\
  \hline
  && m_{i+1,p,j,k} = H_{i,j,k}(-Z_{i,j,k}^{rr} - Z_{i,j,k}^{\xi r} + Z_{i,j-1,k}^{\xi r}+\nonumber\\&& - Z_{i,j,k}^{\eta r} + Z_{i,j,k-1}^{\eta r} ) \nonumber\\
  && m_{i,p,j+1,k} = H_{i,j,k}(-Z_{i,j,k}^{\xi\xi} - Z_{i,j,k}^{r\xi} + Z_{i-1,j,k}^{r \xi}+\nonumber\\&& - Z_{i,j,k}^{\eta\xi} + Z_{i,j,k-1}^{\eta\xi} )
  \nonumber\\
  && m_{i,p,j,k+1} = H_{i,j,k}(-Z_{i,j,k}^{\eta\eta} - Z_{i,j,k}^{r\eta} + Z_{i-1,j,k}^{r\eta}+\nonumber\\&& - Z_{i,j,k}^{\xi\eta} + Z_{i,j-1,k}^{\xi\eta} ) \nonumber\\
  && m_{i-1,p,j,k} = H_{i,j,k}(-Z_{i-1,j,k}^{rr} + Z_{i,j,k}^{\xi r} - Z_{i,j-1,k}^{\xi r}+\nonumber\\&& + Z_{i,j,k}^{\eta r} - Z_{i,j,k-1}^{\eta r}) \nonumber\\
  && m_{i,p,j-1,k} = H_{i,j,k}(-Z_{i,j-1,k}^{\xi\xi} + Z_{i,j,k}^{r\xi} - Z_{i-1,j,k}^{r \xi} +\nonumber\\&& + Z_{i,j,k}^{\eta\xi} - Z_{i,j,k-1}^{\eta\xi} ) \nonumber\\
  && m_{i,p,j,k-1} = H_{i,j,k}(-Z_{i,j,k-1}^{\eta\eta} + Z_{i,j,k}^{r\eta} - Z_{i-1,j,k}^{r\eta}+\nonumber\\&& + Z_{i,j,k}^{\xi\eta} - Z_{i,j-1,k}^{\xi\eta} ) \nonumber\\
  \hline
  && m_{i-1,p,j-1,k} = H_{i,j,k}(-Z_{i-1,j,k}^{r\xi} - Z_{i,j-1,k}^{\xi r}) \nonumber\\
  && m_{i-1,p,j,k-1} = H_{i,j,k}(- Z_{i-1,j,k}^{r\eta} - Z_{i,j,k-1}^{\eta r}) \nonumber\\
  && m_{i-1,p,j,k+1} = H_{i,j,k}(Z_{i,j,k}^{\eta r} + Z_{i-1,j,k}^{r \eta}) \nonumber\\
  && m_{i-1,p,j+1,k} = H_{i,j,k}(Z_{i,j,k}^{\xi r} + Z_{i-1,j,k}^{r \xi}) \nonumber\\
  && m_{i,p,j-1,k-1} =  H_{i,j,k}(-Z_{i,j-1,k}^{\xi\eta} - Z_{i,j,k-1}^{\eta\xi}) \nonumber\\
  && m_{i,p,j-1,k+1} = H_{i,j,k}(Z_{i,j,k}^{\eta\xi} + Z_{i,j-1,k}^{\xi\eta} ) \nonumber\\
  && m_{i,p,j+1,k-1} = H_{i,j,k}(Z_{i,j,k}^{\xi\eta} + Z_{i,j,k-1}^{\eta\xi} ) \nonumber\\
  && m_{i,p,j+1,k+1} = H_{i,j,k}(-Z_{i,j,k}^{\xi\eta} - Z_{i,j,k}^{\eta\xi} ) \nonumber\\
  && m_{i+1,p,j-1,k} = H_{i,j,k}(Z_{i,j,k}^{r\xi} + Z_{i,j-1,k}^{\xi r}) \nonumber\\
  && m_{i+1,p,j,k-1} = H_{i,j,k}(Z_{i,j,k}^{r\eta} + Z_{i,j,k-1}^{\eta r}) \nonumber\\
  && m_{i+1,p,j,k+1} = H_{i,j,k}(-Z_{i,j,k}^{r\eta} - Z_{i,j,k}^{\eta r}) \nonumber\\
  && m_{i+1,p,j+1,k} = H_{i,j,k}(-Z_{i,j,k}^{r\xi} - Z_{i,j,k}^{\xi r}) \nonumber\\
\end{eqnarray}
In the above notation of the matrix, we have to remember that the indices are not those of the matrix (which is 2-dimensional) but represent the indices of the cell, whose temperature the matrix element multiplies, namely they identify the matrix index $\alpha$ in Eq. (\ref{eq:implicit_scheme}), while the other index $l$ refers to the cell ($i,p,j,k)$. In other words, for example the matrix element $m_{i+1, p, j, k}$ couples the cell ($i,p,j,k$) with its first neighbor ($i+1,p,j,k$), while the element $m_{i+1, p, j-1, k}$ couples its second neighbor ($i+1,p,j-1,k$). The horizontal lines in the previous list of equations separate the couplings with the point itself, the first neighbors and the second neighbors. 
We also stress that these matrix elements are dimensionless and they satisfy the following identities, for each ${i,j,k}$:
\begin{eqnarray}
&& m_{i,p,j,k} + m_{i-1,p,j,k} + m_{i+1,p,j,k} + m_{i,p,j-1,k}+\nonumber\\&& + m_{i,p,j+1,k} + m_{i,p,j,k-1} + m_{i,p,j,k+1} = 1
\nonumber\\
\hline
\nonumber\\
&& m_{i-1,p,j-1,k} + m_{i+1,p,j-1,k} + m_{i-1,p,j+1,k}+\nonumber\\&& +  m_{i+1,p,j+1,k} + m_{i,p,j-1,k-1} + m_{i,p,j+1,k-1}+\nonumber\\&& + m_{i,p,j-1,k+1} + m_{i,p,j+1,k+1} = 0~,
\end{eqnarray}
which means that, by construction, the sum of all contributions to the matrix elements is exactly 1 (this is not true anymore if a linearized source is brought in the matrix), and the sum of the contributions from all the second neighbours are 0. 

\section{Initial magnetic field and input Parameters of our simulations.}

\label{sec:initial_Bfield_configuration}

In this section, we detail the formalism used to define the magnetic field initial configuration and, for the sake of reproducibility, the input parameters that define the configurations presented in Table \ref{tab:Bfield_models} in Section \ref{sec:results}.

Generally, a magnetic field $\mathbf{B}$ can be decomposed into a poloidal component $\mathbf{B}_\mathrm{pol}$ and a toroidal component $\mathbf{B}_\mathrm{tor}$, which can be expressed via two scalar functions $\Phi (t,\mathbf{x})$ and $\Psi(t,\mathbf{x})$ as follows:

\begin{align}
    &\mathbf{B}_\mathrm{pol} = \nabla \times \bigl(\nabla \times \Phi \mathbf{k}\bigr)\notag\\
    &\mathbf{B}_\mathrm{tor} = \nabla \times \Psi \mathbf{k},
    \label{eq:poloidal_toroidal_decomposition}
\end{align}
where $\mathbf{k}$ is an arbitrary vector, which in our coordinate system is conveniently identified with the radial versor $\mathbf{k} =\mathbf{r}$. To define the initial magnetic field configuration it is convenient, following \citet{GeppertWiebicke1991}, to expand the poloidal $\Phi$ and toroidal $\Psi$ scalar functions in series of spherical harmonics $Y_{lm}(\theta, \phi)$: 

\begin{align}
    \Phi(t, r, \theta, \phi) = \frac{1}{r}\sum_{l,m} \Phi_{lm}(r,t) Y_{lm} (\theta, \phi)\notag\\
    \Psi(t, r, \theta, \phi)= \frac{1}{r}\sum_{l,m} \Psi_{lm}(r,t) Y_{lm} (\theta, \phi),
\end{align}
where $l=1,..l_\mathrm{max}$ is the degree and $m=-l, ..., l$ is the order of the multipole.

The initial configuration of the magnetic field is fixed by choosing a set of spherical harmonics. In this way, for example, we can construct a general dipolar field by choosing $\Phi_{lm}, \Psi_{lm} \ne 0$ for $l = 1$ and $\Phi_{lm} = \Psi_{lm} = 0$ for $l\ne 1$. 

Regarding the radial dependence of the coefficients $\Phi_{lm}$ and $\Psi_{lm}$, we impose the radial profile of the dipolar poiloidal scalar function $\Phi_{l=1, m}(r)$ as in eq. (8) of \citet{Aguilera2008} (eq. B9 of \citep{Dehman2023_MATINS}):
\begin{equation}
    \Phi_{l=1, m}(r) = \Phi_0 \mu r[a + \tan (\mu R)b]
\end{equation}
where $\Phi_0$ is a normalization and the coefficients $a$ and $b$ are:
\begin{align}
    &a = \frac{\sin (\mu r)}{(\mu r^2)} - \frac{\cos{\mu r}}{\mu r}\notag\\
    &b = -\frac{\cos(\mu r)}{(\mu r)^2} - \frac{\sin (\mu r)}{\mu r}
\end{align}
and $\mu$ is a parameter related to the curvature of the field, calculated for a given stellar radius $R$. This choice of radial dependence allows for a smooth match with the external potential boundary condition. On the other hand, the toroidal field and the $l>1$ poloidal field multiples are confined inside the crust of the star, with the following radial dependence: 

\begin{align}
    &\Phi_{l>1,m}(r) = -\phi_{lm}(R -r)^2(r - R_\mathrm{core})^2\notag \\
    &\Psi_{lm}(r) = -\psi_{lm}(R -r)^2(r - R_\mathrm{core})^2
\end{align}

where $\phi_{lm}$ and $\psi_{lm}$ are normalization input parameters that are set at the beginning of the simulation. 

Finally, the poloidal component of the field is normalized to the maximum (absolute) value of the radial field at the surface and rescaled by the input parameter $B_{\rm pol,init}$, while the toroidal component is normalized to the average root mean square within the volume and rescaled by the input parameter $B_{\rm tor,init}$. In this way the input parameters $B_{\rm pol,init}$, $B_{\rm tor,init}$, $\phi_{lm}$ and $\psi_{lm}$ determine the configuration and the magnitude of the magnetic field in \emph{MATINS}. In Table \ref{tab:Bfield_config_input} we report the input parameters defining the configuration of our simulations. The names of the simulations are the same as the ones reported in Table \ref{tab:Bfield_models} .

\begin{table*}
    \centering
    \begin{tabular}{|c|c|c|l|l|}
    \hline
    Name & $bpol\_init$ [$10^{12}$ G]& $btor\_init$ [$10^{12}$ G] & $\phi_{lm}$  & $\psi_{lm}$   \\
    \hline
    \hline
        Sly4-M1.4-B14-L1 & 100 & 0 & \makecell[l]{$1  \qquad \mathrm{for}\, l=1, m=0$\\ $0 \qquad \mathrm{otherwise}$} & $0 \qquad \forall\, (l, m)$   \\
    \hline
 Sly4-M1.6-B14-L1 & 100 & 0 & \makecell[l]{$1  \qquad \mathrm{for}\, l=1, m=0$\\ $0 \qquad \mathrm{otherwise}$} & $0 \qquad \forall\, (l, m)$  \\
    \hline
    SLy4-M1.4-B14-L2 & 10 & 1000 & \makecell[l]{$0.5 \qquad \mathrm{for}\,l=1, m=-1,0,1$\\  $10 \qquad \mathrm{for}\,l=2, m=-2,1,0,1,2$ \\ $0 \qquad \mathrm{otherwise}$ }  & \makecell[l]{$1 \qquad \mathrm{for}\,l=1, m=-1,0,1$ \\ $1 \qquad \mathrm{for}\,l=2, m=-2, -1,0,1, 2$ \\ $0 \qquad \mathrm{otherwise} $} \\
    \hline
        SLy4-M1.4-B14-L5& 10 & 1000 & \makecell[l]{$0.5 \qquad \mathrm{for}\,l=1, m=-1,0,1$ \\ $10 \qquad \mathrm{for}\,l=2, m=-1,0,1,2$ \\ $10 \qquad \mathrm{for}\,l=3, m=-1,0,1,2,3$\\ $10 \qquad \mathrm{for}\,l=5, m=-1,0,1,2,3$ \\ $0 \qquad \mathrm{otherwise}$}  & \makecell[l]{$1 \qquad \mathrm{for}\,l=1, m=0,1$ \\ $1 \qquad \mathrm{for}\,l=2, m=0,1, 2$ \\ $1 \qquad \mathrm{for}\,l=3, m=0,1,2,3$ \\ $1 \qquad \mathrm{for}\,l=5, m=0,1,2,3$ \\ $0 \qquad 
        \mathrm{otherwise}$}\\
    \hline
        SLy4-M1.4-B14-L10& 10 & 1000 & \makecell[l]{$0.5 \qquad \mathrm{for}\,l=1, m=-1,0,1$ \\ $10 \qquad \mathrm{for}\,l=2, m=-2, -1,0,1,2$ \\ $10 \qquad \mathrm{for}\,l=9, m=-7,-2,0,2,5$\\ $10 \qquad \mathrm{for}\,l=10, m=-6,-5,0,3, 10$ \\ $0 \qquad \mathrm{otherwise}$} & \makecell[l]{$1 \qquad \mathrm{for}\,l=1, m=-1,0,1$ \\ $1 \qquad \mathrm{for}\,l=2, m=-2,-1,0,1,2$ \\ $1 \qquad \mathrm{for}\,l=10, m=-1,0,1$ \\ $0 \qquad \mathrm{otherwise}$}\\
         \hline
        SLy4-M1.4-B14-L2-alt & 30 & 100 & \makecell[l]{$1 \qquad \mathrm{for}\,l=1, m=-1,0,1$\\  $10 \qquad \mathrm{for}\,l=2, m=-2,1,0,1,2$ \\ $0 \qquad \mathrm{otherwise}$ } & \makecell[l]{$1 \qquad \mathrm{for}\,l=1, m=-1,0,1$ \\ $10 \qquad \mathrm{for}\,l=2, m=-2, -1,0,1, 2$ \\ $0 \qquad \mathrm{otherwise} $} \\
    \hline
    \end{tabular}
    \caption{This table includes the input parameters specifying the initial magnetic field configuration for each of the simulations already presented in Table \ref{tab:Bfield_models}. The columns represent the name of the simulation, the poloidal and toroidal normalization parameter, the value of the weight $\phi_{lm}$ of the spherical harmonic $\Phi_{lm}$ for the poloidal function, and the value of the weight $\psi_{lm}$ of the spherical harmonic $\Psi_{lm}$ for the toroidal function.}
    \label{tab:Bfield_config_input}
\end{table*}

\section{Simulation with a different Equation of State}
\label{sec:different_EOS}

\begin{figure*}
    \centering
    \includegraphics[width = \textwidth]{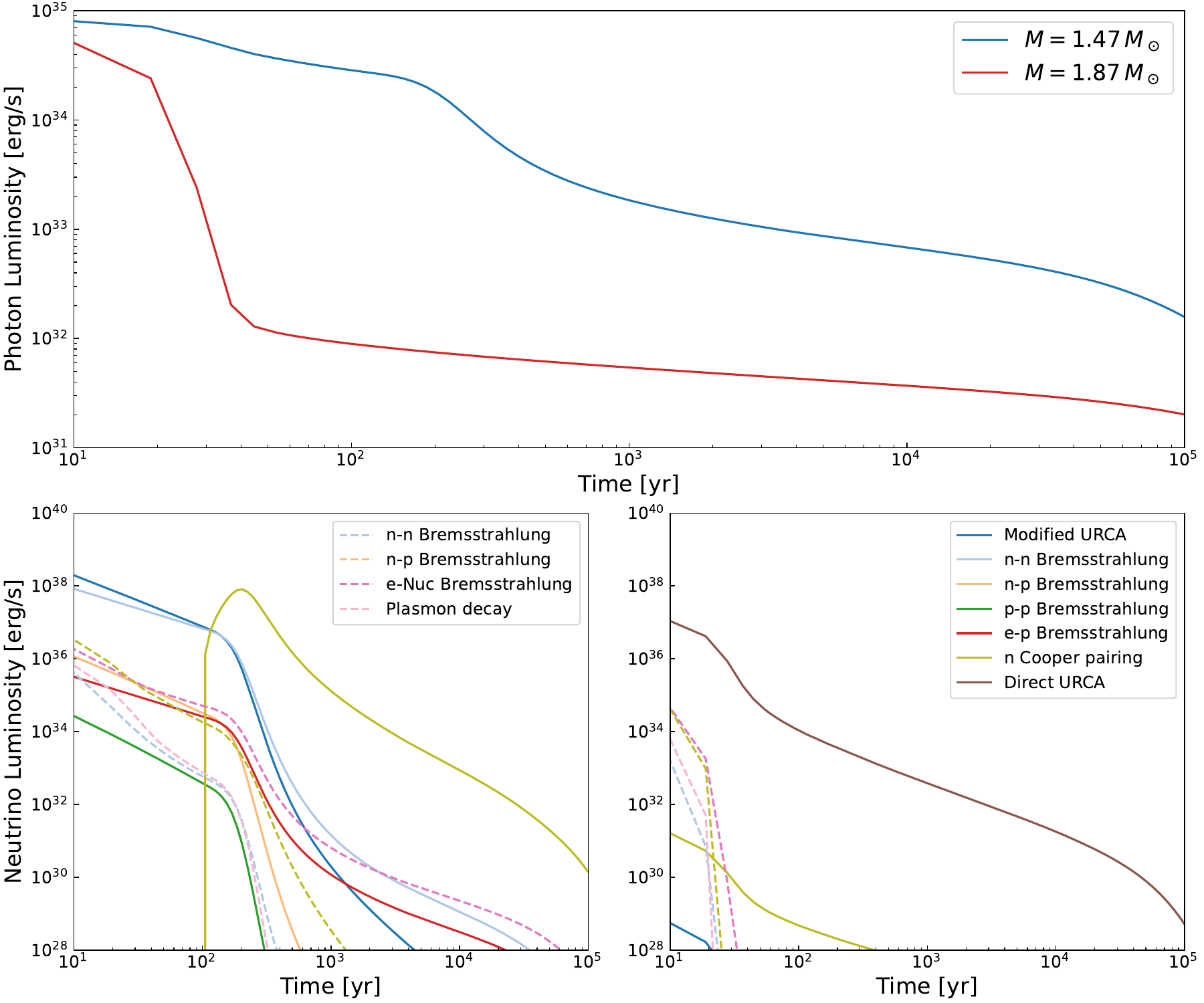}
    \caption{Comparison between two setups characterized by the same \ac{EOS} \emph{BSk24} \citep{Goriely2013} but different NS mass. \emph{Top:} Cooling curves. The blue curve represents the $M = 1.47 \, M_\odot$ setup, the orange one the $M = 1.87 \, M_\odot$ setup. \emph{Bottom left:} Luminosities of the different neutrino emission processes for the $M = 1.47 \, M_\odot$ case. Continuous lines represent processes occurring in the core, and dashed lines represent processes occurring in the crust. \emph{Bottom right:} same as the bottom left pannel, for the $M = 1.87 \, M_\odot$ case.}
    \label{fig:mass_comparison}
\end{figure*}

In this Appendix we present two simulations of a non-magnetized star using the BSk24 \ac{EOS} \citep{Goriely2013}: \emph{BSk24-M1.4-B0} and \emph{BSk24-M1.8-B0}, characterized by $M= 1.47\,M_\odot$ and $M= 1.87\,M_\odot$, respectively. Since these simulations do not present any magnetic field, we adopted for these runs a reduced angular resolution $N_a = 7$, while $N_r = 30$. The aim of this run is to show how with an appropriate \ac{EOS} increasing the mass above a threshold value allows for the activation of Direct URCA processes, which leads to more efficient cooling. This result is shown in Figure \ref{fig:mass_comparison}. In the top panel, the secular evolution of photon luminosity is reported. The blue line represents the case \emph{BSk24-M1.4-B0}, while the orange line represents the case \emph{BSk24-M1.8-B0}. The bottom left panel represents the neutrino luminosity for the case \emph{BSk24-M1.4-B0}, while the bottom right panel is for the case BSk24-M1.8-B0. Solid curves represent processes located in the core, while dashed lines represent those ones occurring in the crust of the star (note that the same process can occur both in the core and in the crust, such as n-n Bremsstrahlung). From Fig. \ref{fig:mass_comparison} we can appreciate how the direct URCA processes are the main contributor to the neutrino luminosity in the \emph{BSk24-M1.8-B0} run, while it is completely absent both in the \emph{BSk24-M1.4-B0} and in the \emph{Sly4-M1.6-B14-L1} case in Fig. \ref{fig:cooling_curve_benchmark} that will be discussed next.  
It is worth noting that in this case most of the energy is radiated by the process within the first years of the life of the \ac{NS}. In contrast, in \emph{BSk24-M1.4-B0}, the dominant mechanism is the modified URCA, which radiates a much smaller amount of energy. 
The enhanced efficiency of the direct URCA with respect to the modified URCA reflects in the photon luminosity, which drops by more than 2 orders of magnitude with respect to the lower mass case before the first $100\,\mathrm{yr}$ of evolution, in agreement with our current knowledge about \ac{NS} cooling \citep{PageApplegate1992}.

\bsp	
\label{lastpage} 
\end{document}